\pdfoutput=1

\documentclass[11pt,twoside,a4paper,cmspaper,final,collab]{cms-tdr}
\def\svnVersion{485217P}\def\cmsCernNoTag{CERN-EP-2018-060}\def\cmsCernDate{\today}\def\cmsMessage{Published in the Journal of High Energy Physics as \href{http://dx.doi.org/10.1007/JHEP11(2018)185}{\doi{10.1007/JHEP11(2018)185.}}}

\begin{document}\cmsNoteHeader{HIG-16-040}

\hyphenation{had-ron-i-za-tion}
\hyphenation{cal-or-i-me-ter}
\hyphenation{de-vices}
\RCS$HeadURL: svn+ssh://svn.cern.ch/reps/tdr2/papers/HIG-16-040/trunk/HIG-16-040.tex $
\RCS$Id: HIG-16-040.tex 475524 2018-09-19 22:58:01Z ferrif $

\newlength\cmsFigWidth
\ifthenelse{\boolean{cms@external}}{\setlength\cmsFigWidth{0.85\columnwidth}}{\setlength\cmsFigWidth{0.4\textwidth}}
\ifthenelse{\boolean{cms@external}}{\providecommand{\cmsLeft}{top\xspace}}{\providecommand{\cmsLeft}{left\xspace}}
\ifthenelse{\boolean{cms@external}}{\providecommand{\cmsRight}{bottom\xspace}}{\providecommand{\cmsRight}{right\xspace}}

\cmsNoteHeader{HIG-16-040}
\title{Measurements of Higgs boson properties in the diphoton decay channel in proton-proton collisions at \texorpdfstring{$\sqrt{s} = 13\TeV$}{sqrt(s) = 13 TeV}}

\date{\today}

\newcommand{\m}{\ensuremath{m}} 
\newcommand{\mgg}{\ensuremath{{\m}_{\Pgg\Pgg}}\xspace}
\newcommand{\mH}{\ensuremath{{\m}_{\PH}}\xspace}
\newcommand{\mHhat}{\ensuremath{\widehat{{\m}}_{\PH}}\xspace}
\newcommand{\muhat}{\ensuremath{\widehat{\mu}}}
\newcommand{\Zee}{\ensuremath{{\PZ\to\EE}}\xspace}
\newcommand{\ttH}{\ensuremath{\ttbar\PH}\xspace}
\newcommand{\cttH}{\ensuremath{\cPqt\cPqt\PH}\xspace}

\newcommand{\ptg}{\ensuremath{\pt^\Pgg}}
\newcommand{\ptgg}{\ensuremath{\pt^{\Pgg\Pgg}}}
\newcommand{\ptgo}{\ensuremath{\pt^{\Pgg 1}}}
\newcommand{\ptgt}{\ensuremath{\pt^{\Pgg 2}}}
\newcommand{\ptvi}{\ensuremath{{\ptvec}^{\,i}}}
\newcommand{\ptvgg}{\ensuremath{{\ptvec}^{\,\Pgg\Pgg}}}

\newcommand{\kappaf}{\ensuremath{\kappa_{\mathrm{f}}}}
\newcommand{\kappaV}{\ensuremath{\kappa_{\mathrm{V}}}}
\newcommand{\kappagamma}{\ensuremath{\kappa_{\Pgg}}}
\newcommand{\kappagluon}{\ensuremath{\kappa_{\mathrm{g}}}}

\newcommand{\ggH}{\ensuremath{\Pg\Pg\PH}\xspace}
\newcommand{\VH}{\ensuremath{\mathrm{V}\PH}\xspace}
\newcommand{\WH}{\ensuremath{\PW\PH}\xspace}
\newcommand{\ZH}{\ensuremath{\PZ\PH}\xspace}
\newcommand{\VBF}{\ensuremath{\mathrm{VBF}}\xspace}

\newcommand{\Ich}{\ensuremath{\mathcal{I}_{\text{ch}}}\xspace}
\newcommand{\alphas}{\ensuremath{\alpha_\mathrm{s}}}

\abstract{Measurements of Higgs boson properties in the $\HGG$ decay channel are reported. The analysis is based on data collected by the CMS experiment in proton-proton collisions at $\sqrt{s}=13\TeV$ during the 2016 LHC running period, corresponding to an integrated luminosity of 35.9\fbinv. Allowing the Higgs mass to float, the measurement yields a signal strength relative to the standard model prediction of $1.18^{+0.17}_{-0.14}=1.18\ ^{+0.12}_{-0.11}\stat ^{+0.09}_{-0.07}\syst ^{+0.07}_{-0.06}\thy$, which is largely insensitive to the exact Higgs mass around 125\GeV. Signal strengths associated with the different Higgs boson production mechanisms, couplings to bosons and fermions, and effective couplings to photons and gluons are also measured.}

\hypersetup{%
pdfauthor={CMS Collaboration},%
pdftitle={Measurements of Higgs boson properties in the diphoton decay channel in proton-proton collisions at sqrt(s) = 13 TeV},%
pdfsubject={CMS},%
pdfkeywords={CMS,physics,Higgs,diphoton}}

\maketitle

\section{Introduction}
\label{sec:intro}

The standard model of particle physics
(SM)~\cite{SheldonL1961579,PhysRevLett.19.1264,SalamNobel}
has been very successful in explaining the interactions between
elementary particles. During the Run~1 period (2010-2012) of the CERN
LHC, with proton-proton collisions at centre-of-mass energies of $7$ and $8\TeV$,
a new particle was discovered by the ATLAS~\cite{Aad:2012tfa} and
CMS~\cite{Chatrchyan:2012xdj,Chatrchyan:2013lba} Collaborations.
The discovery was followed by a comprehensive set of studies of the
properties of this new boson in the decay channels and production modes
accessible with the LHC Run~1 data set. Measurements from ATLAS and
CMS~\cite{ref:2015zhl,ref:2016vau} have shown that the properties
of the new boson are consistent with expectations for the SM Higgs
boson~\cite{Englert:1964et,Higgs:1964ia,Higgs:1964pj,Guralnik:1964eu,Higgs:1966ev,Kibble:1967sv}.

Despite the small branching fraction predicted by the SM (${\approx}0.2\%$),
the $\HGG$ decay channel provides a clean final
state with an invariant mass peak that can be reconstructed with high
precision. As a consequence, $\HGG$ was one of the most important
channels for the discovery of the Higgs boson and first
measurements of its properties~\cite{Khachatryan:2014ira,Aad:2014eha}.
In Run~2, with proton-proton collisions at $\sqrt{s} = 13\TeV$, this
channel remains one of the most sensitive to continue the precise
characterization of the Higgs boson.

In this paper, measurements of the Higgs boson production rates with respect
to the SM prediction (signal strength modifiers) are presented, along with
measurements of the coupling modifiers to fermions and bosons, and
effective coupling modifiers to photons and gluons, in the so-called
$\kappa$ framework~\cite{Heinemeyer:2013tqa}. Improved precision on
these parameters constrains possible deviations in the Higgs sector of
the SM.
The analysis is based on proton-proton collision data collected at
$\sqrt{s} = 13\TeV$ by the CMS experiment in 2016, corresponding to an
integrated luminosity of 35.9\fbinv.

\section{The CMS detector}
\label{sec:detector}

The central feature of the CMS apparatus is a superconducting solenoid,
13\unit{m} in length and with an inner diameter of 6\unit{m}, which
provides an axial magnetic field of 3.8\unit{T}.
Within the solenoid volume are a silicon pixel and strip tracker, a
lead tungstate crystal electromagnetic calorimeter (ECAL), and a brass
and scintillator hadron calorimeter (HCAL), each composed of a barrel
and two endcap sections. Forward calorimeters extend the pseudorapidity
($\eta$) coverage provided by the barrel and endcap detectors. Muons are
detected in gas-ionization chambers embedded in the steel flux-return
yoke outside the solenoid.

Charged-particle trajectories are measured by the silicon pixel and
strip tracker, with full azimuthal coverage within $\abs{\eta} < 2.5$.
The ECAL and HCAL surround the tracking
volume and cover the region $\abs{\eta} < 3.0$.
The ECAL barrel extends to $\abs{\eta} < 1.48$, while the endcaps cover the
region $1.48 < \abs{\eta} < 3.0$.
A lead/silicon-strip preshower detector is located in front of the
ECAL endcap in the region $1.65 < \abs{\eta} < 2.6$.
The preshower detector includes two planes of silicon sensors measuring
the $x$ and $y$ coordinates of the impinging particles.
A steel/quartz-fibre Cherenkov forward calorimeter extends the
calorimetric coverage to $\abs{\eta} < 5.0$.
In the region $\abs{\eta} < 1.74$, the HCAL cells have widths of 0.087 in
both pseudorapidity and azimuth ($\phi$).
In the $(\eta, \phi)$ plane, and for $\abs{\eta} < 1.48$, the HCAL cells map
on to $5{\times}5$
ECAL crystal arrays to form calorimeter towers projecting radially
outwards from points slightly offset from the nominal interaction point.
In the endcap, the ECAL arrays matching the HCAL cells contain fewer
crystals. The calibration of the ECAL uses the azimuthal symmetry of the
energy flow in minimum-bias events, $\pi^{0}\to\Pgg\Pgg$, $\eta\to\Pgg\Pgg$,
$\PW\to\Pe\nu$, and $\Zee$ decays.
Changes in the response of the ECAL crystals due to irradiation during
the LHC running periods and their subsequent recovery are monitored
continuously and corrected for, using light injected from a laser
system. More details on the methods employed are given in Ref.~\cite{Chatrchyan:2013dga}.

The global event reconstruction algorithm, also called particle-flow event
reconstruction~\cite{Sirunyan:2017ulk}, attempts to reconstruct and identify
individual particles using an optimized combination of information
from the various elements of the CMS detector. The energy of photons is
directly obtained from the ECAL measurement with a procedure described in
greater detail in Section~\ref{sec:phoene}.
The energy of electrons is determined from a combination of the electron
momentum at the primary interaction vertex as determined by the tracker,
the energy of the corresponding ECAL cluster, and the energy sum of all
bremsstrahlung photons spatially compatible with originating from the
electron track.
The energy of muons is obtained
from the curvature of the corresponding track.
The energy of charged hadrons is determined from a combination of their
momentum measured in the tracker and the matching ECAL and HCAL energy
deposits, corrected for zero-suppression effects and for the response
function of the calorimeters to hadronic showers.
Finally, the energy of neutral hadrons is obtained from the
corresponding corrected ECAL and HCAL energy.

Hadronic jets are clustered from these reconstructed particles using
the infrared- and collinear-safe anti-$\kt$
algorithm~\cite{Cacciari:2008gp}, with a distance parameter of 0.4.
The jet momentum is determined as the vectorial sum of all particle momenta
in the jet.
An offset correction is applied to jet energies to take into account the
contribution from additional proton-proton interactions within the
same or nearby bunch crossings. Jet energy corrections are derived
from simulation, and are confirmed with in situ measurements of the
energy balance in dijet, multijet, photon $+$ jet, and leptonically
decaying $\PZ + \text{jets}$ events~\cite{Khachatryan:2016kdb}.
The jet momentum is found from simulation to be within 5 to 10\% of the
true momentum over the entire jet transverse momentum (\pt) spectrum and
detector acceptance.
Additional selection criteria are applied to each event to remove
spurious jet-like features originating from isolated noise patterns in
certain HCAL regions.

To identify jets originating from the hadronization of bottom quarks,
the combined secondary vertex (CSV) $\PQb$ tagging algorithm is
used~\cite{Chatrchyan:2012jua, Sirunyan:2017ezt}. The algorithm tags jets from
\cPqb\ hadron decays by their displaced decay vertex,
providing a numerical discriminant value that is higher for
jets likely to be initiated by \cPqb\ quarks.
Two tagging algorithm working points, medium and loose, are used
in this analysis: the medium (loose) point provides an efficiency
for identifying \cPqb quark jets of about 70\%\,(85\%) and a
misidentification probability for jets from light quarks and gluons of
about 1\%\,(10\%).

The missing transverse momentum vector is taken as the negative vector sum of
all reconstructed particle candidate transverse momenta in the event
reconstruction, and its magnitude is referred to as $\ptmiss$.

A more detailed description of the CMS detector, together with a
definition of the coordinate system used and the relevant kinematic
variables, can be found in Ref.~\cite{Chatrchyan:2008zzk}.

\section{Analysis strategy}
\label{sec:strategy}

The dominant Higgs boson production mechanism in proton-proton
collisions is gluon-gluon fusion (\ggH), with additional contributions from
vector boson fusion (\VBF), and production in association with a vector boson
(\VH) or with a top quark pair (\ttH).

To maximise the sensitivity of the analysis, specific production modes
with reduced background contamination are targeted. Events are
categorized by requiring specific features in the final state: forward
jets for \VBF, top decay products such as muons, electrons, missing
transverse energy from neutrinos, jets arising from the hadronization of
\cPqb\ quarks for \ttH, and vector-boson decay products such as muons,
electrons, missing transverse energy, or dijets with a characteristic
invariant mass for \VH production.
The events with no specific features, mostly coming from \ggH, are
categorized according to their expected probability to be signal rather
than background.

Several multivariate techniques are used in the analysis. An initial set
is used to improve the event reconstruction, and particularly the photon
energy estimate, the photon identification, the identification of the
diphoton primary vertex and the estimate of its probability of being the
true diphoton vertex. In the subsequent steps of the analysis, the event
classification benefits from multivariate techniques to categorize \ggH
events, to enhance the identification of forward jets in \VBF events and the
separation of such events from \ggH events, to enhance the \cPqb\ tagging and
the separation of \ttH jets in events with multiple jets.

Measurements are extracted by a simultaneous maximum-likelihood fit
to the diphoton invariant mass distributions in all event categories.
Simulated samples are used to derive the signal model, while the
background is obtained from the fit to the data. The latter aspect is
particularly important, as it makes the use of simulated samples
only relevant to the optimization of the multivariate classifiers used
in the different steps of the analysis. While imperfect simulation
might induce suboptimal performance, the use of multivariate inputs
uncorrelated with the diphoton invariant mass ensures that no bias is
introduced. The impact of the choice of the event generator on the
multivariate discriminators has also been checked and found to be
negligible.

\section{Data sample and simulated events}
\label{sec:samples}

The events used in this analysis were selected by diphoton triggers with
asymmetric transverse energy ($\ET$) thresholds of 30 and 18\GeV.
The trigger selection requires a loose calorimetric identification using
the shape of the electromagnetic showers, a loose isolation requirement,
and a selection on the ratio of the HCAL and ECAL deposits of the photon
candidates.
The $\RNINE$ shower shape variable is used in the trigger to identify
photons that convert to an $\EE$ pair in the tracker material before
reaching the ECAL surface.
The $\RNINE$ variable is defined as the energy sum of the $3\times3$
crystals centred on the most energetic crystal in the candidate
electromagnetic cluster divided by the energy of the candidate. The
electromagnetic showers from photons that convert before reaching the
calorimeter have wider transverse profiles and lower values of $\RNINE$
than those of unconverted photons.
The trigger efficiency is measured from $\Zee$ events using the
tag-and-probe technique~\cite{tagNprobe}. Efficiencies in simulation are
corrected to match those measured in data.

Simulated signal events are generated using \MGvATNLO v2.2.2 at
next-to-leading order (NLO)~\cite{Alwall:2014hca} in perturbative
quantum chromodynamics (QCD) with FxFx merging~\cite{Frederix:2012ps},
the parton level samples being
interfaced to $\PYTHIA 8.205$~\cite{Sjostrand:2007gs} for parton showering
and hadronization. The CUETP8M1 \PYTHIA underlying event tune parameter set is
used~\cite{Khachatryan:2015pea}. Events produced via the gluon fusion
mechanism are weighted as a function of the Higgs boson \pt and the number of
jets in the event, to match the prediction from the {\textsc{nnlops}}
program~\cite{Hamilton:2013fea}. Parton distribution functions (PDFs)
are taken from the NNPDF3.0~\cite{Ball:2014uwa} set. The signal cross sections and
branching fraction recommended by the LHC Higgs cross section working
group are used~\cite{LHCHXSWG}.

The dominant background to \HGG\ consists of the irreducible prompt
diphoton production, and the reducible backgrounds from $\GAMJET$ and
dijet events where the jets are misidentified as isolated photons.
Background events, used for the trainings of multivariate discriminants
and for category optimization, have been simulated using various event
generators.
The diphoton background is modeled with the \SHERPA
v.2.2.1~\cite{Gleisberg:2008ta} generator. It includes the Born
processes with up to 3 additional jets as well as the box processes at
leading order.
Multijet and \GAMJET\ backgrounds are modeled with $\PYTHIA$, with a
filter applied to enhance the production of jets with a large fraction
of electromagnetic energy.
The $\PW\Pgg$ and $\PZ\Pgg$ samples are generated with
\MGvATNLO at leading order, while Drell--Yan events
are simulated with the same generator at NLO precision.

The detailed response of the CMS detector is simulated using the
\GEANTfour~\cite{Agostinelli:2002hh} package.
This includes the simulation of the multiple proton-proton interactions
taking place in each bunch crossing, referred to as pileup. These
can occur at the nominal bunch crossing (in-time pileup) or at the
crossing of previous and subsequent bunches (out-of-time pileup), and
the simulation accounts for both.
Simulated events are weighted to reproduce the distribution of the
number of interactions in data.
The average number of pileup interactions measured in data amounts to
23, with a root-mean-square (\textsc{rms}) of about 6.

\section{Photon reconstruction and identification}
\label{sec:recoid}

Photon candidates are reconstructed as part of the global
event reconstruction, as described in Section~\ref{sec:detector}.
Photons are identified as ECAL energy clusters not linked to the
extrapolation of any charged-particle trajectory to the ECAL.
The clustering algorithm allows an almost complete collection of the
energy of the photons, even for those converting in the material
upstream of the calorimeter.
First, cluster ``seeds'' are identified as local energy maxima above a
given threshold.
Second, clusters are grown from the seeds by aggregating crystals with
at least one side in common with a clustered crystal and with an energy
in excess of a given threshold. This threshold represents about two
standard deviations of the electronic noise in the ECAL and amounts
to 80\MeV in the barrel and, depending on $\abs{\eta}$, up to 300\MeV in
the endcaps. The energy of each crystal can be shared among adjacent
clusters assuming a Gaussian transverse profile of the electromagnetic
shower.
Finally, clusters are merged into ``superclusters'', to allow good
energy containment, accounting for geometrical variations of the
detector along $\eta$, and optimizing robustness against pileup.

\subsection{Photon energy}
\label{sec:phoene}

The energy of photons is computed from the sum of the energy of the
clustered crystals, calibrated and corrected for changes in the response over
time~\cite{Chatrchyan:2013dga} and considered in the clustering procedure.
The preshower energy is added to that of the superclusters in the region
covered by this detector.
To optimize the resolution, the photon energy is corrected for the
containment of the electromagnetic shower in the superclusters and
the energy losses from converted photons~\cite{Khachatryan:2015iwa}.
The correction is computed with a multivariate regression technique
that estimates simultaneously the energy of the photon and its 
uncertainty.
This regression is trained on simulated photons using as the target
the ratio of the true photon energy and the sum of the energy of the
clustered crystals. The inputs are shower shapes and position variables
-- both sensitive to shower containment and possible unclustered energy --
preshower information, and global event observables sensitive to
pileup.

\begin{figure}[h]
\centering
  \includegraphics[width=0.48\textwidth]{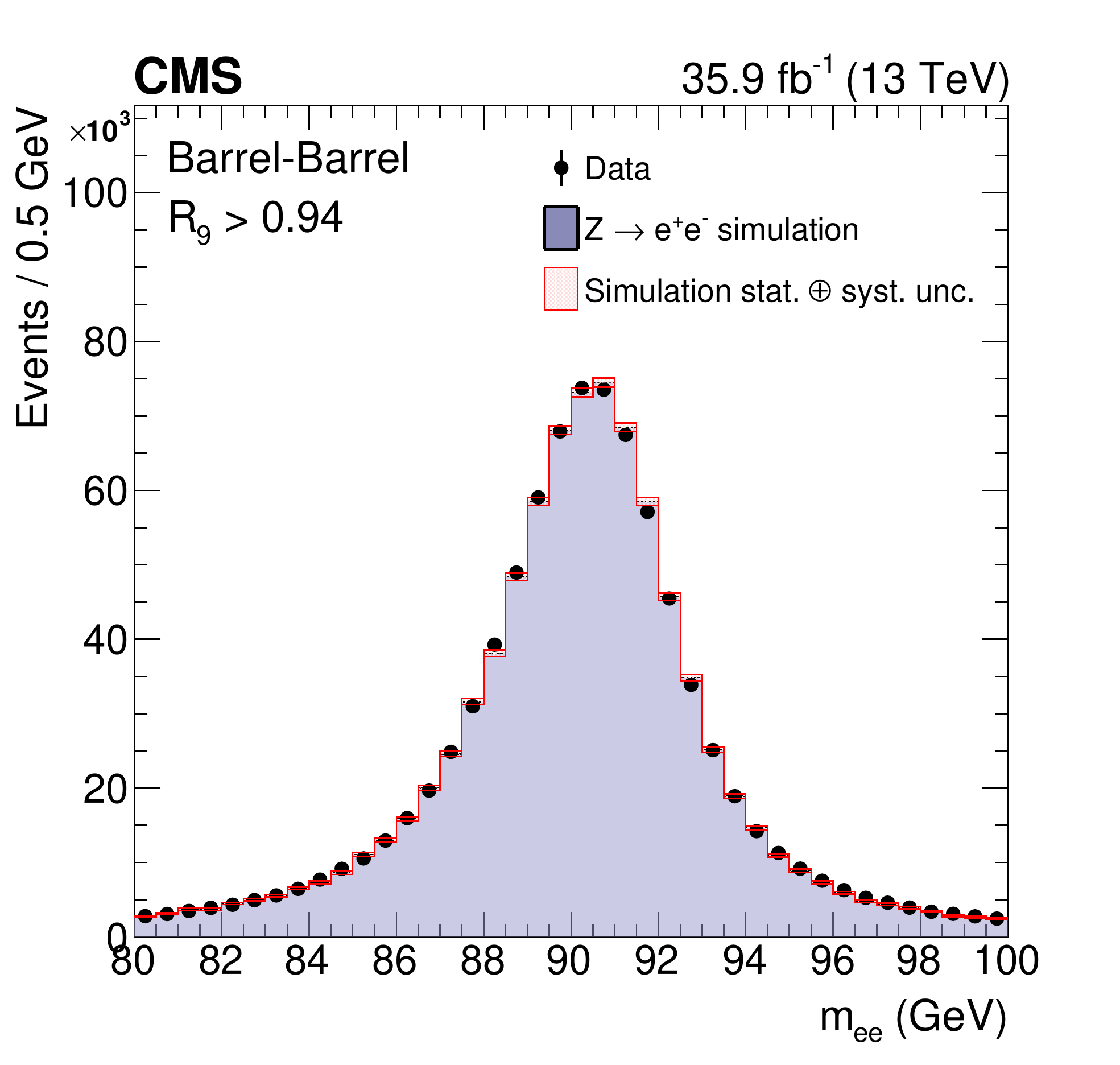}
  \includegraphics[width=0.48\textwidth]{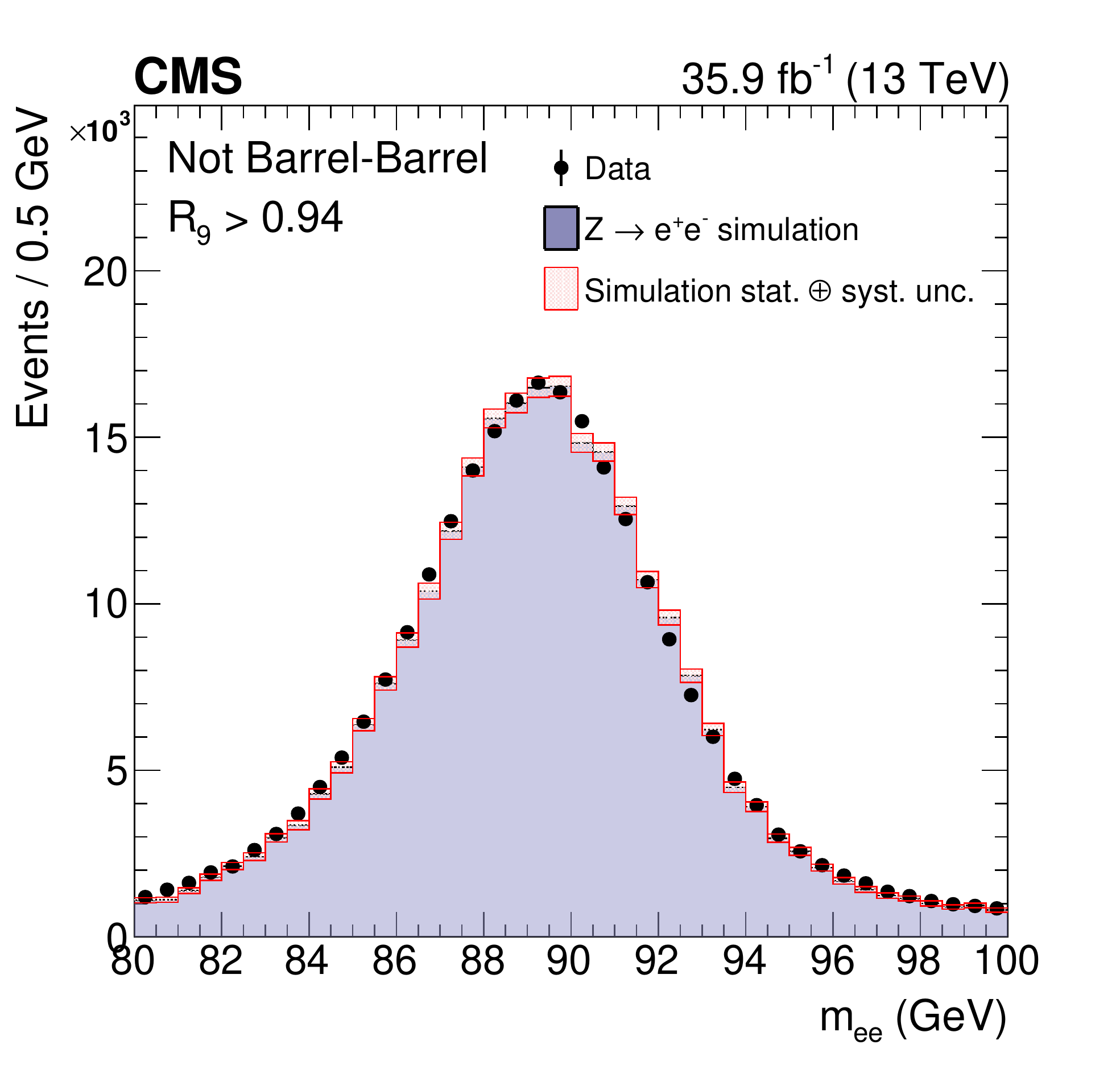}\\
\caption{Comparison of the dielectron invariant mass distributions in data and
  simulation (after energy smearing) for $\Zee$ events where
  electrons are reconstructed as photons.
  The comparison is shown requiring $\RNINE > 0.94$ for both ``photons'' and
  for (left) events with both photons in the barrel, and (right) the
  remaining events.
  The simulated distributions are normalized to the integral of the data
  distribution in the range $87 < m_{\Pe\Pe} < 93\GeV$ to
  highlight the agreement in the bulk of the distributions.}
\label{fig:Zee_mass_cats_diphoBDT}
\end{figure}

A multistep procedure has been implemented to correct the energy scale
in data, and to determine the additional smearing to be applied to the
reconstructed photon energy in simulated events so as to reproduce the energy
resolution observed in data.
First, the energy scale in data is equalized with that in simulated
events, and residual long-term drifts in the response are corrected,
using $\PZ\to\EE$ decays in which the electron showers are reconstructed as
photons. Then, the photon energy resolution predicted by the simulation
is improved by adding a Gaussian smearing determined from the
comparison between the $\PZ\to\EE$ line-shape in data and
simulation (Fig.~\ref{fig:Zee_mass_cats_diphoBDT}).
The corrections to the energy scale are extracted differentially in
time, $\abs{\eta}$ (two categories in the barrel and two in the endcaps) and
$\RNINE$ (two categories). They range from about 0.1 to about 0.3\% in
the barrel and from about 0.2 to about 2\% in the endcap, depending on the category.
The amount of smearing required is extracted differentially in the same $\abs{\eta}$
and $\RNINE$ categories as the energy scale corrections and ranges from about
0.1 to about 2.7\%, depending on the category.

\subsection{Photon preselection}

The photons considered further in this analysis are required to satisfy
preselection criteria similar to, but slightly more stringent than, the
trigger requirements. The preselection requirements consist of:

\begin{itemize}
        \item $\pt^{\Pgg 1} > 30\GeV$ and $\pt^{\Pgg 2} > 20\GeV$, where
                $\pt^{\Pgg 1}$ and  $\pt^{\Pgg 2}$ are the transverse
                momenta of the leading (in $\pt$) and subleading photons,
                respectively;
        \item $\abs{\eta}<2.5$, excluding the barrel-endcap transition
                region $1.44 < \abs{\eta} < 1.57$, where the photon
                energy reconstruction is affected by a suboptimal
                containment of the electromagnetic shower;
         \item a selection on the $\RNINE$ variable and on
               $\sigma_{\eta \eta}$ -- the lateral extension of the
               shower, defined as the energy-weighted spread within the
               $5{\times}5$ crystal matrix centred on
               the crystal with the largest energy deposit in the
               supercluster -- to reject ECAL energy deposits
               incompatible with a single isolated electromagnetic
               shower, such as those coming from neutral mesons;
       \item a selection on the ratio of the energy in the HCAL cells behind
               the supercluster to the energy in the supercluster (H/E),
               to reject hadrons;
       \item  an electron veto, which rejects the photon candidate if its
         supercluster is matched to an electron track with no missing hits
         in the innermost tracker layers;
       \item a requirement on the photon isolation ($\mathcal{I}_{\text{ph}}$),
	       defined as the sum of the transverse energy of the particles
	       identified as photons and falling inside a cone of radius
               ${R=\sqrt{(\Delta\eta)^2 + (\Delta\varphi)^2}}=0.3$
	       around the photon candidate direction;
               the sum is corrected for the contribution of the
               pileup estimated from the median energy density in the
               event~\cite{Cacciari:2011ma};
       \item a requirement on the track isolation in a hollow cone
	       ($\mathcal{I}_{\text{tk}}$), the sum of the transverse momenta
	       of all tracks in a cone of radius $R=0.3$
	       around the photon candidate direction (with tracks in an
	       inner cone of size $R=0.04$ not included in the sum);
               the cone is hollow to use the same isolation definition also
               for electrons;
       \item a loose requirement on charged-hadron isolation
	       ($\Ich$), the sum of the transverse momenta of
	       charged particles inside a cone of radius $R = 0.3$ around the
	       photon candidate; this requirement is added to the one on
               track isolation to match the selection applied to photon
               candidates as part of data reconstruction;
       \item a loose requirement on the photon identification (as described in
               Section~\ref{sec:phoId}).
\end{itemize}

The selection thresholds are reported in
Table~\ref{table:preselcuts}. Additionally, both photons must satisfy
either (a)~$\RNINE > 0.8$ and $\Ich <20\GeV$, or
(b)~$\Ich /\ptg<0.3$.

\begin{table}[htbp]
  \centering
  \newlength\cmsTabSkip\setlength{\cmsTabSkip}{2ex}
    \topcaption{Schema of the photon preselection requirements.}
    \label{table:preselcuts}
    \begin{tabular}{lccccc}
      \hline
      \multicolumn{1}{c}{} & \RNINE & H/E     & $\sigma_{\eta \eta}$ & $\mathcal{I}_{\text{ph}}$ (\GeVns) & $\mathcal{I}_{\text{tk}}$ (\GeVns) \\
      \hline
      \multirow{2}{*}{Barrel}
      & $[0.5, 0.85]$  & $<$0.08 & $<$0.015 & $<$4.0 & $<$6.0
      \\
      & $>$0.85   & $<$0.08 & \NA       & \NA & \NA
      \\[\cmsTabSkip]
      \multirow{2}{*}{Endcaps}
      & $[0.8, 0.90]$  & $<$0.08 & $<$0.035 & $<$4.0 & $<$6.0
      \\
      & $>$0.90     & $<$0.08 & \NA       & \NA & \NA
      \\ \hline
    \end{tabular}
\end{table}

The efficiency of all preselection criteria, except the electron
veto requirement, is measured with a tag-and-probe
technique using $\Zee$ events. The efficiency for
photons to satisfy the electron veto requirement, which cannot be measured
with $\Zee$ events, is obtained from
$\PZ\to\MM\Pgg$ events, in which the photon is produced by final-state
radiation and provides a sample of prompt photons with a purity higher than 99\%.
The photon \pt in this sample ranges from about 20 to about 60\GeV.
The measured efficiency for photons to satisfy the electron veto
requirement has no dependency on the photon \pt within about $\pm 1\%$,
and is well reproduced in simulated events.

Table~\ref{tab:PhotEff} shows the preselection efficiencies
measured in data, $\epsilon_{\text{data}}$, and simulation,
$\epsilon_{\text{MC}}$, along with their ratio
$\epsilon_{\text{data}} / \epsilon_{\text{MC}}$.
Statistical and systematic uncertainties are included both in the
efficiencies and in their ratio. The measured ratios are used to correct
the signal efficiency in simulated signal samples and the associated
uncertainties are propagated to the expected signal yields.

\begin{table}[htbp]
\centering
\topcaption{Photon preselection efficiencies as measured in four photon
categories, obtained with tag-and-probe techniques using $\Zee$ and
$\PZ\to\MM\Pgg$ events. The quoted uncertainties include the
statistical and systematic components.}
\label{tab:PhotEff}
\begin{tabular}{lccc}
\hline
Preselection category
& $\epsilon_{\text{data}}$ (\%) & $\epsilon_{\text{MC}}$ (\%)
& ${\epsilon_{\text{data}}}/{\epsilon_{\text{MC}}}$ \\

\hline
Barrel; $\RNINE>0.85$ & $94.2\pm 0.9$ & $94.7\pm 0.9$ & $0.995\pm 0.001$\\
Barrel; $\RNINE<0.85$ & $82.5\pm 0.7$ & $82.5\pm 0.7$ & $1.000\pm 0.003$\\
Endcap; $\RNINE>0.90$ & $90.1\pm 0.2$ & $91.3\pm 0.1$ & $0.987\pm 0.005$\\
Endcap; $\RNINE<0.90$ & $49.7\pm 1.4$ & $53.8\pm 1.5$ & $0.923\pm 0.010$\\

\hline
\end{tabular}
\end{table}

\subsection{Photon identification}
\label{sec:phoId}

A boosted decision tree (BDT) is used to separate prompt photons from
photon candidates that arise from misidentified jet fragments, but which
satisfy the preselection.
This photon identification BDT is trained using simulated $\GAMJET$
events where prompt photons are considered as signal and non-prompt
photons as background.

The photon identification BDT is trained with the following input variables:

\begin{itemize}
        \item shower shape observables, corrected to mitigate data and
                simulation discrepancies;
        \item isolation variables, $\mathcal{I}_{\text{ph}}$ and
                $\Ich$; two kinds of $\Ich$ are computed,
                including hadrons associated with the chosen primary
                vertex (described in Section~\ref{sec:vtx}), and including hadrons
                associated with the vertex providing the largest
                isolation sum; the latter is effective in rejecting
                misidentified photon candidates originating from jets
                coming from a vertex other than the chosen one;
              \item photon $\eta$ and energy, which are correlated with the
                shower topology and isolation variables;
              \item the median energy density per unit area in the event, $\rho$, to
                minimize the impact of pileup on the above inputs.
\end{itemize}

\begin{figure}[hptb]
\centering
\includegraphics[width=0.48\textwidth]{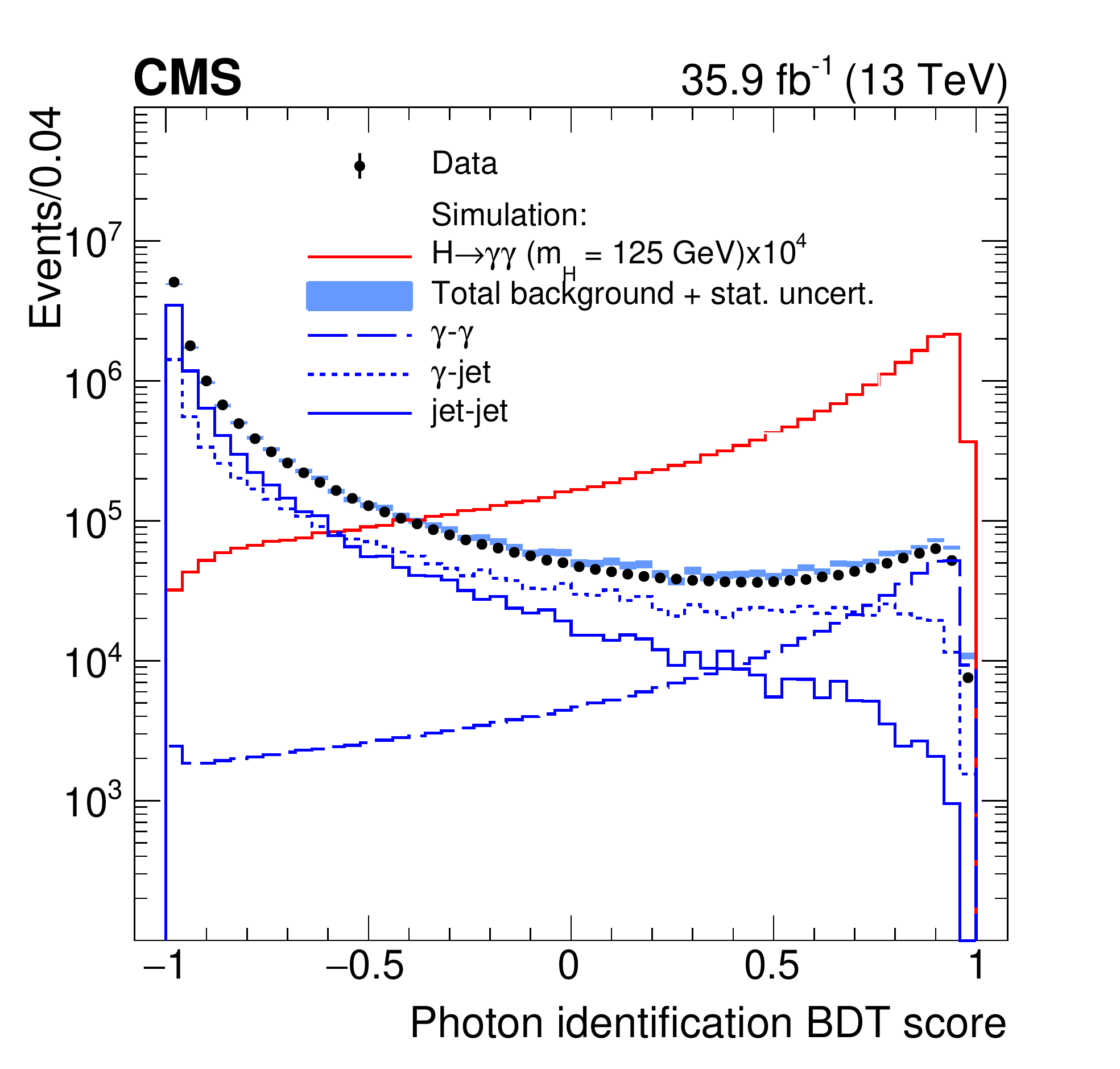}
\includegraphics[width=0.48\textwidth]{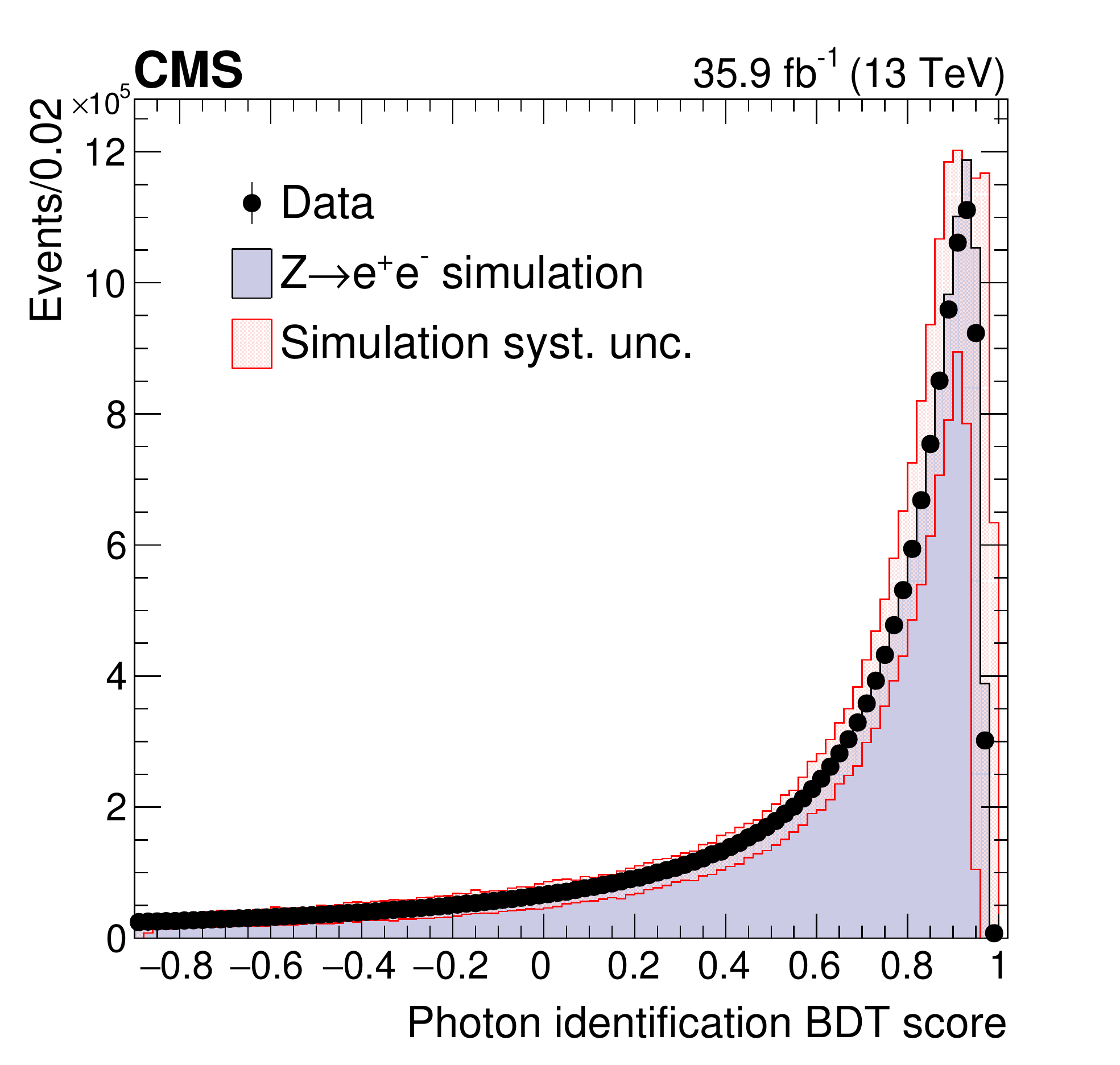}
\caption{(Left) Distribution of the photon identification BDT score of
the lowest scoring photon of diphoton pairs with an invariant mass in
the range $100 < \mgg < 180\GeV$, for events passing the preselection in
the 13~TeV data set (points), and for simulated background events (blue
histogram). Histograms are also shown for different components of the
simulated background. The sum of all background distributions is scaled
up to data. The red histogram corresponds to simulated Higgs boson
signal events.
(Right) Distribution of the photon identification BDT score for
$\PZ\to\EE$ events in data and simulation, where the electrons are
reconstructed as photons. The systematic uncertainty applied to the
shape from simulation (hashed region) is also shown.
}
\label{fig:MVA_DPS}
\end{figure}

Figure~\ref{fig:MVA_DPS} (left) shows the photon identification BDT
score of the lowest-scoring photon from all diphoton pairs with an
invariant mass in the range $100 < \mgg < 180\GeV$, for events passing
the preselection in data and simulated background events.

The photon identification BDT score is also shown in
Fig.~\ref{fig:MVA_DPS} (right) for electrons reconstructed as photons in
$\PZ\to\EE$ events, in data and simulation.
The systematic uncertainty in the photon identification score,
represented by the hashed region, is conservatively assigned to cover
the largest observed discrepancy between data and simulation for
electrons in the ECAL endcaps.

\section{Diphoton vertex}
\label{sec:vtx}

The determination of the primary vertex from which the two photons
originate has a direct impact on the diphoton invariant mass resolution.
If the position along the beam axis ($z$) of the interaction producing
the diphoton is known to better than about 10\mm, the invariant mass
resolution is dominated by the photon energy resolution.
For comparison, the distribution in $z$ of the position of the vertices reconstructed
from the observed tracks has an \textsc{rms} spread of about 3.4\cm.

The diphoton vertex assignment relies on a BDT (the vertex identification BDT)
whose inputs are observables related to tracks
recoiling against the diphoton system:
\begin{itemize}
        \item $\sum_{i}\abs{\ptvi}^{2}$,
        \item $-\sum_{i} {\ptvec}^{\,i}\cdot \ptvgg / \abs{\ptvgg}$,
        \item $(\abs{\sum_{i}\ptvi} - \ptgg) / (\abs{\sum_{i}{\ptvi}} + \ptgg)$,
\end{itemize}
where $\vec{\pt}^{i}$ is the transverse momentum of the $i$th track
associated with a given vertex and $\ptvgg$ is the
transverse momentum of the diphoton system measured with respect to the
same vertex. The sum runs over all charged particle-flow candidates
associated with the given vertex.

In the presence of tracks from photons converted in the tracker
material, two additional input variables are used:
\begin{itemize}
        \item the number of conversions,
        \item the pull $\abs{z_{\text{vtx}} - z_\Pe} /\sigma_{z}$ between the
                longitudinal position of the reconstructed vertex,
                $z_{\text{vtx}}$, and the longitudinal position of the
                vertex estimated using conversion track(s), $z_\Pe$,
                where the variable $\sigma_{z}$ denotes the uncertainty
                in $z_\Pe$.
\end{itemize}

A second vertex-related multivariate discriminant (vertex probability BDT),
used in the diphoton BDT (discussed in Section~\ref{sec:classification}),
is designed to estimate, event-by-event, the probability for the vertex
assignment to be within 10\mm of the diphoton interaction point.
The vertex probability BDT is trained on simulated $\HGG$ events using
the following input variables:
\begin{itemize}
        \item the number of vertices in each event;
        \item the values of the vertex identification BDT score for
                the three most probable vertices in each event;
        \item the distances between the chosen vertex and the second and
                third choices;
        \item the magnitude of the transverse momentum of the diphoton system,
                $\ptgg$;
        \item the number of photons with an associated conversion track.
\end{itemize}

\begin{figure}[hbtp]
\centering
\includegraphics[width=0.5\textwidth]{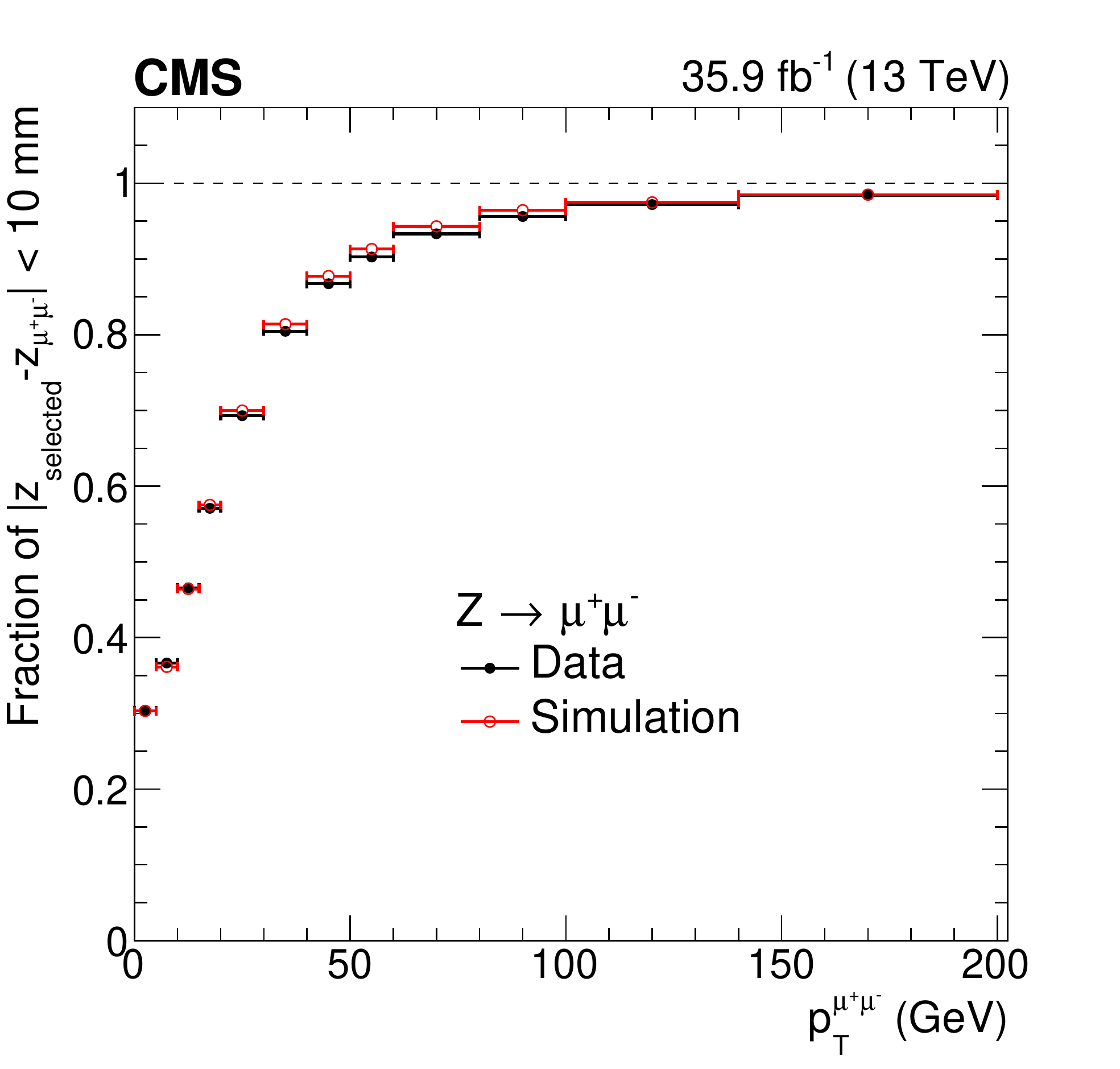}%
\caption{Validation of the $\HGG$ vertex identification algorithm on
        $\PZ\to\MM$ events omitting the muon tracks. Simulated events are
        weighted to match the distributions of pileup and location of
        primary vertices in data.
}
\label{fig:vtxZmumu}
\end{figure}

The performance of the vertex identification BDT is
validated using $\PZ\to\MM$ events (Fig.~\ref{fig:vtxZmumu}), where
the vertices are fitted omitting the muon tracks to mimic a diphoton
system.
In addition, the use of tracks from converted photons to locate the
vertex is validated using $\GAMJET$ events. Discrepancies between data
and simulation are corrected for in the analysis and a corresponding
uncertainty is considered.

In the simulated samples the width of the beam spot
was about a factor 1.5 larger than
what was subsequently observed in data. To correct for this,
simulated events in which the selected vertex is more than 1\mm
away from the generated one are weighted such that the width of the
distribution of the primary vertices is the same as the beam spot
 width in data.

\begin{figure}[hbtp]
\centering
\includegraphics[width=0.5\textwidth]{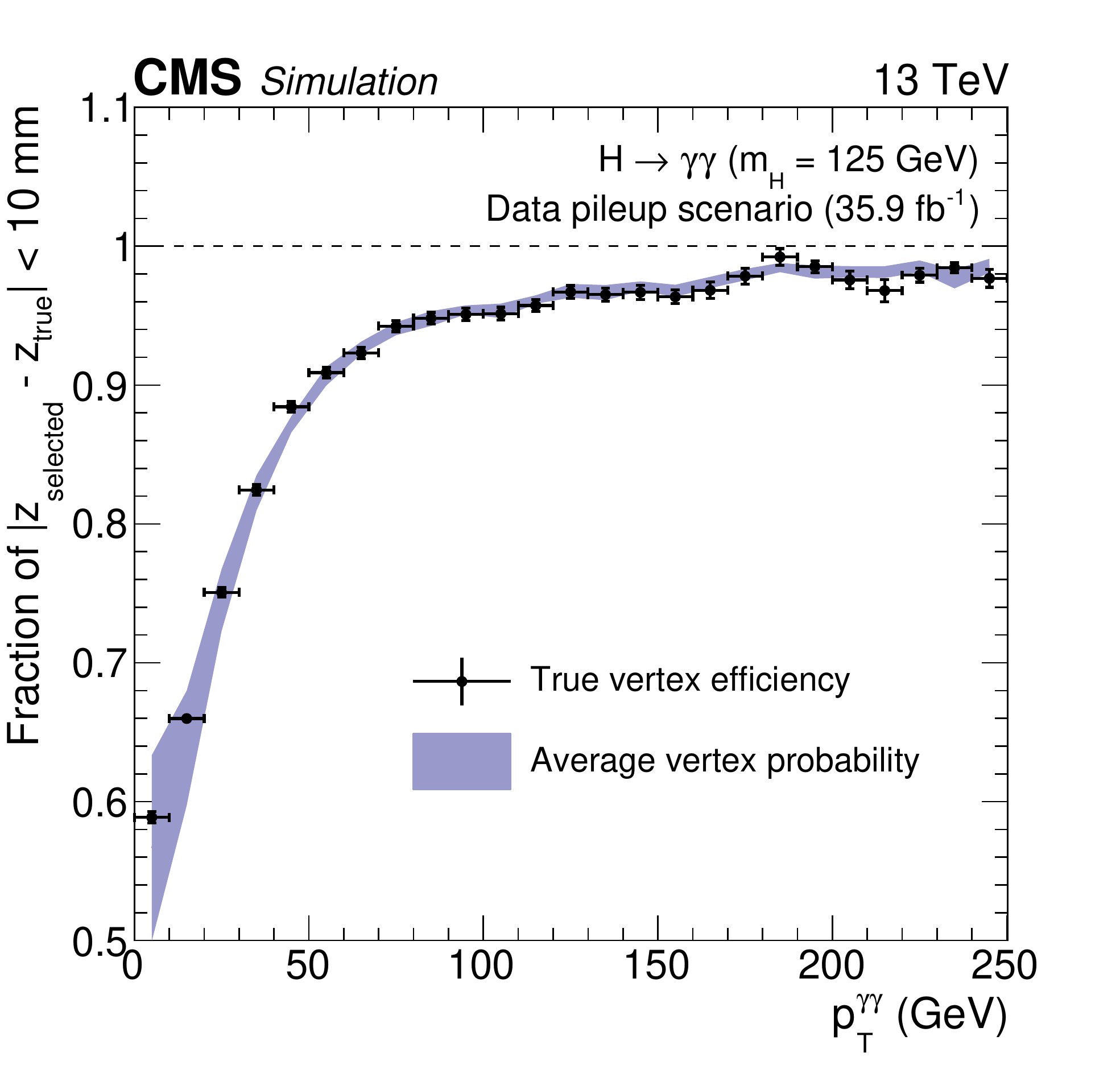}%
\includegraphics[width=0.5\textwidth]{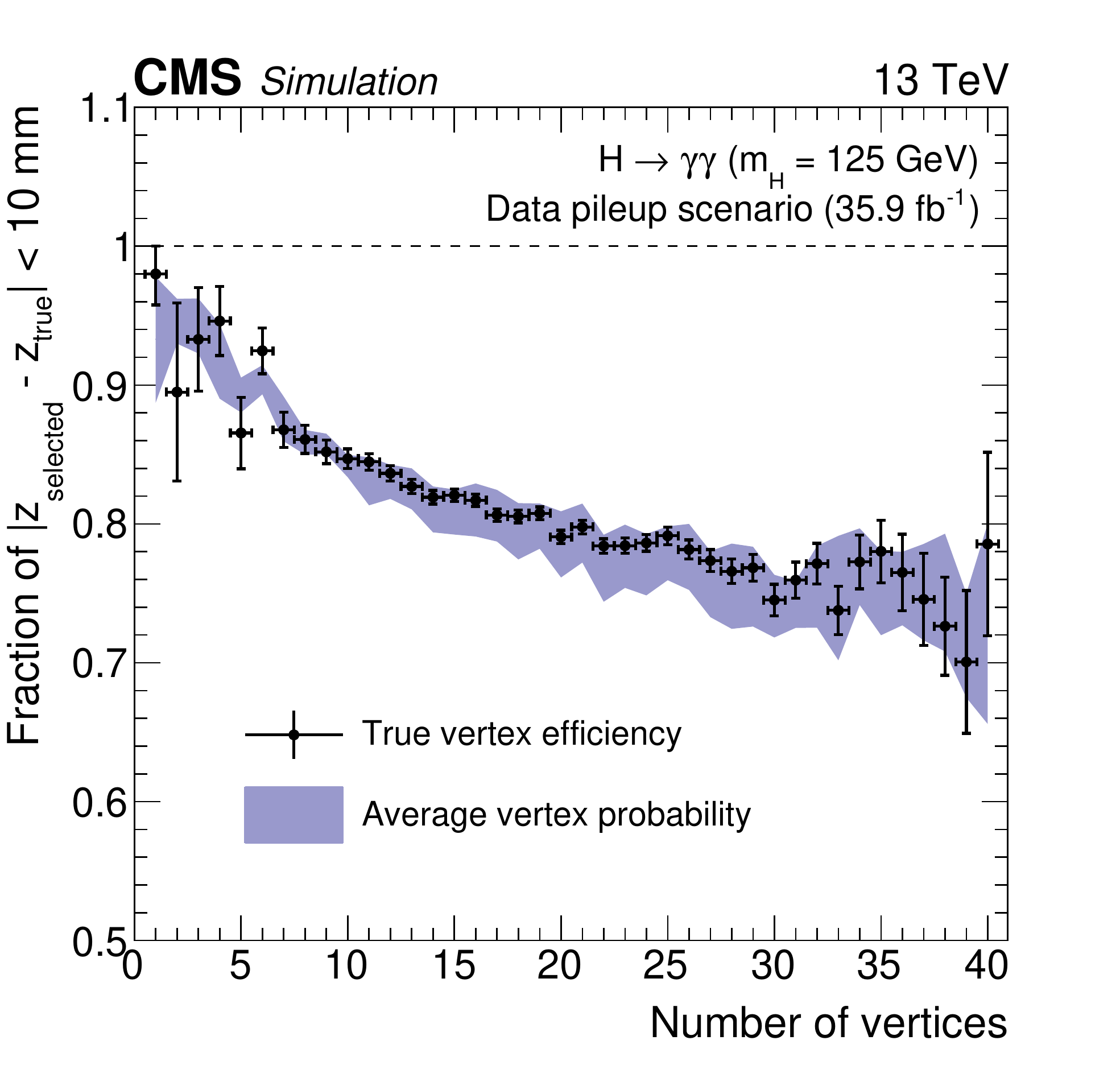}
\caption{Comparison of the true vertex identification efficiency and the
  average estimated vertex probability as a function
  of the reconstructed diphoton \pt (left) and of the number of primary
  vertices (right) in simulated $\HGG$ events with
  $m_{\PH} = 125\GeV$. Events are weighted according to the cross sections
  of the different production modes and to match the distributions of pileup
  and location of primary vertices in data.
}
\label{fig:hhgvtxeff}
\end{figure}

The efficiency of correctly assigning the diphoton vertex to be within
10\mm of the true vertex in \HGG\ simulated events is shown in
Fig.~\ref{fig:hhgvtxeff} as a function of the \pt of the diphoton
pair and as a function of the number of primary vertices in the event,
and compared with the average estimated vertex probability BDT.
The overall efficiency is about 81\%.

\section{Event classification}
\label{sec:classification}

The event selection requires two preselected photon candidates with
$\ptgo > \mgg/3$ and $\ptgt > \mgg /4$, in the mass
range $100 < \mgg < 180\GeV$.
The use of \pt thresholds scaled by $\mgg$ prevents a distortion of
the low end of the invariant mass spectrum.
The requirement on the photon $\pt$ is applied after the vertex assignment.

To improve the sensitivity of the analysis, events are classified
targeting different production mechanisms and according to their mass
resolution and predicted signal-to-background ratio. In each category, the
selections are optimized to maximize the significance of the expected
signal with respect to the background.
As the first step of the classification, exclusive event categories are
defined by dedicated selections on additional reconstructed objects to select
Higgs boson production mechanisms other than \ggH: \VBF, \VH or \ttH.

All objects are reconstructed as described in Section~\ref{sec:detector}
and (for photons) Section~\ref{sec:recoid}.
In addition, electrons are required to be within $\abs{\eta} < 2.5$ and
outside the barrel-endcap transition region. Muons are required to be
within $\abs{\eta} < 2.4$.

\begin{figure}[hptb]
\centering
\includegraphics[width=0.49\textwidth]{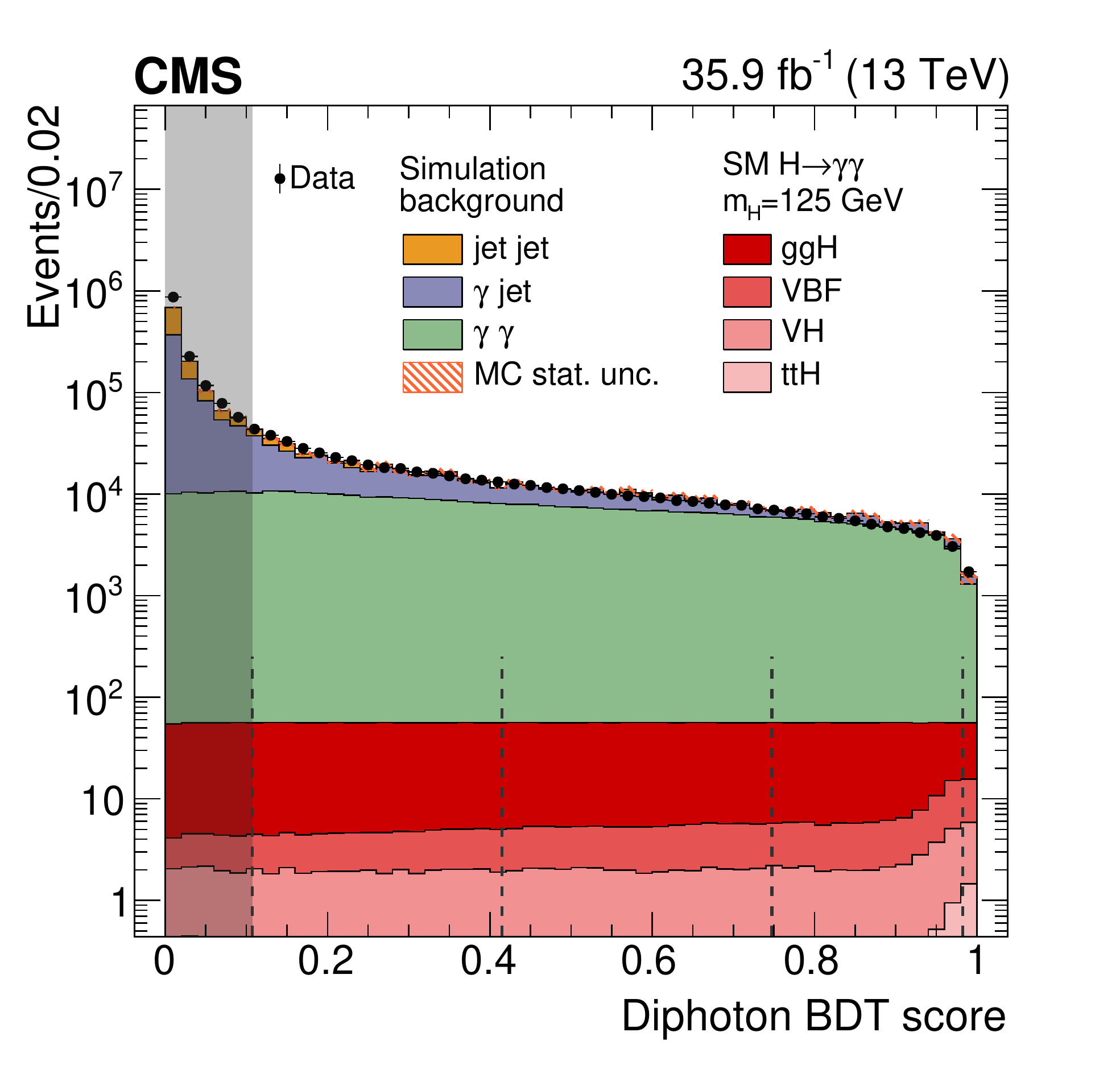}
\includegraphics[width=0.49\textwidth]{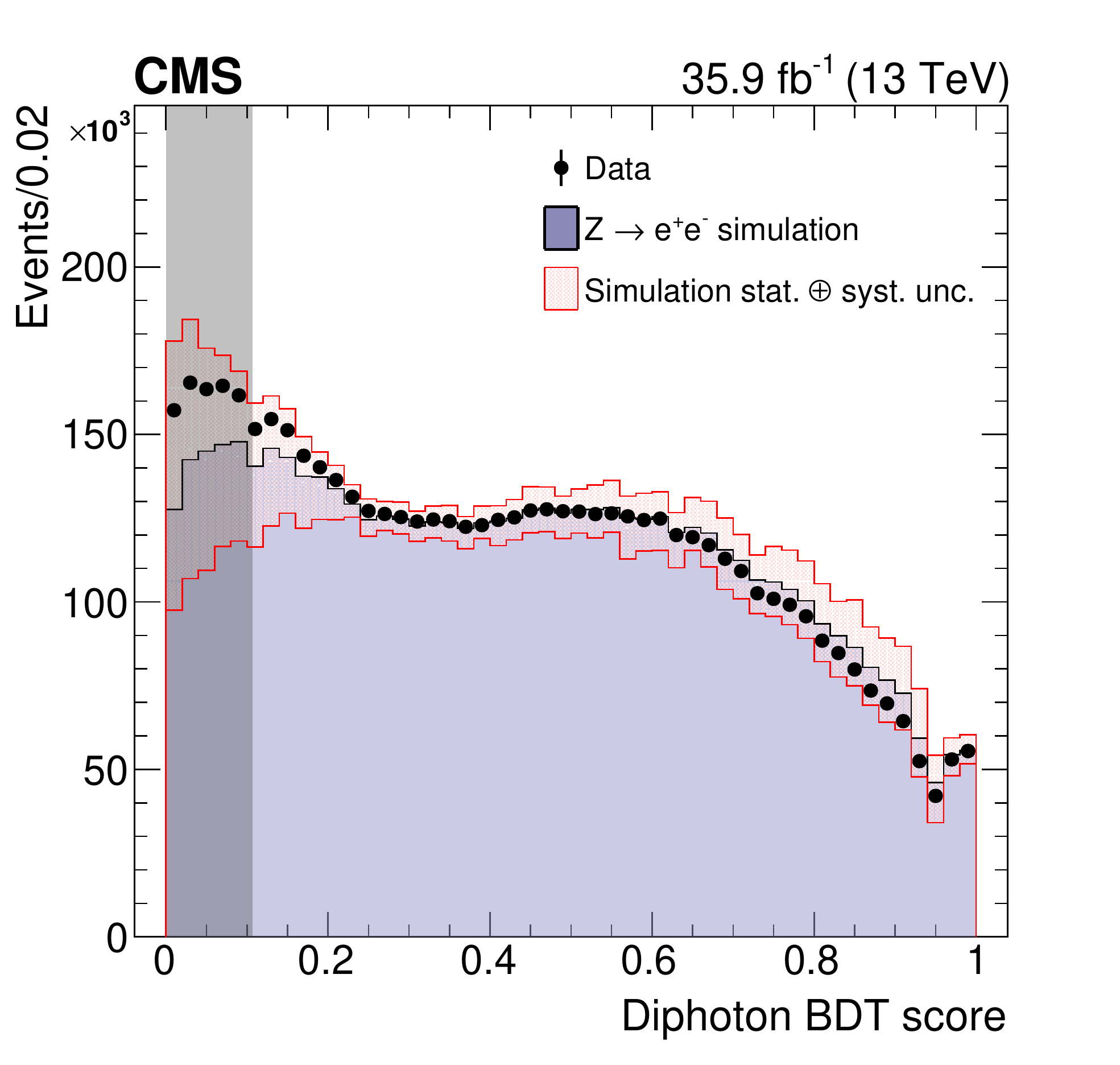}%
\caption{
(Left) Transformed score distribution from the diphoton multivariate
classifier for events with two photons satisfying the preselection
requirements in data (points), simulated signal (red shades), and
simulated background (coloured histograms). Both signal and background
are stacked together.
The vertical dashed lines show the boundaries of the untagged
categories, the grey shade indicates events discarded from the analysis.
(Right) Score distribution of the diphoton multivariate classifier in $\Zee$
events where the electrons are reconstructed as photons. The points show
the distribution for data, the histogram shows the distribution for
simulated Drell--Yan events. The pink band indicates the statistical and
systematic uncertainties in simulation. The grey shade indicates events
discarded from the analysis.
}
\label{fig:diphotonMVA}
\end{figure}

A dedicated diphoton BDT is used in the event categorization.
The diphoton BDT assigns a high score to events with photons
showing signal-like kinematics, good mass resolution, and high photon
identification BDT score. The input variables to the classifier are:
\begin{itemize}
        \item $\ptg / \mgg$ for each photon;
        \item the pseudorapidity of the two photons;
        \item the cosine of the angle between the two photons in the
                transverse plane;
        \item photon identification BDT scores for both photons;
        \item two per-event relative mass resolution estimates,
                 one under the
                hypothesis that the mass has been reconstructed using the
                correct primary vertex, and the other under the hypothesis
                that the mass has been reconstructed using an incorrect vertex;
        \item the per-event probability estimate that the correct primary
                vertex has been assigned to the diphoton.
\end{itemize}
The relative mass resolution is computed from the propagation of the photon
energy resolution estimates, assuming the functional forms of the photon
resolutions are Gaussian.
Figure~\ref{fig:diphotonMVA} (left) shows the transformed score from
the diphoton multivariate classifier for data and simulated signal and
backgrounds,
 for events with two photons satisfying the preselection requirements.
The classifier score has been transformed such that the sum of signal events from
all the production modes has a uniform distribution.
A validation of the score from the diphoton multivariate classifier
obtained in $\Zee$ events, where the electrons are reconstructed as
photons, is shown in Fig.~\ref{fig:diphotonMVA} (right) for data and simulation.

\subsection{Event categories for \texorpdfstring{\ttH}{ttbar H} production}

\begin{figure}[hptb]
\centering
\includegraphics[width=0.5\textwidth]{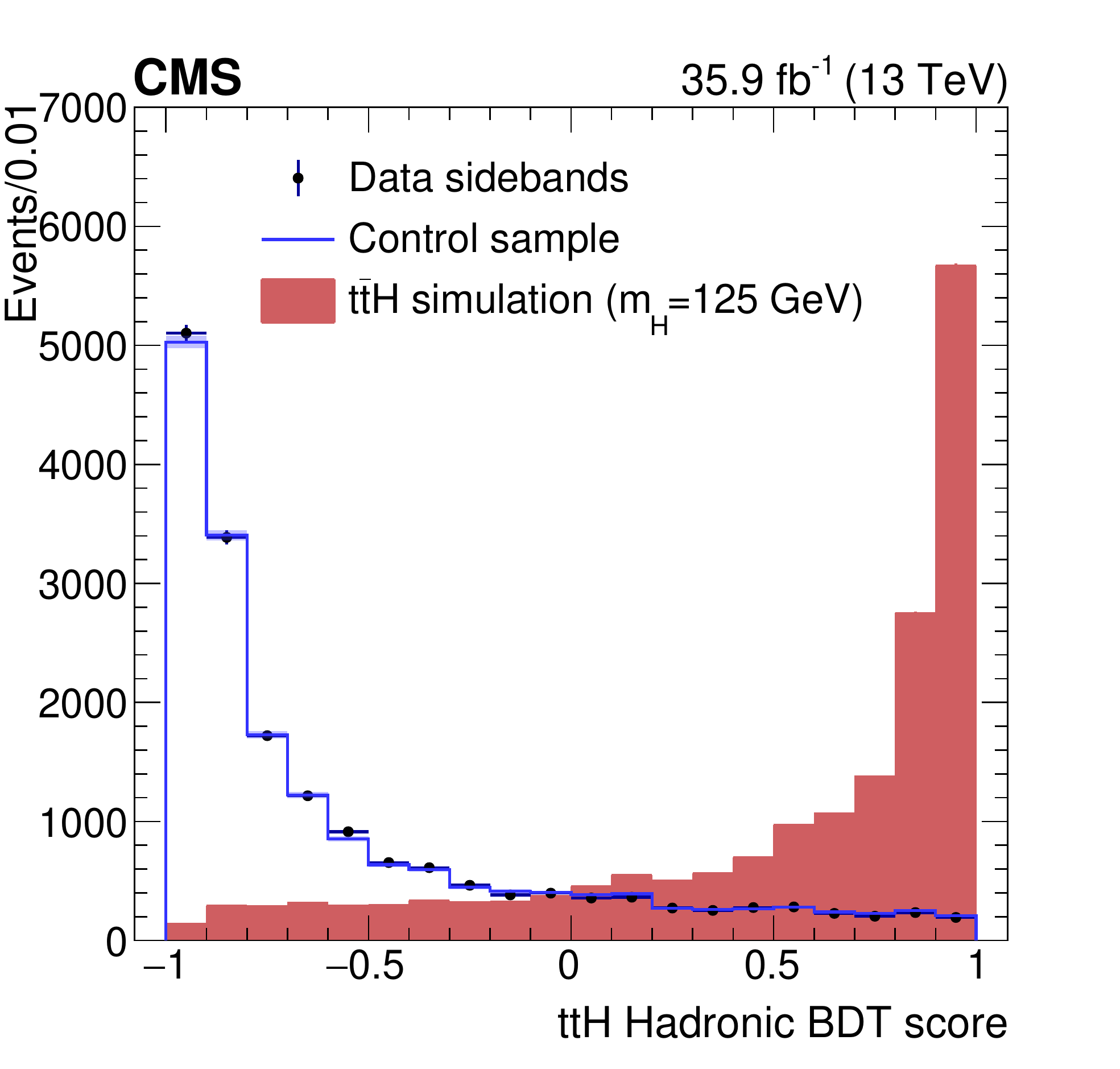}
\caption{Score distribution of the jet multivariate discriminant used to
        enhance jet tagging in the ttH multijet category. The points show the
        distribution for data in the signal region sidebands, $\mgg < 115\GeV$
        or $\mgg > 135\GeV$; the histogram shows the distribution for events in
        the data control sample; the filled histogram shows the distribution
        for simulated signal events. The distributions in the simulated and
        control samples are scaled as to match the integral of that from the
        data sidebands.
}
\label{fig:ttHMVA}
\end{figure}

Events produced in association with a top quark pair feature two \cPqb\ quarks from
the decay of the top quarks, and may be accompanied by charged leptons
or additional jets. In the latter case, to enhance the tagging of \ttH
multijet events, a multivariate discriminant is built upon the following
inputs:
\begin{itemize}
        \item the number of jets with $\pt > 25\GeV$;
        \item the leading jet $\pt$;
        \item the two highest scores from the \cPqb tag CSV discriminator.
\end{itemize}
The output of this discriminant is shown in Fig.~\ref{fig:ttHMVA}. The
threshold on the discriminant is optimized jointly with the requirement
on the diphoton BDT score by maximizing the expected sensitivity to
\ttH production.

To cross-check the performance of this BDT observable, a control sample
in data is defined by selecting events with a pair of photons, one of
which passes the preselection and photon identification requirements,
while the other has no preselection applied and an inverted criterion
on the score from the photon identification BDT. As the efficiency for
selecting such photons is not the same as for the signal region, events
in the control samples are weighted according to the $\eta$ and $\pt$
of the photons so as to obtain a control sample with similar kinematic
properties as the signal region, but statistically independent.

Depending on the type of the top quark decay, the following categories
are defined:
\begin{itemize}
        \item semileptonic top quark decays (\cttH Leptonic):
        \begin{itemize}
                \item leading photon $\pt > \mgg / 2$,
                        subleading photon $\pt > \mgg / 4$;
                \item diphoton classifier BDT score greater than 0.11;
                \item at least one lepton with $\pt> 20\GeV$; electrons
                        must satisfy loose requirements on the same
   			observables as described in Ref.~\cite{CMS-DP-2015-067};
                        muons are required to pass a tight
                        selection based on the quality of the track, the
                        number of hits in the tracker and muon system,
                        and the longitudinal and transverse impact
                        parameters of the track with respect to the muon
                        vertex, and to satisfy a requirement on the relative
                        isolation (after correction for pileup) based on
                        transverse energy of the charged hadrons,
                        neutral hadrons, and photons, in a cone around
                        the muon with a radius between 0.05 and 0.2,
                        depending on the $\pt$ of the muon;
                \item all selected leptons $\ell$ are required to have
                        $R(\ell, \Pgg) > 0.35$, where $R$ is the distance
                        between the objects in the $\eta-\phi$ plane;
                \item specifically for electrons: $\abs{\m_{\Pe,
                        \Pgg} - \m_\PZ} > 5\GeV$, where
      			$m_{\Pe,\Pgg}$ is the invariant mass of any pair of
			electron and photon and
                        $\m_\PZ$ refers to the mass of the $\PZ$ boson;
                \item at least two jets in the event with $\pt > 25\GeV$,
                        $\abs{\eta} < 2.4$, and
                        $R(\text{jet}, \Pgg) > 0.4$ and
                        $R(\text{jet}, \ell) > 0.4$;
                \item at least one of the jets in the event
                        identified as a \cPqb\ jet
                        according to the CSV tagger medium
                        requirement;
        \end{itemize}
        \item hadronic top quark decays (\cttH Hadronic):
        \begin{itemize}
                \item leading photon $\pt > \mgg/3$, subleading photon $\pt>\mgg/4$;
                \item diphoton classifier BDT score greater than 0.58;
                \item no leptons, defined according to the criteria of
                        the \cttH Leptonic category;
                \item at least three jets in the event with $\pt > 25\GeV$ and $\abs{\eta} < 2.4$;
                \item at least one of the jets in the event identified as a
                        \cPqb\ jet according to the CSV tagger loose requirement;
                \item score from the \cttH Hadronic multivariate
                        discriminant greater than 0.75.
        \end{itemize}
\end{itemize}

\subsection{Event categories for \VH production}

The selection criteria targeting the associated production of the Higgs
boson with a vector \PW\ or \PZ\ boson exploit the presence of
leptons, missing transverse momentum, and jets. To reduce contamination
from Drell--Yan events with an electron misreconstructed as a photon,
or with photons radiated in the final state, photon candidates are
required to be separated in angle from the closest lepton. The criteria
are the following:
\begin{itemize}
        \item leptonic $\PZ$ decays (\ZH Leptonic):
        \begin{itemize}
                \item leading photon $\pt > 3 \mgg / 8$, subleading photon $\pt>\mgg/4$;
                \item diphoton classifier BDT score greater than 0.11;
                \item two same-flavour leptons within the fiducial region,
                        $\pt > 20\GeV$; electrons and muons are required to
                        satisfy the same identification criteria as for the
                        \cttH Leptonic category;
                \item dilepton invariant mass $\m_{\ell\ell}$
                        in the range $70 < \m_{\ell\ell} < 110\GeV$;
                \item $R(\Pgg, \Pe) > 1.0$, $R(\Pgg, \PGm) > 0.5$, for each
                        of the leptons;
                \item in addition, a conversion veto is applied to the electrons to
                  reduce the number of electrons originating from photon conversions,
                  by requiring that, when an electron and a photon candidate share a
                  supercluster, the electron track is well separated from the centre
                  of the supercluster:\\ $R(\mathrm{supercluster},\Pe \text{-track}) > 0.4$.
        \end{itemize}
        \item leptonic \PW\ decays (\WH Leptonic):
        \begin{itemize}
                \item leading photon $\pt > 3 \mgg / 8$, subleading photon $\pt>\mgg/4$;
                \item diphoton classifier BDT score greater than 0.28;
                \item at least one lepton with $\pt > 20\GeV$;
                        electrons and muons are required to satisfy the same
                        identification criteria as for the \ZH Leptonic category;
                \item $R(\Pgg, \ell) > 1.0$ and conversion veto as in
                        the \ZH Leptonic category;
                \item missing transverse momentum $\ptmiss > 45\GeV$;
                \item up to two jets each satisfying $\pt > 20\GeV$,
                        $\abs{\eta} < 2.4$, $R(\text{jet}, \ell) > 0.4$,
                        and $R(\text{jet}, \Pgg) > 0.4$;
        \end{itemize}
        \item \PW\ or \PZ\ leptonic decays, relaxed selection
                (\VH LeptonicLoose):
        \begin{itemize}
                \item as for \WH Leptonic with the requirement on the missing
                        transverse momentum to be $\ptmiss < 45\GeV$;
        \end{itemize}
        \item \PW\ or \PZ\ leptonic decays, with at least one missing lepton
                (\VH MET):
        \begin{itemize}
                \item leading photon $\pt > 3 \mgg / 8$, subleading photon $\pt > \mgg / 4$;
                \item diphoton classifier BDT score greater than 0.79;
                \item missing transverse momentum $\ptmiss > 85\GeV$;
                \item angle in the transverse plane between the
                        direction of the diphoton and the $\ptvecmiss$
                        $\Delta\phi(\Pgg\Pgg, \ptvecmiss) > 2.4$;
        \end{itemize}
        \item hadronic decays of \PW\ and \PZ\ (\VH Hadronic):
        \begin{itemize}
                \item leading photon $\pt > \mgg / 2$,
                        subleading photon $\pt > \mgg / 4$;
                \item diphoton classifier BDT score greater than 0.79;
                \item at least two jets, each with $\pt > 40\GeV$ and $\abs{\eta}
                        < 2.4$, $R(\text{jet}, \Pgg) > 0.4$;
                \item dijet invariant mass in the range $60 < \m_{jj} <
                        120\GeV$;
                \item $\abs{\cos\theta^\star} < 0.5$, where $\theta^\star$
                        is the angle that the diphoton system makes, in
                        the diphoton-dijet centre-of-mass frame, with
                        respect to the direction of motion of the
                        diphoton-dijet system in the lab frame. The
                        distribution of this variable is rather uniform for \VH
                        events, while it is strongly peaked at 1 for
                        background and events from \ggH production.
        \end{itemize}
\end{itemize}

\subsection{Event categories for \VBF production}

\begin{figure}[hptb]
\centering
\includegraphics[width=0.49\textwidth]{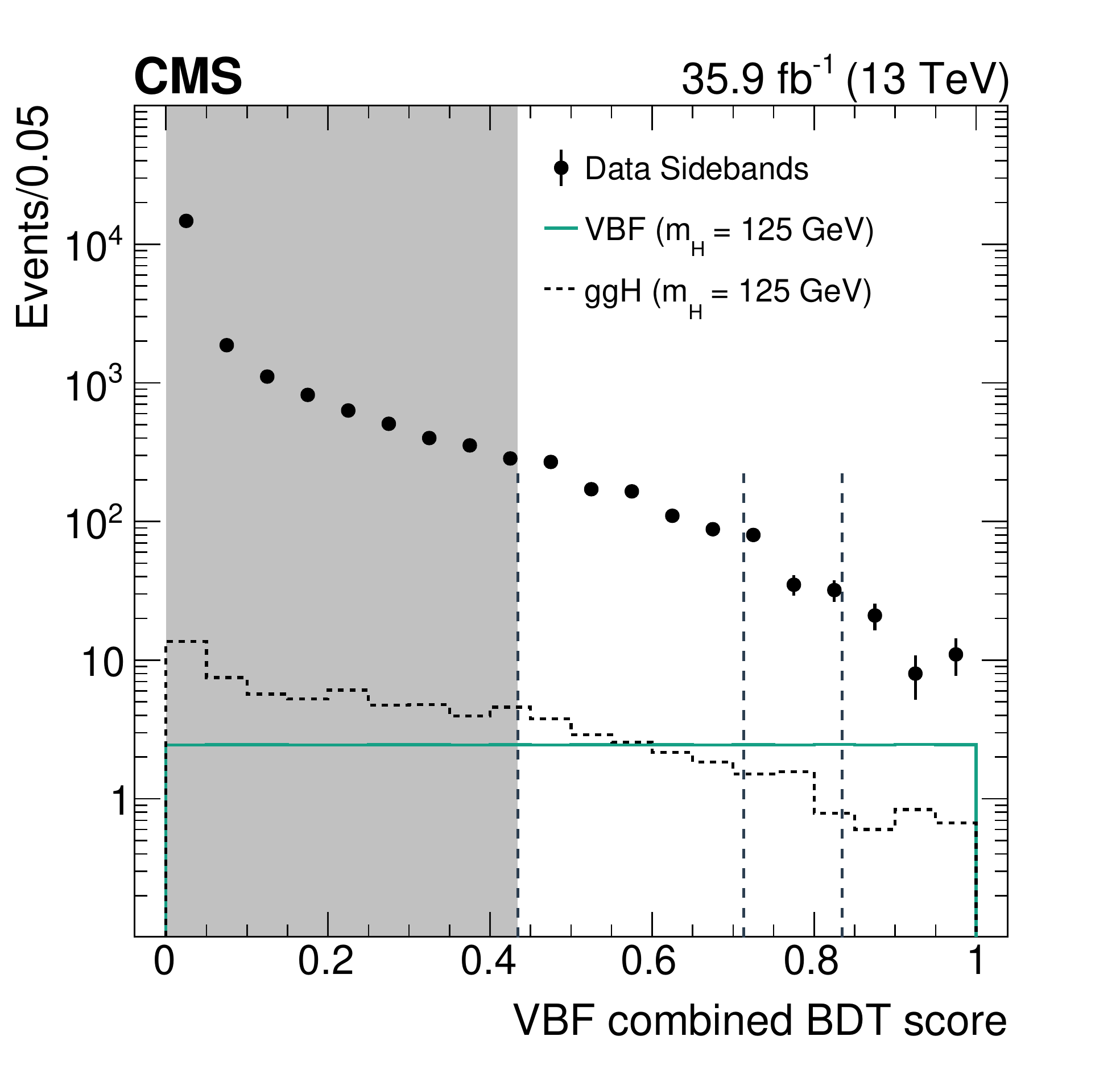}
\includegraphics[width=0.49\textwidth]{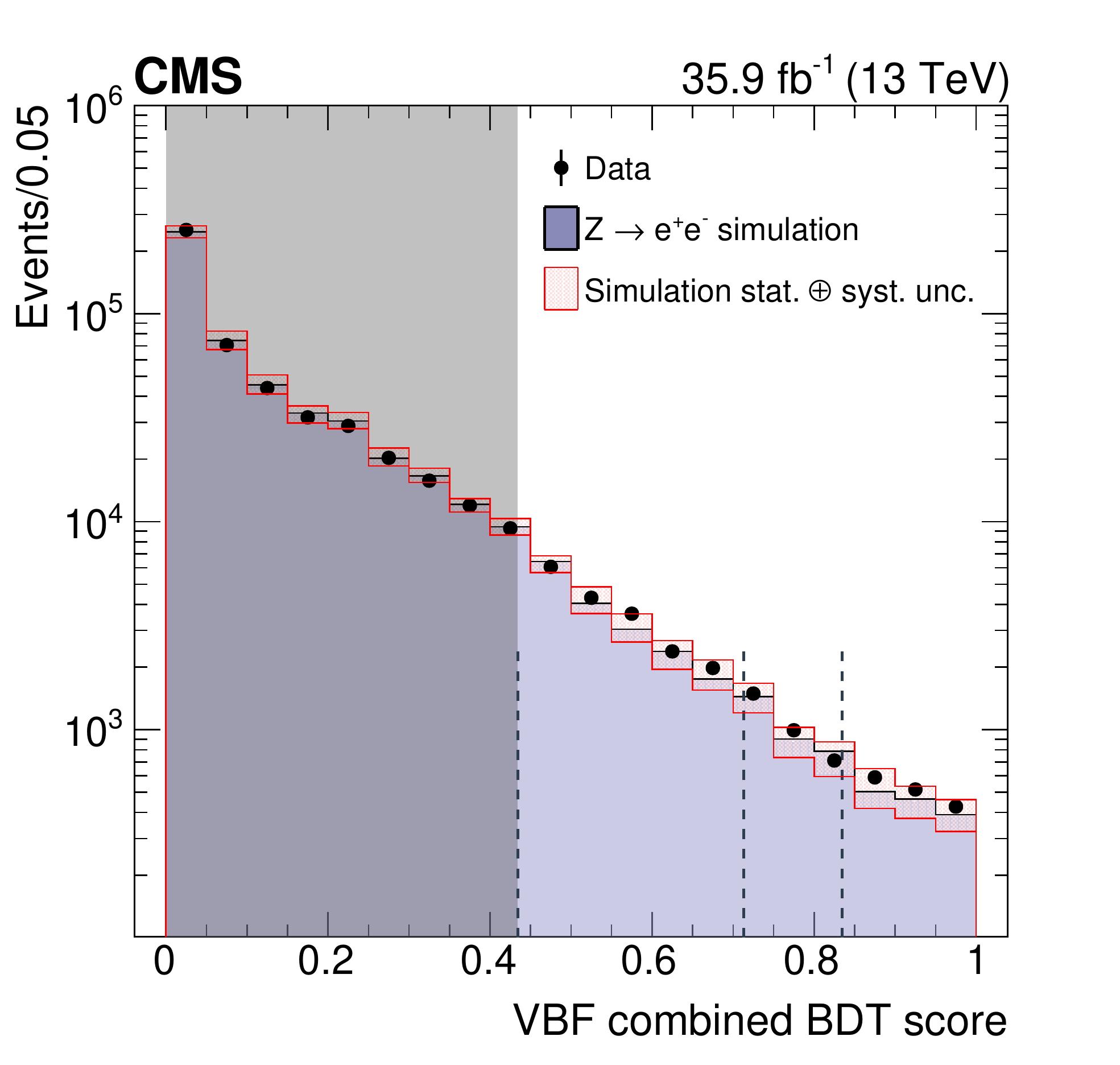}%
\caption{Score distribution from the \VBF combined BDT for
(left) \ggH and \VBF signal distributions, compared to background taken from
data in the mass sideband regions, and
(right) $\Zee$ + jets events.
On the left, the signal region selection is applied to the simulated \ggH and \VBF
events; these are compared to points representing the background,
as determined from data using the signal region selection in mass sidebands.
On the right, the signal selection is applied to electrons reconstructed
as photons,
with points showing the distribution for data and the histogram showing
the distribution for simulated Drell--Yan events, including statistical and
systematic uncertainties (pink band).
In both plots, dotted lines delimit the three \VBF categories, while the
grey region is discarded from the analysis.
}
\label{fig:vbfMVA}
\end{figure}

Events produced via the \VBF process feature two jets in the final
state separated by a large rapidity gap.
A multivariate discriminant is trained to tag the distinctive kinematics
of the \VBF jets, considering as background the production process
of \ggH + jets. This discriminant is given as input to an additional
multivariate classifier (\VBF combined BDT) along with the score from the
diphoton BDT, and the ratio $\ptgg / \mgg$.
Figure~\ref{fig:vbfMVA} (left) shows the transformed score from the
\VBF combined BDT for data in the mass sideband regions from
105--115\GeV and 135--145\GeV, along with the predicted \VBF and \ggH
distributions.
The \VBF combined BDT
score has been transformed such that the signal events from the \VBF
production mode has a uniform distribution.
A validation of the score from the combined multivariate classifier
obtained in $\Zee$ + jets events, where the electrons are reconstructed
as photons and at least two jets satisfy the requirements listed below
to enter the \VBF category, is shown in Fig.~\ref{fig:vbfMVA} (right) for
data and simulation.

The selections targeting the \VBF production mechanism are the following:

\begin{itemize}
        \item leading photon $\pt > \mgg / 3$, subleading photon $\pt>\mgg/4$;
        \item photon identification BDT score greater than $-0.2$, to
                provide additional rejection against background events
                whose kinematics yield a high diphoton BDT score
                despite  one reconstructed photon with a relatively
                low photon identification BDT score;
        \item one jet with $\pt > 40\GeV$ and one with $\pt > 30\GeV$, both
                with $\abs{\eta} < 4.7$ and with a tight requirement on the
                pileup jet identification;
        \item invariant mass of the two jets $\m_{jj} > 250\GeV$;
        \item \VBF combined multivariate discriminant greater than 0.43.
\end{itemize}

Three categories are defined using the score from the combined
discriminant, and are optimized to maximize the expected signal
significance in the \VBF production channel.

\subsection{Event categories for \ggH production}

Events not passing any exclusive category are classified using the multivariate
discriminator described in the introduction of this section.
The score from this classifier is used to select and divide the events
into four ``untagged'' categories according to the diphoton mass
resolution and predicted signal over background ratio. The number of
categories is determined by maximizing the expected signal
significance. The boundaries of these categories are shown in
Fig.~\ref{fig:diphotonMVA}.

\subsection{Final classification}

Each event is classified exclusively by applying the category selections
in order and choosing the highest-priority category satisfied by the
event. Category selections targeting specific production processes are
applied first, ranked by expected signal significance, then untagged
categories. The final ordering is thus \cttH Leptonic, \cttH Hadronic, \ZH
Leptonic, \WH Leptonic, \VH LeptonicLoose, \VBF categories, \VH MET, \VH
Hadronic, and untagged.
The fraction of events with multiple diphoton pairs satisfying one or
more category selections is less than $2\times 10^{-4}$. In this case,
the diphoton in the highest-priority category is selected or, in case of
ambiguities, the diphoton pair with the highest sum of photon $\pt$ is
selected.

\section{Signal model}
\label{sec:signal}

The signal shape for the diphoton invariant mass distribution in each
category and for a nominal Higgs boson mass $\mH$ is constructed from
simulation using events from the different production modes.

The simulation includes the tuning of the photon shower variables to
the data, and accounts for trigger, reconstruction and identification
efficiencies as measured with data-driven techniques (as discussed in
Section~\ref{sec:recoid}). It also weights the events so that the
distribution of the number of interactions and the primary vertex
location reproduce those observed in data, as explained in
Sections~\ref{sec:samples} and~\ref{sec:vtx}.

Since the shape of the $\mgg$ distribution changes considerably
depending on whether the vertex associated with the candidate diphoton
was correctly identified within 10\mm, distributions for the correct
vertex and wrong vertex assignments are fit separately when
constructing the signal model. For each process, category, and vertex
scenario, the $\mgg$ distributions are fitted using a sum
of at most five Gaussian functions.

For each process, category, and vertex scenario, a simultaneous fit
of signal samples at mass values in the range from 120 to 130\GeV is
performed to obtain parametric variations of the Gaussian function
parameters used in the signal model fit. Polynomials are used to
describe these variations.

\begin{figure} [hbtp]
	\centering
		{\includegraphics[width=0.48\textwidth]{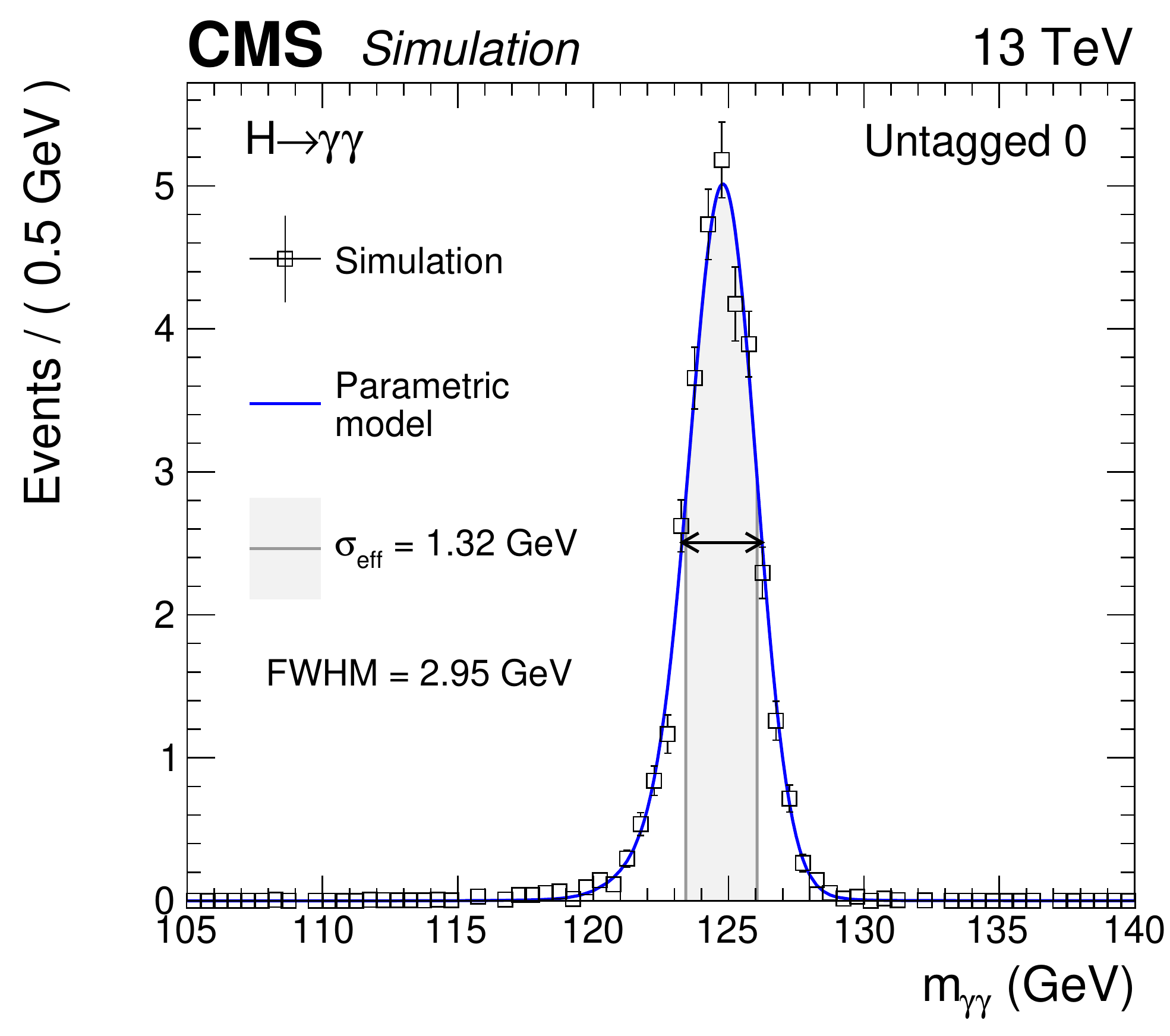}}
		{\includegraphics[width=0.48\textwidth]{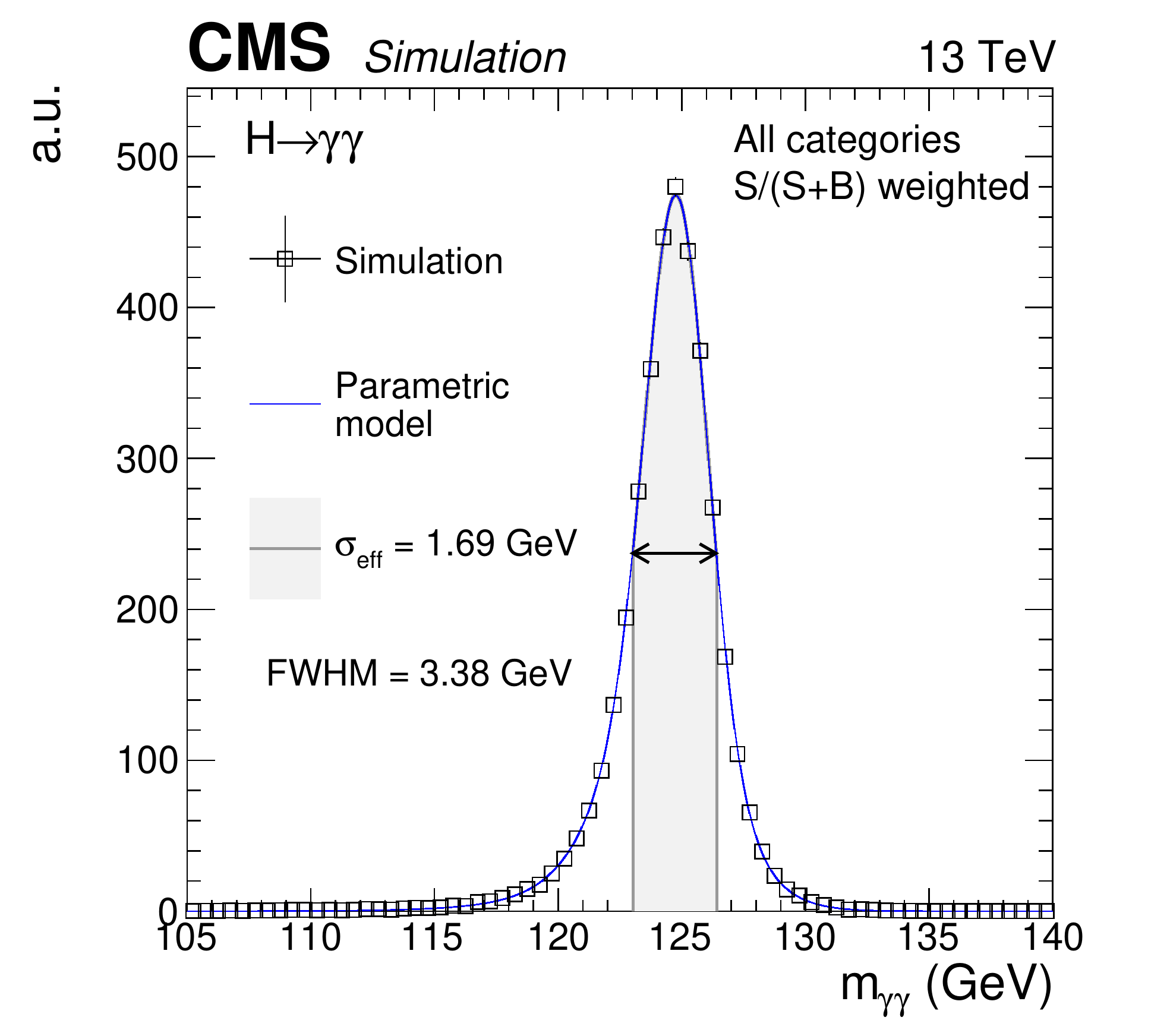}}\\
  \caption{Parametrized signal shape for the best resolution category
  (left) and for all categories combined together and weighted by the
  $\textrm{S}/(\textrm{S}+\textrm{B})$ ratio (right) for a simulated
  $\HGG$ signal sample with $\mH = 125\GeV$. The open
  squares represent weighted simulated events and the blue lines are the
  corresponding models. Also shown are the $\sigma_{\text{eff}}$ value
  (half the width of the narrowest interval containing 68.3\% of the
  invariant mass distribution) and the corresponding interval as a grey
  band, and the full width at half maximum (FWHM) and the corresponding
  interval as a double arrow.}
		\label{fig:statAnalysisSigPlots}
\end{figure}

The final fit function for each category is obtained by summing the
functions for all production modes normalized to the expected signal
yields in that category.
Figure~\ref{fig:statAnalysisSigPlots} shows the signal model corresponding
to $\mH =125\GeV$ for the best resolution category and also for all
categories combined together, weighted by the
$\textrm{S}/(\textrm{S}+\textrm{B})$ ratio, where $\textrm{S}$
is the number of signal events, and $\textrm{B}$ the number of background events
in a window around the signal peak, in each category.

\begin{figure}[hbtp]
	\centering
		\includegraphics[width=0.5\textwidth]{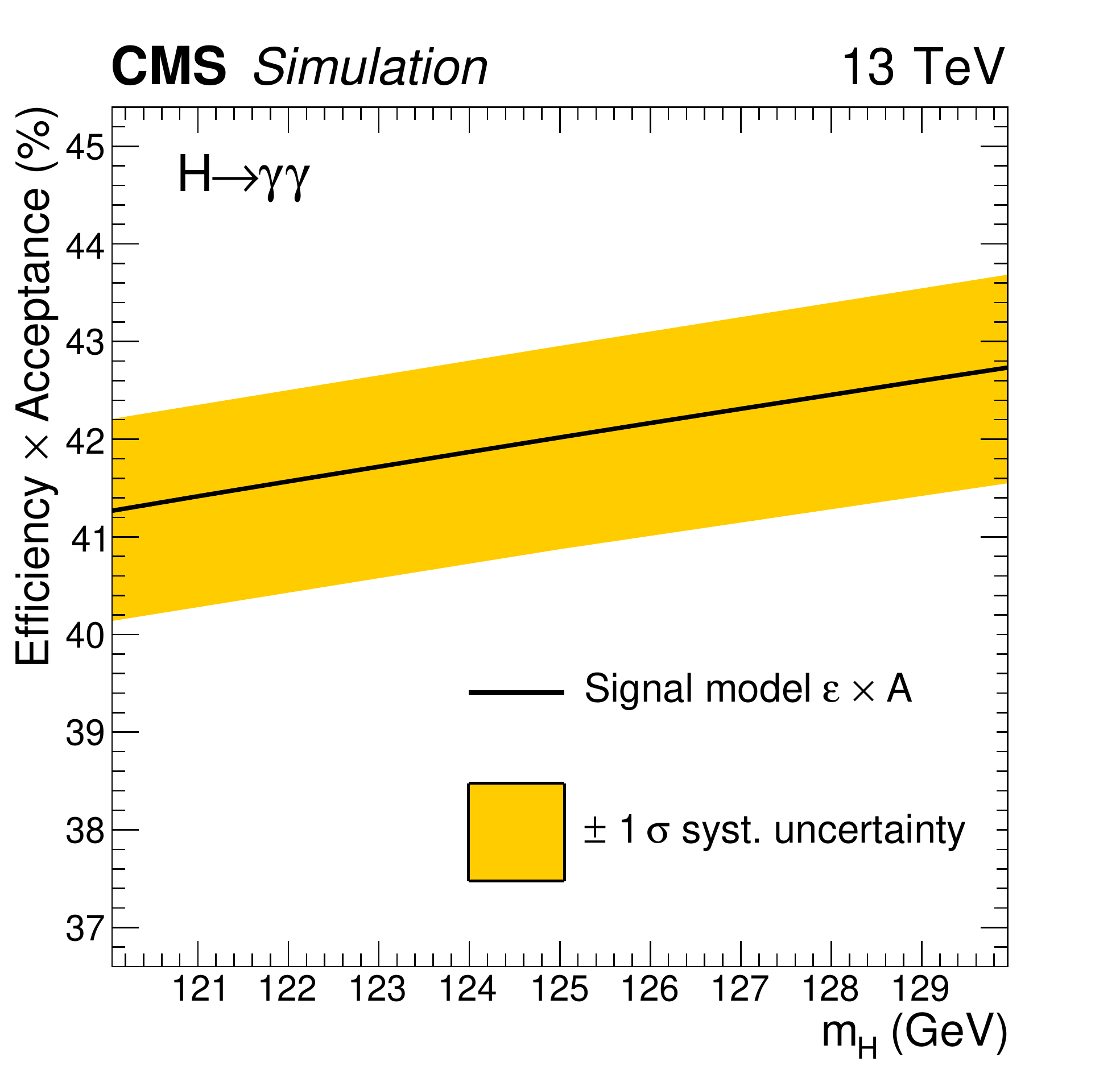}
  \caption{The product of efficiency and acceptance of the signal model as a
  function of $\mH$ for all categories combined.
The black line represents the yield from the
signal model. The yellow band indicates the effect of the $\pm$1 standard
deviation of the systematic uncertainties for trigger, photon
identification and selection, photon energy scale and modelling
of the photon energy resolution, and vertex identification (described in
Section~\ref{sec:syst}).
}
		\label{fig:effxAcc}
\end{figure}

The product of efficiency and acceptance of the signal model as a function of
$\mH$ for all categories combined is shown in Fig.~\ref{fig:effxAcc}.

\section{Background model}
\label{sec:background}

The model used to describe the background is extracted from data with
the discrete profiling method~\cite{DiscreteProfilingMethod}
as implemented in Ref.~\cite{Khachatryan:2014ira}. This
technique was designed as a way to estimate the systematic uncertainty
associated with choosing a particular analytic function to fit the
background $\mgg$ distribution. The method treats the choice of the
background function as a discrete nuisance parameter in the likelihood
fit to the data.

No assumptions are made about the particular processes composing the
background nor the functional form of their smoothly falling diphoton
invariant mass distribution.
A large set of candidate function families is considered, including
exponentials, Bernstein polynomials, Laurent series, and power law
functions. For each family of functions, an F-test~\cite{ftest} is
performed to determine the maximum order to be used,
while the minimum order is determined
by requiring a reasonable fit to the data.
The background is assumed to be a smoothly falling distribution; this
is supported by the shape of background distributions both in simulated
events and in data, in the latter case those of the events rejected by
the diphoton BDT.

When fitting these functions to the background $\mgg$ distribution,
the value of twice the negative logarithm of the likelihood (2NLL)
is minimized. A penalty of $n_\mathrm{p}$ is added to 2NLL to take into account the
number of floating parameters $n_\mathrm{p}$ in each candidate function and avoid favouring
functions with a greater number of free parameters. When making
a measurement of a given parameter of interest, the discrete profiling
method determines the minimal 2NLL by considering all allowed functions
for each value of the parameter.

\section{Systematic uncertainties}
\label{sec:syst}

The systematic uncertainties are treated differently depending on how
they affect the $\mgg$ signal distribution.
The parameters of the signal model shape are allowed to vary,
within the constraints set by the measurements described in
Section~\ref{sec:phoene}, to account for systematic uncertainties in the
photon scale and resolution. Additional nuisance parameters are included
to account for systematic uncertainties which affect the overall rate
and migration of signal events between the categories, and are log-normal constrained.
For cases where the systematic uncertainty has an effect on the input
to one of the classification discriminants, the uncertainty takes the
form of a variation in the category yield, representing event migration
between categories.

\subsection{Theoretical uncertainties}
Theoretical uncertainties in the signal yield associated with QCD calculations
typically have an overall normalization uncertainty, taken from Ref.~\cite{LHCHXSWG},
along with an additional uncertainty accounting for the migration of
events between the analysis categories.  The category migration uncertainties
are factorized from the overall yield uncertainty by scaling them
appropriately so that the overall yield (including events outside
the acceptance of the analysis) is unchanged. The uncertainties computed in this
way are:
\begin{itemize}
\item \textit{QCD scale uncertainty}: related to variations of the
        renormalization and factorization scales, has two nuisance
        parameters affecting the overall normalization uncertainty and
        depending on the number of jets in the event. Variations are found to
        be typically less then 5\%.
\item \textit{PDF uncertainties}:
        have an overall normalization from the PDF4LHC
        prescription~\cite{Butterworth:2015oua,LHCHXSWG},
        while the bin-to-bin migrations are calculated from the
        NNPDF3.0~\cite{Ball:2014uwa} PDF set
        using the {\sc MC2hessian} procedure~\cite{Carrazza:2015aoa}.
        The category migrations
        are found to be typically less than 1\%, depending on the category.
\item \textit{$\alphas$ uncertainty}: the uncertainty in the value of
        the strong force coupling constant $\alphas$ is evaluated following
        the PDF4LHC prescription.
        The overall variation in the relative event yield
        due to the $\alphas$ uncertainty is at most 2.6\%.
\end{itemize}

Further theoretical uncertainties are:
\begin{itemize}
\item \textit{Underlying event and parton shower uncertainty}: is
        obtained using samples where the choice and tuning of the generator
        has been modified. This systematic uncertainty is treated as an event
        migration systematic as it will mainly affect the jets in the analysis.
        The possibility that an event could move from one \VBF category to another or
        from any \VBF category to an inclusive category is assigned a systematic
        uncertainty of 7 and 9\%, respectively.
\item \textit{Gluon fusion contamination in the ttH
        tagged categories}: the theoretical predictions for gluon fusion
        are less reliable in a regime where the Higgs boson is produced in
        association with a large number of jets. The systematic
        uncertainty in the gluon fusion contamination in the
        ttH tagged categories has been estimated taking
        into account several contributions:
        \begin{itemize}
                \item uncertainty due to the limited size of the
                        simulated sample: 10\%.
                \item uncertainty from the jet
                        modelling. This uncertainty is estimated as the
                        observed difference in the jet multiplicity
                        between \MGvATNLO predictions
                        and data in $\ttbar + \mathrm{jets}$
                        events (which are dominated by gluon fusion
                        production $\Pg\Pg \to \ttbar$),
                        with fully leptonic \ttbar decays. This
                        uncertainty is about 35\% in the bins with the
                        largest discrepancy ($N_{\textrm{jets}}\geq
                        5$).
                \item uncertainty in the gluon splitting
                        modelling. This is estimated by scaling
                        the fraction of events from gluon fusion
                        with real \cPqb\ jets by the observed difference
                        between data and simulation in the ratio
                        $\sigma(\ttbar\bbbar)/\sigma(\ttbar\mathrm{jj})$
                        at 13 TeV~\cite{Sirunyan:2017snr}. This
                        uncertainty implies a variation of about 50\% in
                        the yield of gluon fusion events.
        \end{itemize}
\item \textit{Gluon fusion contamination in categories with additional jets
        and a high-\pt Higgs boson}: particularly important for
        estimating the yield in the \VBF categories. A total of seven
        nuisance parameters account for different systematic effects:
        \begin{itemize}
                \item uncertainties in jet multiplicities: two nuisance
                parameters account for missing higher-order corrections
                and two for migrations between categories with different
                jet multiplicity.
                These are based on the STWZ~\cite{Stewart:2013faa} and
		BLPTW~\cite{Stewart:2013faa,Liu:2013hba,Boughezal:2013oha}
		predictions.
                \item uncertainties in the Higgs boson \pt modelling: two
                nuisance parameters include migrations between regions
                with \pt in the range between 60 and $120\GeV$ and above
                $120\GeV$. A third nuisance parameter accounts for the
                impact of top quark mass effects, which are negligible
                for a Higgs boson \pt below $150\GeV$ and rise to about
		35\% at $500\GeV$;
                these impact primarily the tightest untagged and \VBF
                categories, where the resulting uncertainty in the
                predicted gluon fusion yield is 6--8\%.
                \item uncertainties in the acceptance of gluon fusion events
                in the \VBF categories, due to missing higher-order QCD
                effects in the calculations: these are estimated by
                variations of the renormalization and factorization scales in
                MCFM 5.8~\cite{Campbell:2010ff}.
                Two nuisance parameters account for the uncertainty in
                the overall normalizations of Higgs boson events with 2 extra
                jets, or with 3 or more extra jets, allowing one to
                propagate the impact of jet suppression from the
                kinematic selections in the \VBF BDT scores. An extension
                of the Stewart--Tackmann method~\cite{Stewart:2011cf,Gangal:2013nxa}
                is used. The impact on the yield of gluon fusion events
                in \VBF categories is 8--13\%.
	\end{itemize}
\item \textit{Uncertainty in the $\HGG$ branching fraction}:
  is estimated to be about 2\%~\cite{LHCHXSWG}.
\end{itemize}

\subsection{Experimental uncertainties in the photon energy scale}
The experimental uncertainties in the photon energy scale and resolution
are propagated through to the signal model in the final statistical fit,
allowing the shape to vary. These uncertainties are:
\begin{itemize}
\item \textit{Energy scale and resolution}:
  The uncertainties in the overall photon energy scale and resolution corrections
  are assessed with $\Zee$ events and applied to photons.
  These uncertainties account for
        varying the $\RNINE$ distribution, the regression training
        (using electrons instead of photons) and the electron selection used
        to derive the corrections. The uncertainty in the additional energy
        smearing is assigned propagating the uncertainties in the various
        $\abs{\eta}$ and $\RNINE$ bins to the Higgs boson signal phase space. In
        both cases dedicated nuisance parameters are included as additional
        systematic terms in the signal model and amount to a 0.15 to 0.5\%
        effect on the photon energy depending on the photon category. The effect
        on the measurement of the inclusive signal strength modifier is found to
        be about 2.5\%.
      \item \textit{Nonlinearity of the photon energy}:
        An additional uncertainty accounts for the possible residual
        differences in the linearity of the energy scale between data and simulation.
        This effect is studied using Lorentz-boosted \PZ\ boson dielectron decays.
        The effect is found to be at most 0.1\% on the photon energy in all categories,
        except in the untagged category with highest signal-to-background ratio,
        for which it is 0.2\%.
\end{itemize}
Additional uncertainties are assigned based on studies accounting for
  differences between electrons and photons on the following points.
\begin{itemize}
\item \textit{Nonuniformity of the light collection}:
  The uncertainty in the modelling of the fraction of scintillation light reaching
  the photodetector as a function of the longitudinal depth in the crystal at which it was emitted.
        The uncertainty has been slightly
        increased with respect to Run~1 to account for the larger loss in
        transparency of the ECAL crystals. The size of the effect on the
        photon energy scale for 2016 data is estimated to be 0.07\%.
\item \textit{Electromagnetic shower modelling}: A further small uncertainty is added to
        account for imperfect electromagnetic shower simulation in
        \GEANTfour. A simulation made with a previous version of the
        shower description, not using the Seltzer--Berger model for the
        bremsstrahlung energy spectrum~\cite{Seltzer:1974zz}, changes the energy scale
        for both electrons and photons. Although mostly consistent
        with zero, the variation is interpreted as a limitation on our
        knowledge of the correct simulation of the showers, leading to a
        further uncertainty of 0.05\% in the photon energy.
\item \textit{Modelling of the material budget}: The uncertainty in the material
        budget between the interaction point and the ECAL, which affects the
        behaviour of electron and photon showers, is estimated with specially
        simulated samples where the material budget is uniformly varied by $\pm
        5$~\%. This accounts for the difference in the estimate of the material
        budget between data and simulation, using methods based on electron
        bremsstrahlung, multiple scattering of pions, and energy flow in
        ECAL. The effect on the energy scale is at most 0.24\%.
\item \textit{Shower shape corrections}: The uncertainty deriving from the
        imperfect shower shape modelling in simulation. It is estimated using
        simulation with and without the corrections on the shower shape variables
        applied to mitigate discrepancies between data and simulation (as
        described in Section~\ref{sec:phoId}).
        This uncertainty in the energy scale is at most 0.01--0.15\%,
	depending on the photon category.
\end{itemize}

\subsection{Additional experimental uncertainties}
Other experimental uncertainties are accounted for by propagating
the uncertainties in the efficiencies, scale factors, and selection variables
through the analysis and applying them to the per-category signal yield:
\begin{itemize}
\item \textit{Trigger efficiency}: the trigger efficiency is measured from
        $\Zee$ events using the tag-and-probe technique; the impact
        on the event yields is at most 0.1\%.
\item \textit{Photon preselection}: the systematic uncertainty is taken as
        the uncertainty in the ratio between the efficiency measured in data and
        in simulation; it ranges from 0.1 to 0.7\%, according to the photon
        category, and results in an event yield variation from 0.2 to
        0.5\%, depending on the event category.
\item \textit{Photon identification BDT score}: to cover
        the observed discrepancies between data and simulation, the uncertainty
        in the signal yields in the different categories of the analysis is
        estimated conservatively by propagating the uncertainty described in
        Section~\ref{sec:recoid} through the diphoton BDT and categorization.
\item \textit{Per-photon energy resolution estimate}: this uncertainty is
        parameterized conservatively as a rescaling of the resolution by
        $\pm 5\%$ about its nominal value, to cover all differences
        between data and simulation in the output distribution of the
        estimator.
        The variation is propagated through the diphoton BDT and
        categorization procedure.
\item \textit{Jet energy scale and smearing corrections}: this
        uncertainty is implemented as migration within \VBF categories, within
        ttH categories, within \VH categories, and from tagged to untagged
        categories. Jet energy scale corrections account for an 8 to 18\%
        migration between the \VBF categories and 11\% from the \VBF to untagged
        categories. The migration due to the energy scale is about 5\% in
        \ttH categories and up to about 15\% in \VH categories. The jet energy
        resolution has an impact on the event migration of less than 3\% in all
        categories except VH, for which the effect can be as large as 20\%.
        However, the processes contributing to the \VH categories and showing
        the largest migrations represent a marginal fraction of events, so that
        their effect on the results is negligible. Processes contributing to
        the majority of the events in the \VH categories show migrations of about 3\%.
\item\textit{Missing transverse energy}: this uncertainty is computed by
        shifting the
        reconstructed $\pt$ of the particle candidates entering the
        computation of $\ptmiss$
        within the momentum scale and resolution uncertainties appropriate
        to each type of reconstructed object,
        as described in Ref.~\cite{CMS-PAS-JME-16-004}.
        It results in a 10 to 15\% migration from the \ggH categories
        into the \VH MET category.
\item \textit{Pileup jet identification}: this uncertainty is estimated by
        comparing in data and simulation the identification score of
        jets in events with a \PZ boson and one balanced jet. The full
        discrepancy between data and simulation is used to estimate the
        event migration, which is of the order of 1\% or less.
\item \textit{Lepton isolation and identification}: for both electrons and muons
        the uncertainty is computed by varying the ratio of the
        efficiency measured in data and simulation within its uncertainty.
        The measurement is done using the tag-and-probe technique on \PZ\
        events. The resulting differences in the selection efficiency
        are less than 1\% for the \cttH Leptonic category, 1.5\% for the
        \WH Leptonic category, and 3\% for the \ZH Leptonic category.
\item \textit{\cPqb\ tagging efficiency}: uncertainties have been evaluated
  by comparing data and simulated distributions for the CSV \cPqb\ tagging
  discriminant, as described in Section~\ref{sec:detector}.  The
        uncertainties include the statistical component in the
        estimate of the fraction of heavy- and light-flavour jets
        in data and simulation, and the corresponding mutual
        contaminations.
        These are propagated differently for the hadron-tagged category
        and the lepton-tagged category, because the former uses the \cPqb\
        tagging discriminant distribution as input to a specialized ttH
        BDT, whereas the latter uses a fixed working point, as described
        in Section~\ref{sec:classification}. For the lepton-tagged
        category, the uncertainty is evaluated by varying the measured
        \cPqb\ tagging efficiencies in data and simulation within their
        uncertainties~\cite{Chatrchyan:2012jua}.
        For the hadron-tagged category, the uncertainty is evaluated
        by modifying the shape of the \cPqb\ tagging discriminant in the
        simulation. The resulting uncertainty in the signal yields is
        about 2\% in the lepton-tagged category and less than 5\% in the
        hadron-tagged category.
\item \textit{Vertex finding efficiency}: the largest contribution to this
        uncertainty comes from the modelling of the underlying event, plus
        the uncertainty in the ratio of data and simulation obtained using
        $\PZ\to\mu^{+}\mu^{-}$ events. It is handled as an additional
        nuisance parameter built into the signal model, which allows the fraction
        of events in the correct and wrong vertex scenario to change. The size
        of the uncertainty in the vertex selection efficiency is 2\%.
\item \textit{Integrated luminosity}: it amounts to
        a 2.5\% uncertainty in the signal yield~\cite{CMS-PAS-LUM-17-001}.
\end{itemize}
The choice of the background parametrization is handled using the
discrete profiling method, described in Section~\ref{sec:background},
which propagates the uncertainty on the choice of function through the fits.

\begin{figure}[ht]
\centering
  \includegraphics[width=0.85\textwidth]{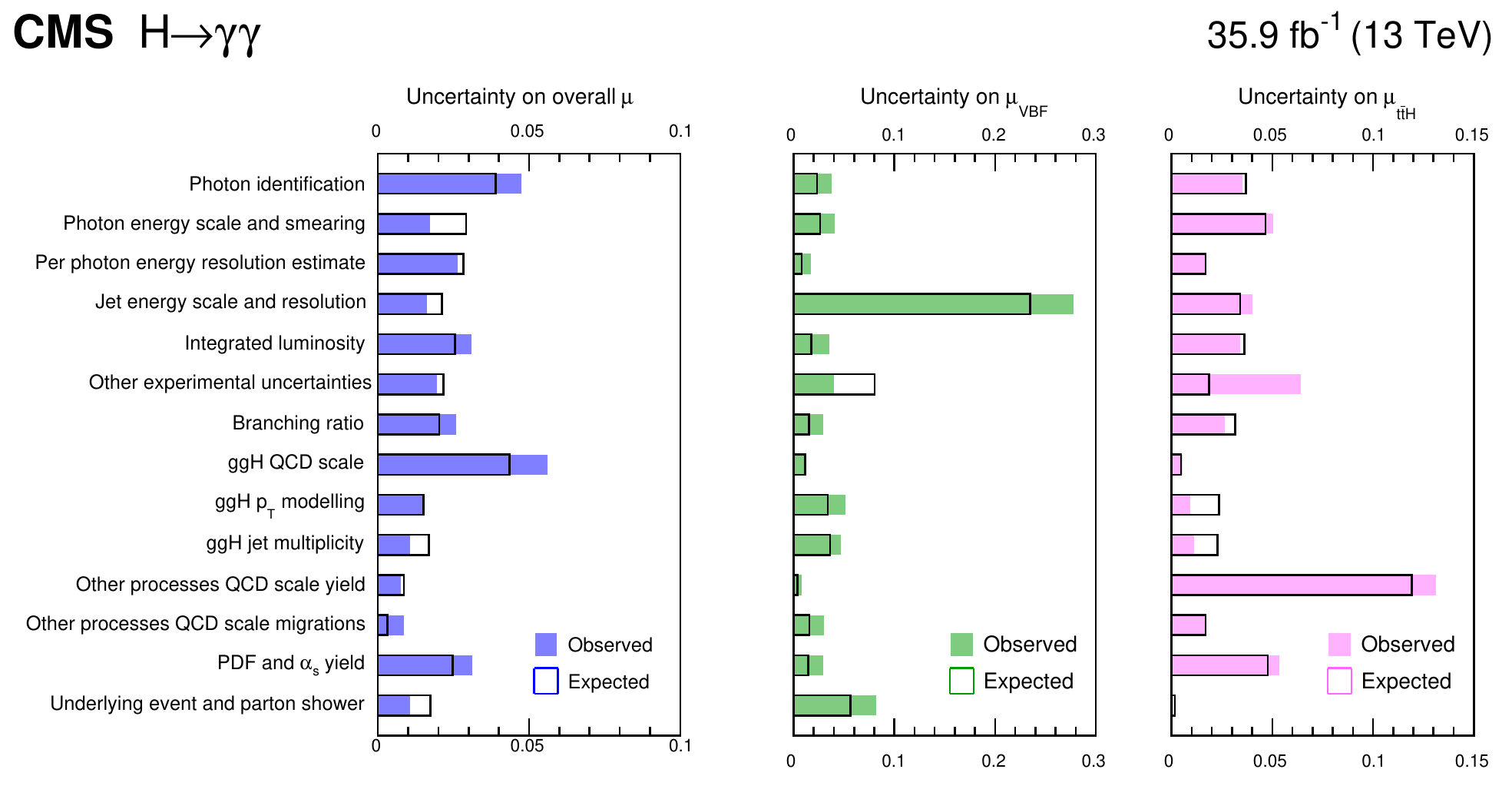}
\caption{Summary of the impact of the different
systematic uncertainties on the overall signal strength modifier and on the
signal strength modifiers for the \VBF and \ttH production
processes. The observed (expected) results are shown by the solid (empty)
bars.}
\label{fig:systSummary}
\end{figure}

The dominant systematic uncertainties on the signal strengths and
couplings are the photon shower shape modelling (which affects the
photon identification and per-photon energy resolution estimate), the
photon energy scale and smearing, the jet energy scale, the integrated
luminosity. The most important theoretical uncertainties are the
branching fraction, and the renormalization and factorization scale
uncertainties. Each of these uncertainties has an impact of a few percent
on the overall signal strength, with some dependence on the targeted
production mechanism, as shown in Fig.~\ref{fig:systSummary}.

\section{Results}
\label{sec:res}

\begin{figure}[htbp]
 \centering
 {\includegraphics[width=0.45\textwidth]{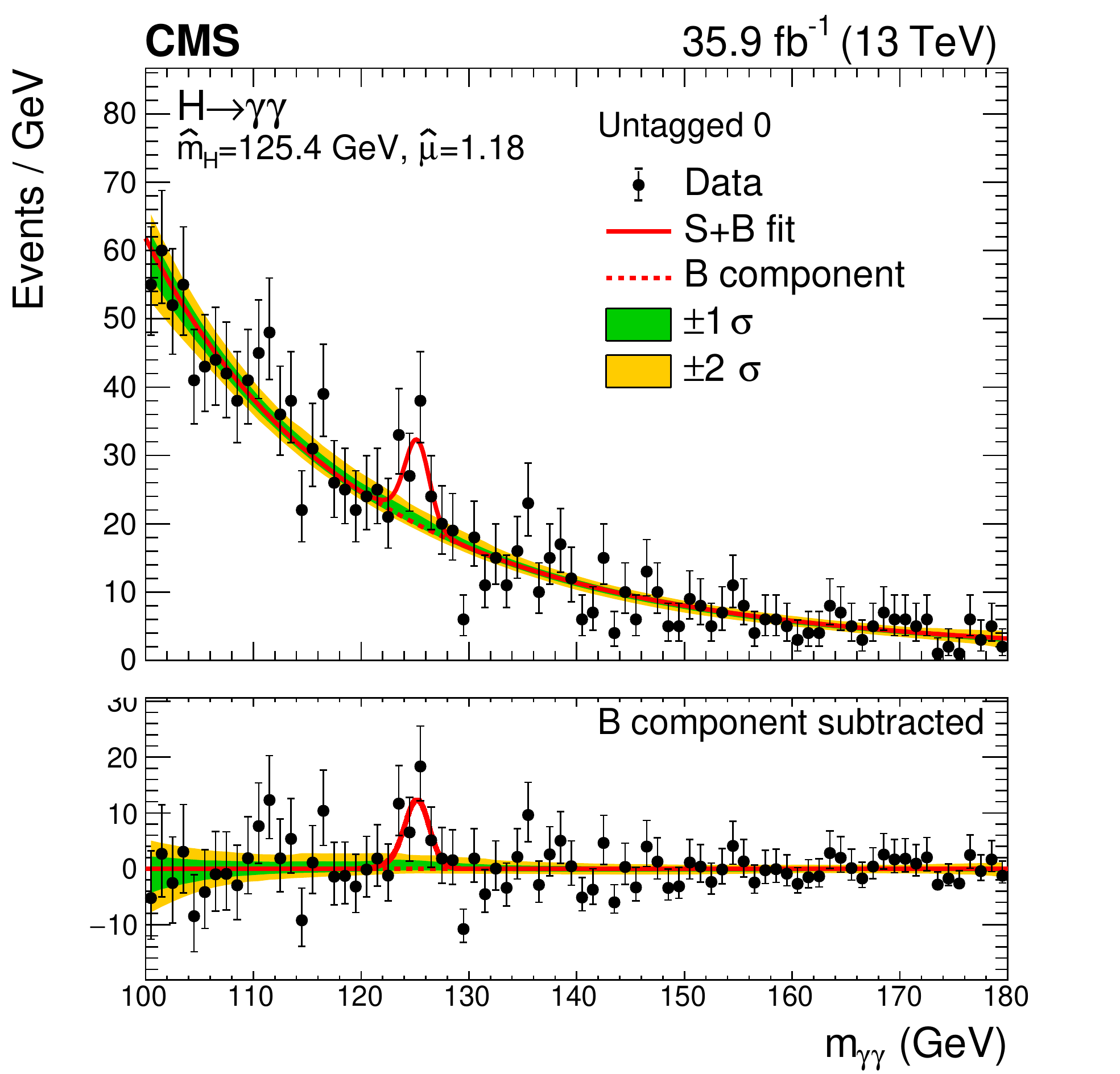}}
 {\includegraphics[width=0.45\textwidth]{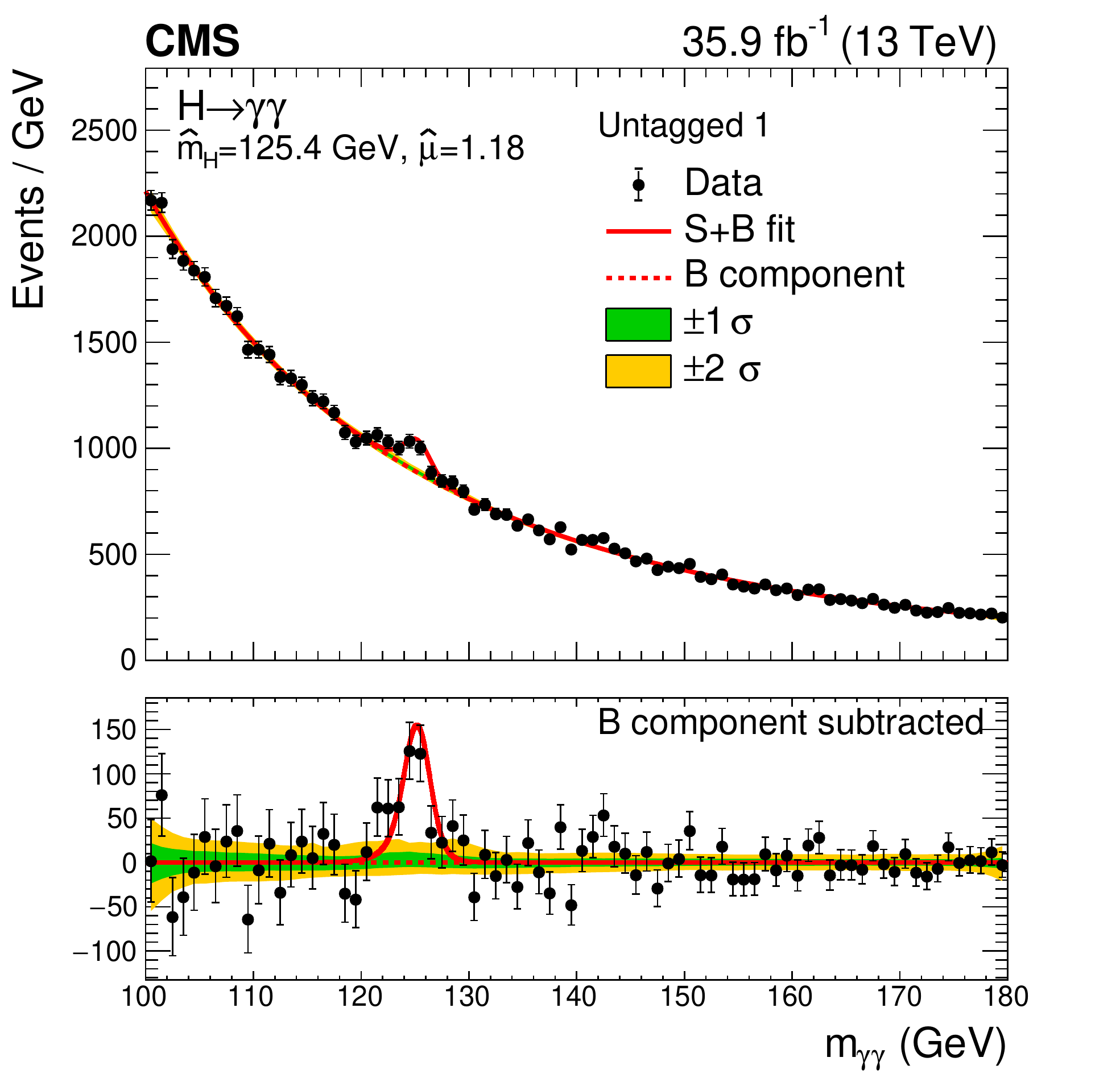}}\\
 {\includegraphics[width=0.45\textwidth]{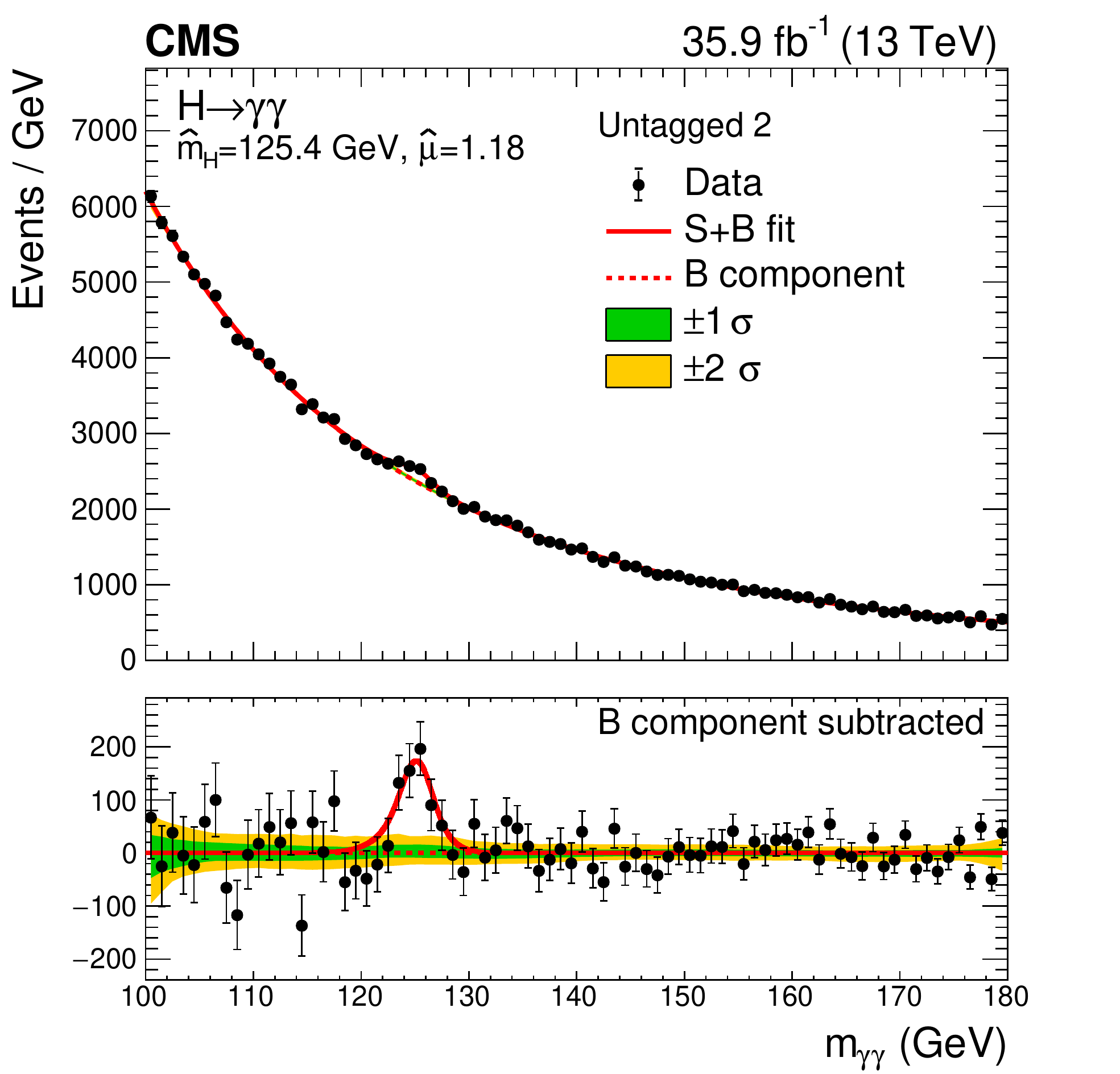}}
 {\includegraphics[width=0.45\textwidth]{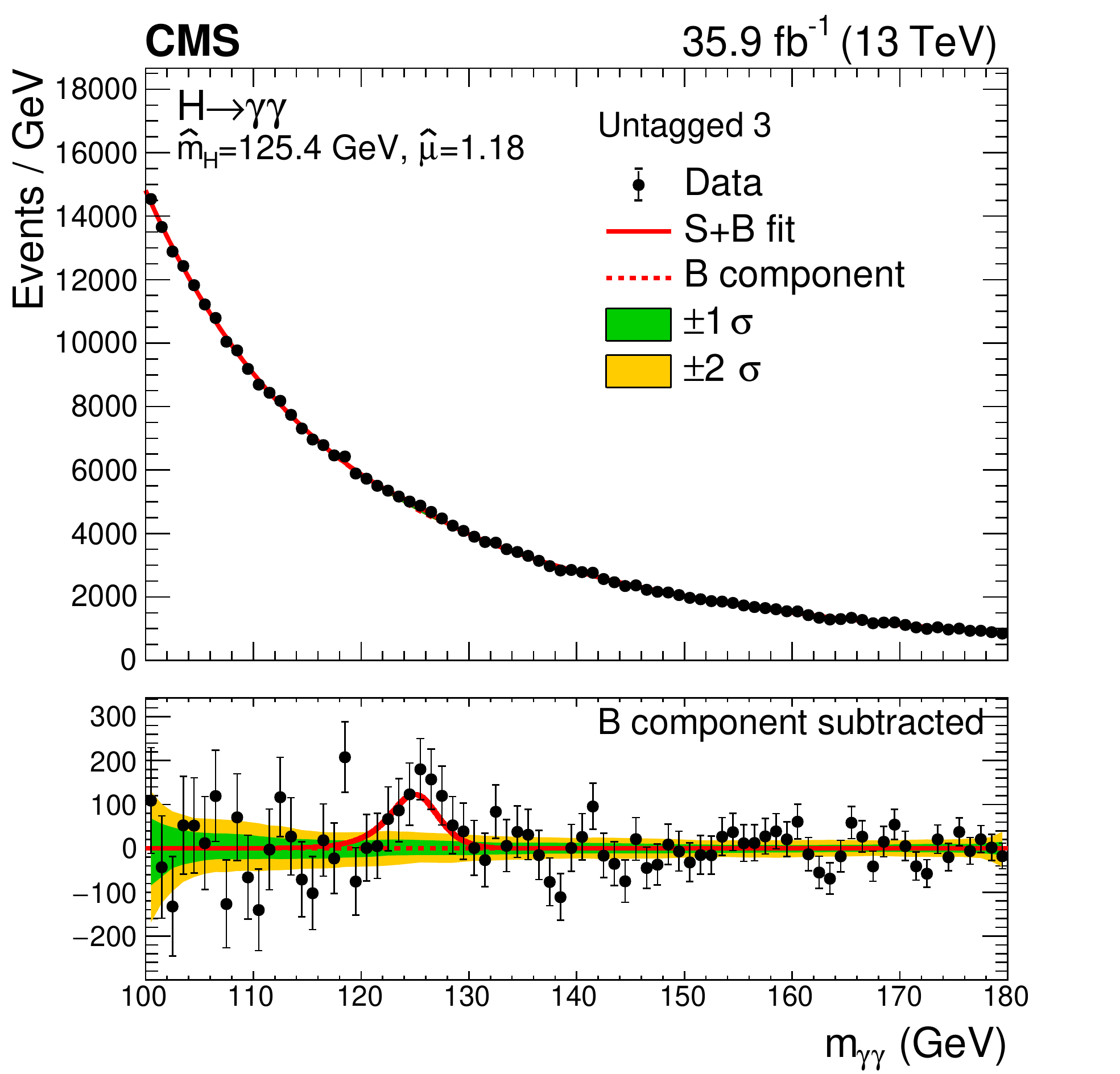}}
  \caption{Data and signal-plus-background model fits in the four
untagged categories are shown. The one (green) and two (yellow) standard
deviation bands include the uncertainties in the background component
of the fit. The lower panel in each plot shows the residuals after the
background subtraction.}
 \label{fig:statAnalysisBkgValidationPlots1}
\end{figure}

\begin{figure}[htbp]
 \centering
 {\includegraphics[width=0.45\textwidth]{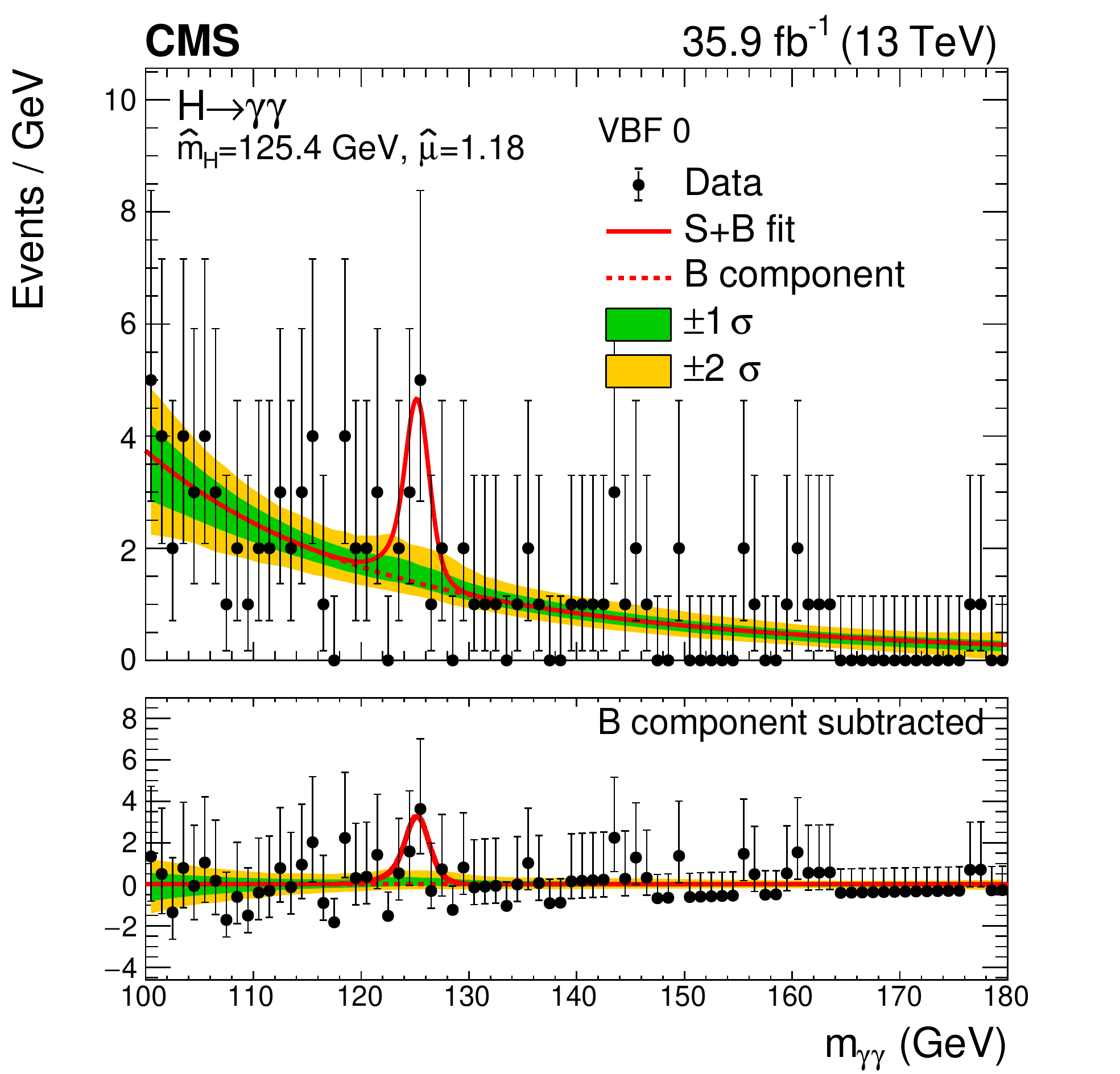}}
 {\includegraphics[width=0.45\textwidth]{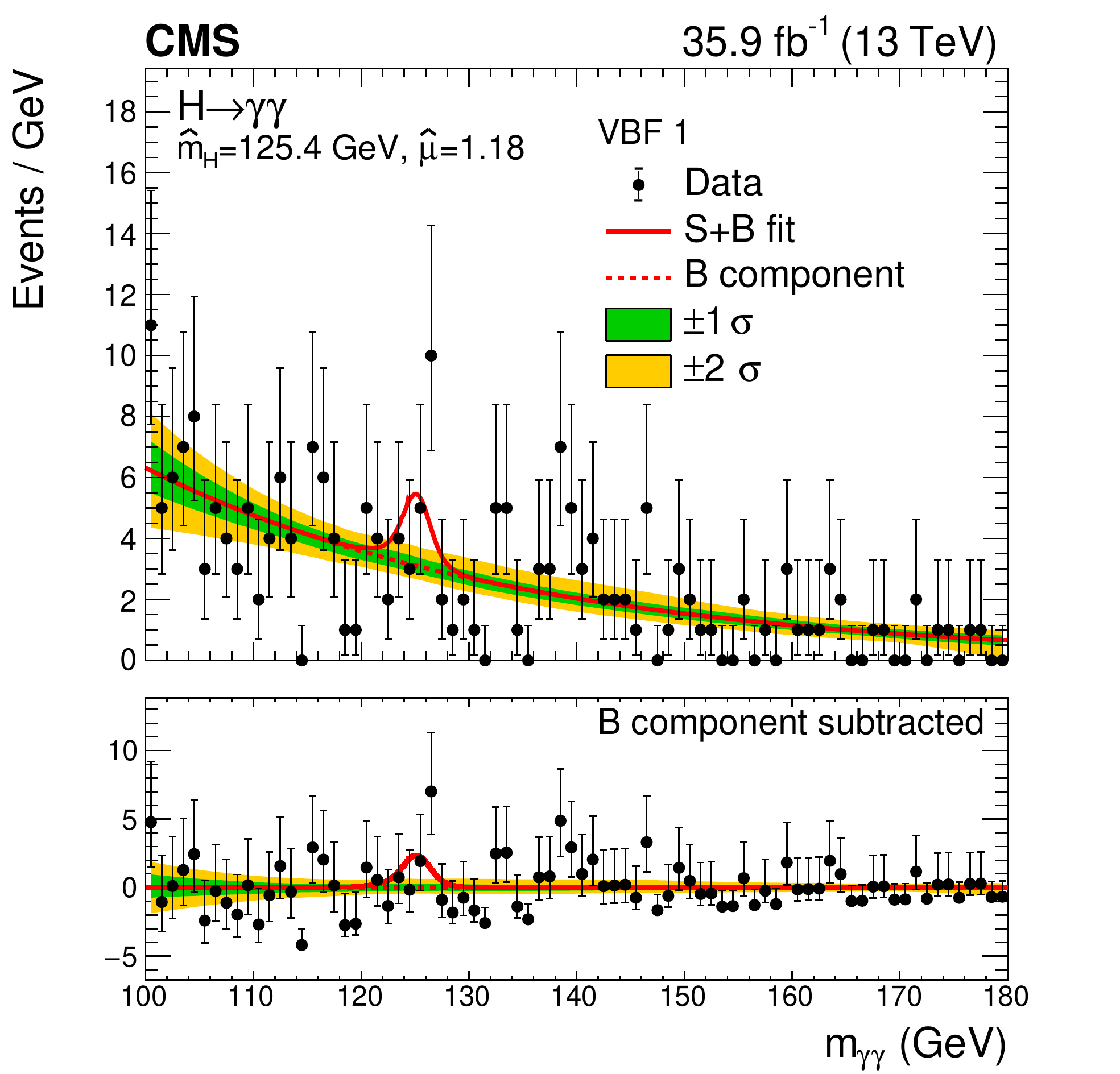}}\\
 {\includegraphics[width=0.45\textwidth]{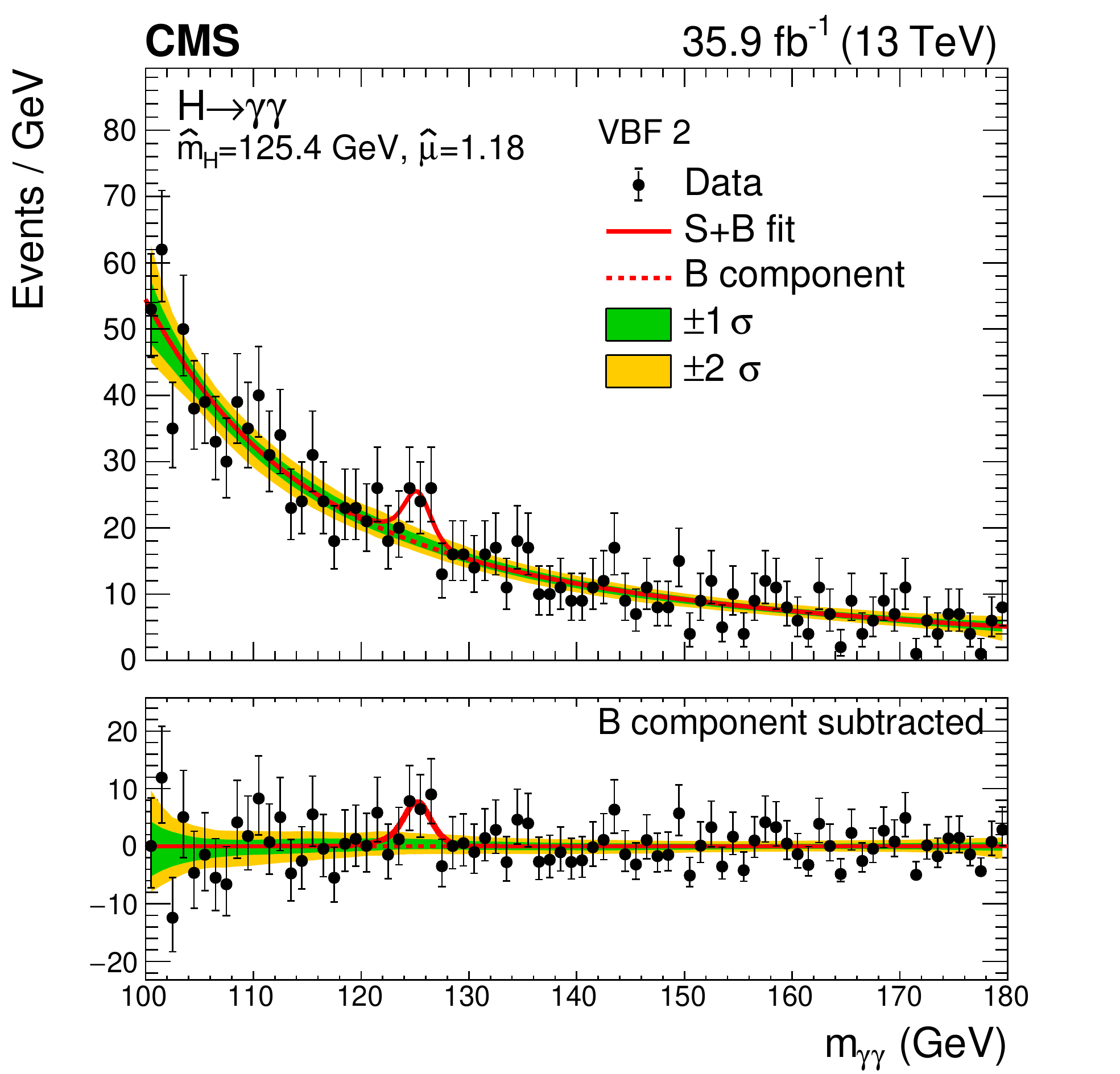}}\\
 {\includegraphics[width=0.45\textwidth]{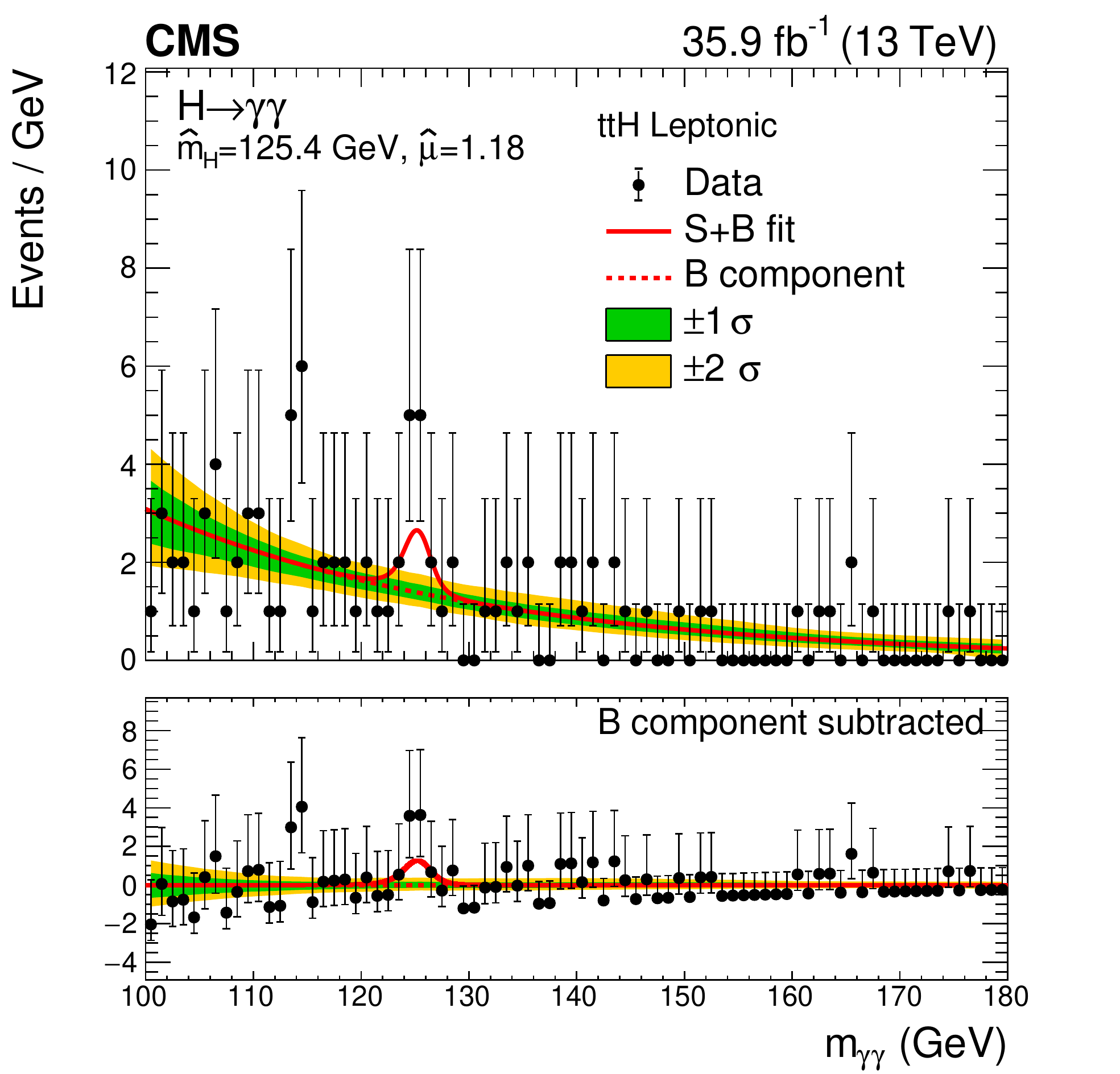}}
 {\includegraphics[width=0.45\textwidth]{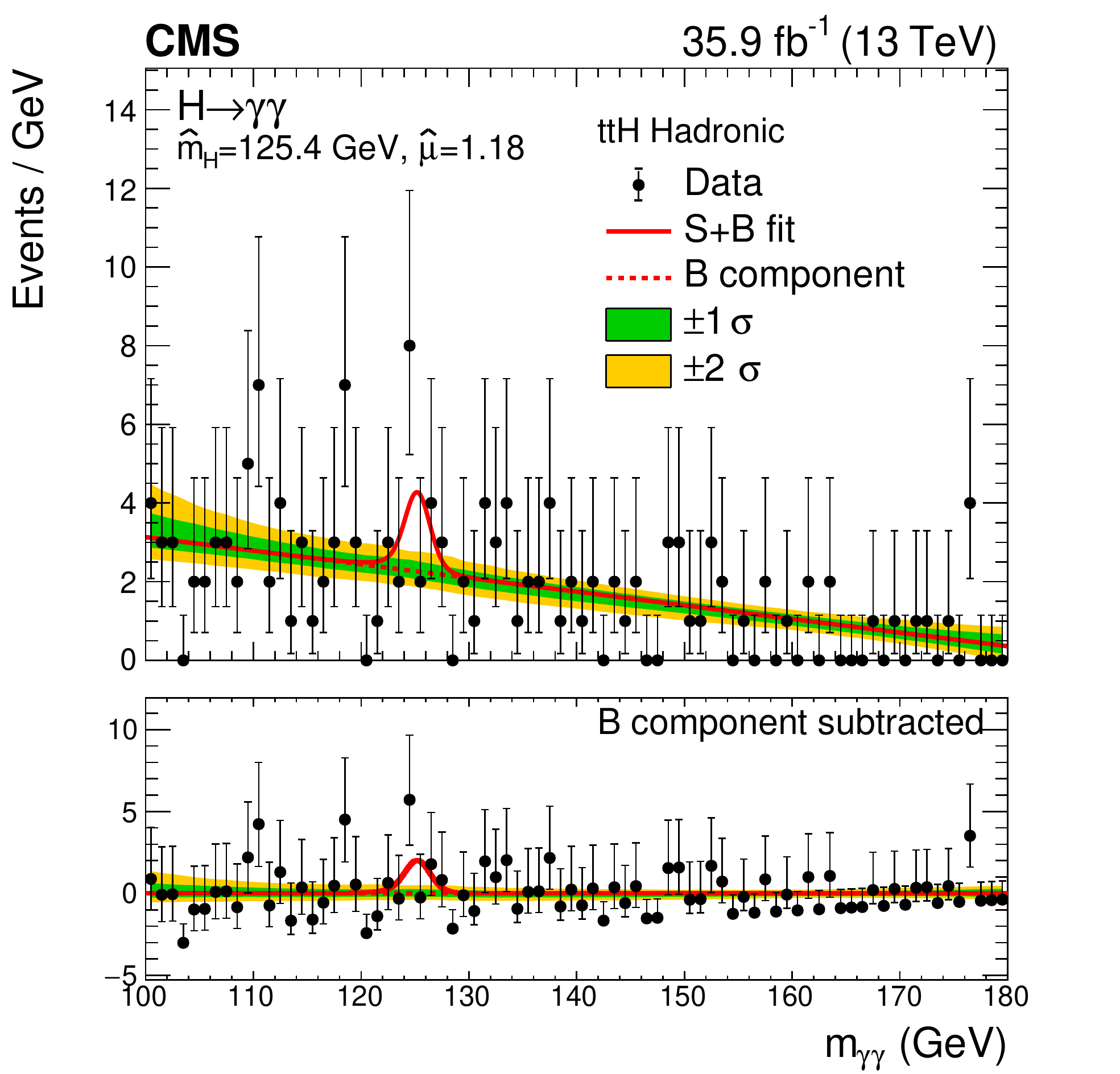}}
\
  \caption{Data and signal-plus-background model fits in \VBF and \ttH
  categories are shown. The one (green) and two (yellow) standard
  deviation bands include the uncertainties in the background component
  of the fit. The lower panel in each plot shows the residuals after the
  background subtraction.}
 \label{fig:statAnalysisBkgValidationPlots2}
\end{figure}

\begin{figure}[htbp]
 \centering
 {\includegraphics[width=0.45\textwidth]{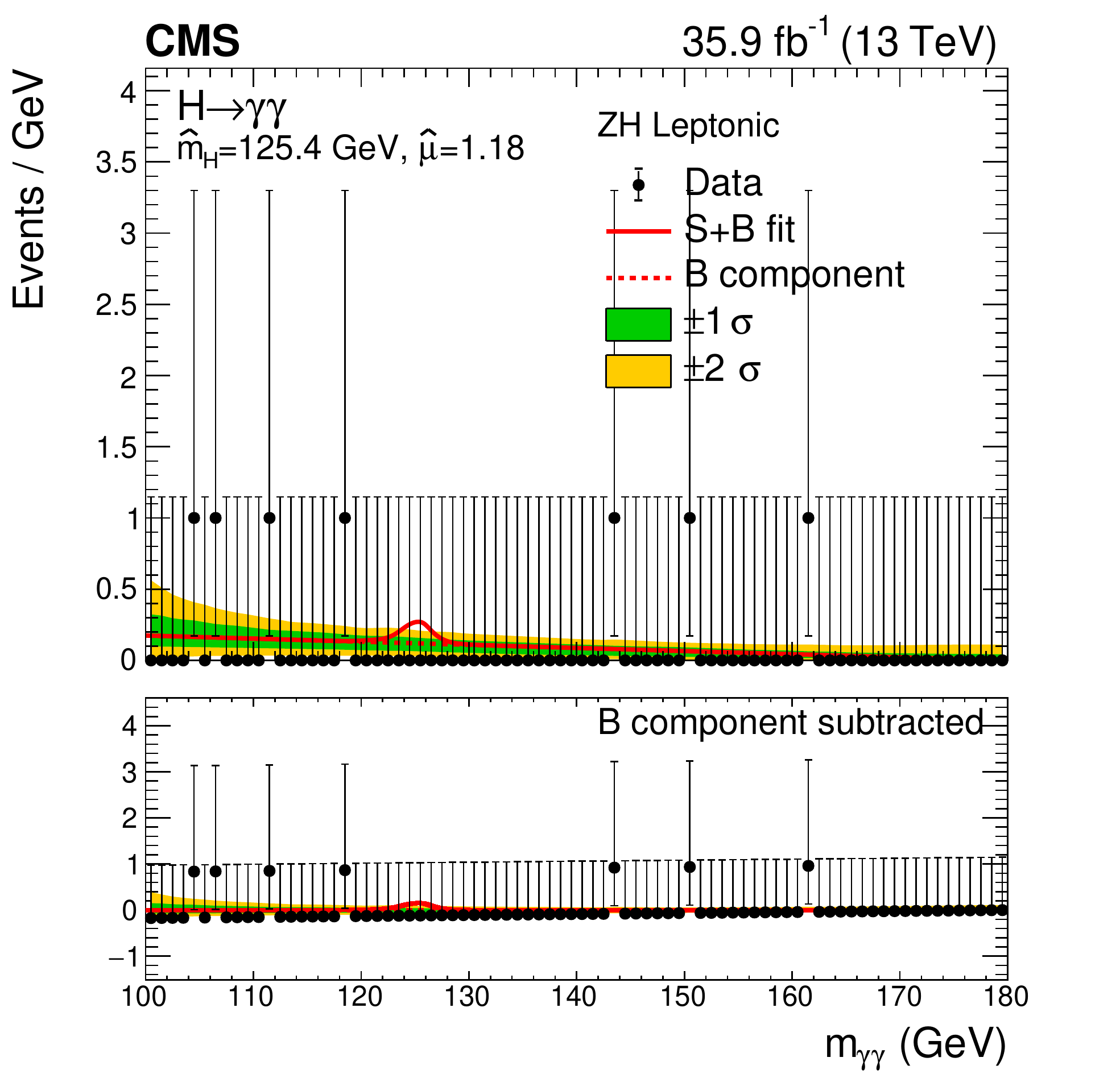}}
 {\includegraphics[width=0.45\textwidth]{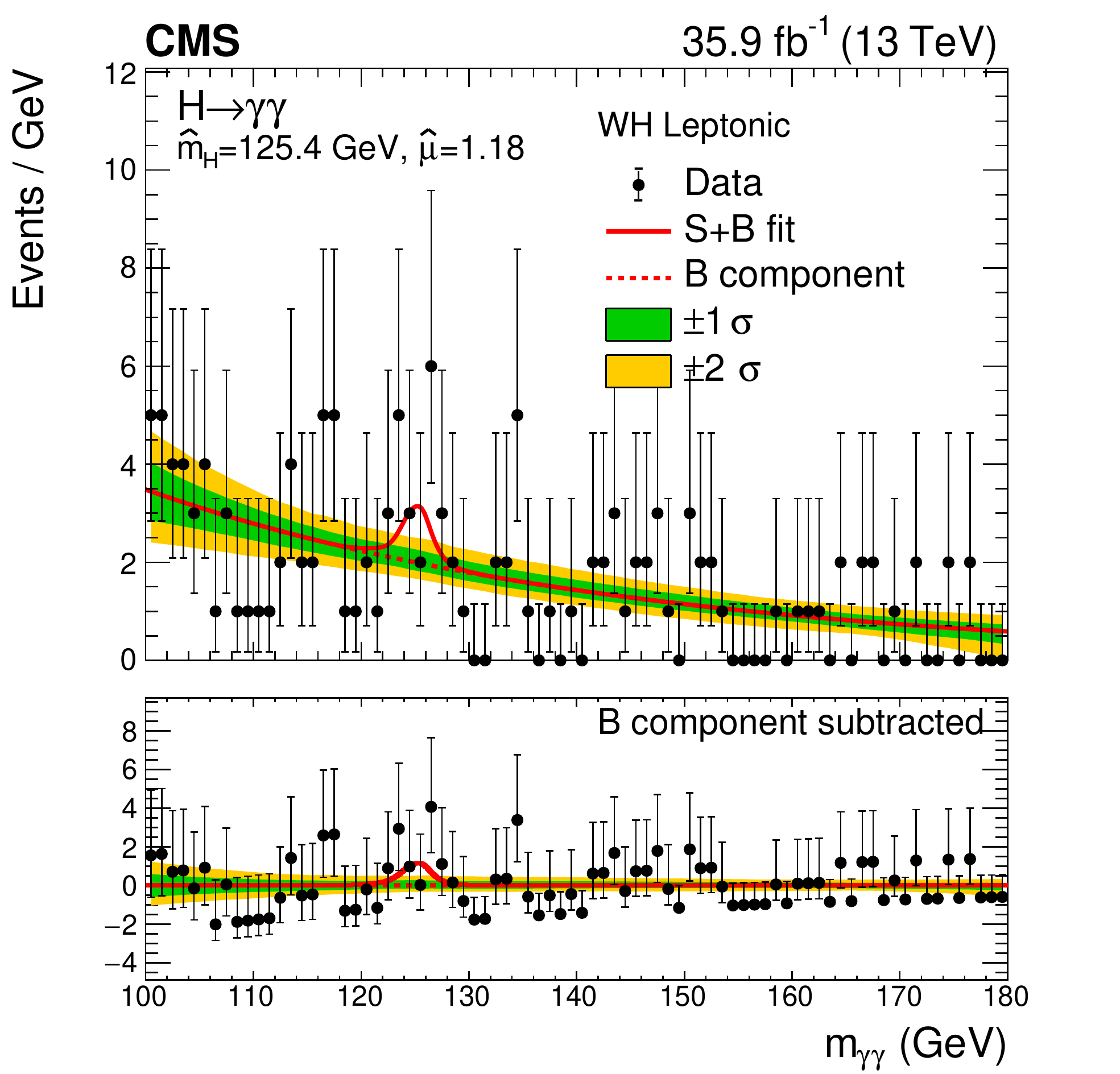}}\\
 {\includegraphics[width=0.45\textwidth]{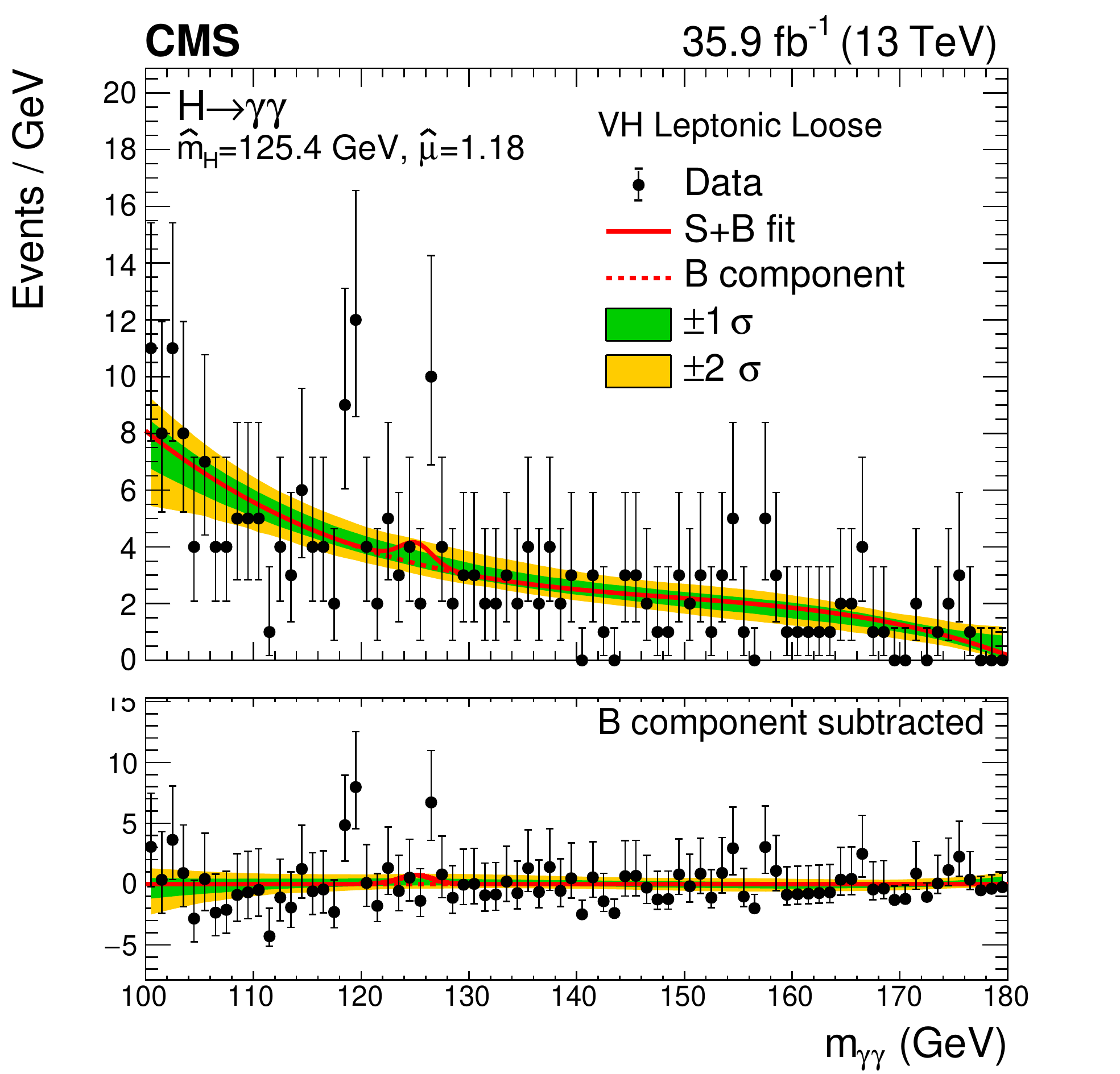}}\\
 {\includegraphics[width=0.45\textwidth]{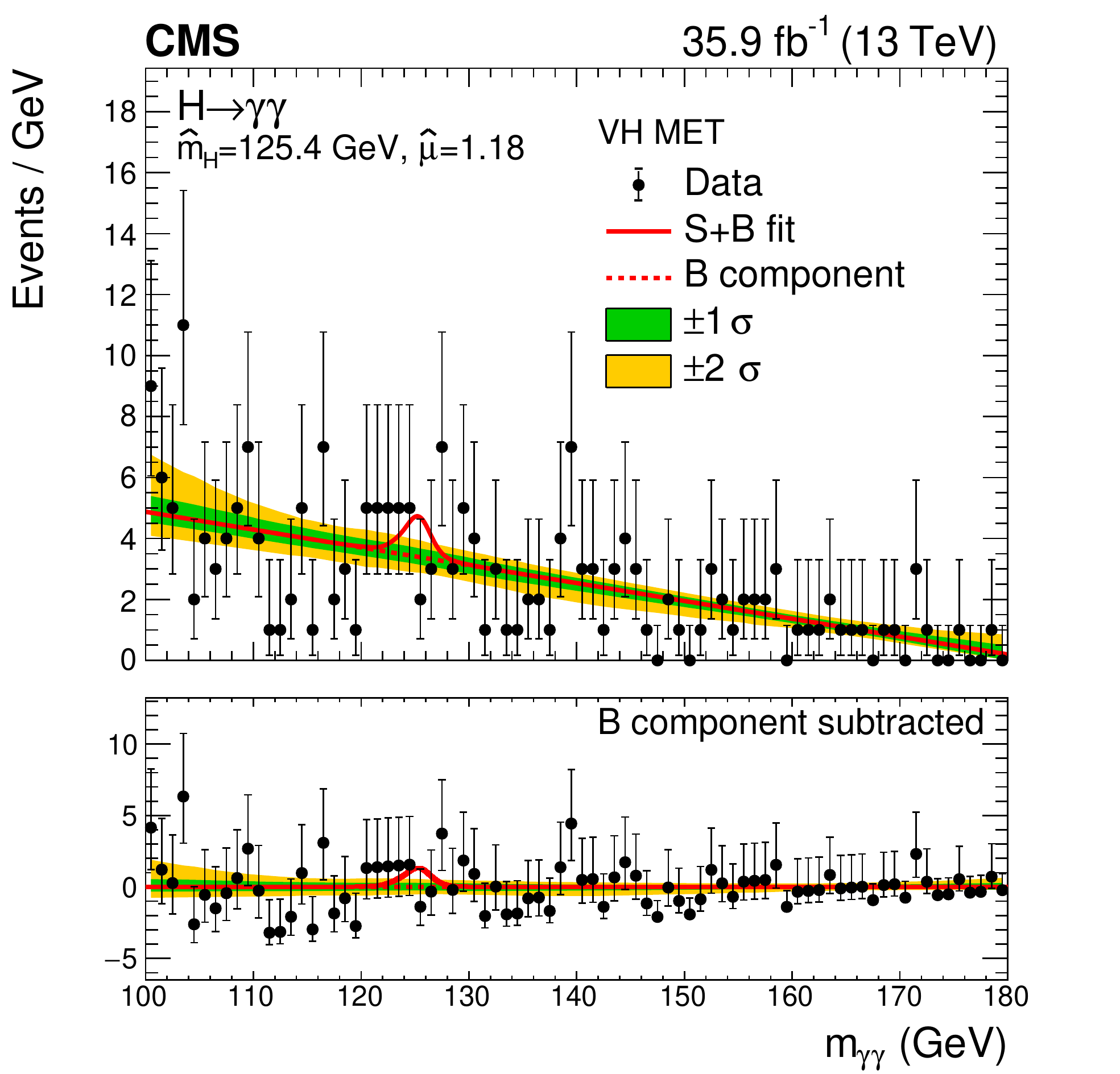}}
 {\includegraphics[width=0.45\textwidth]{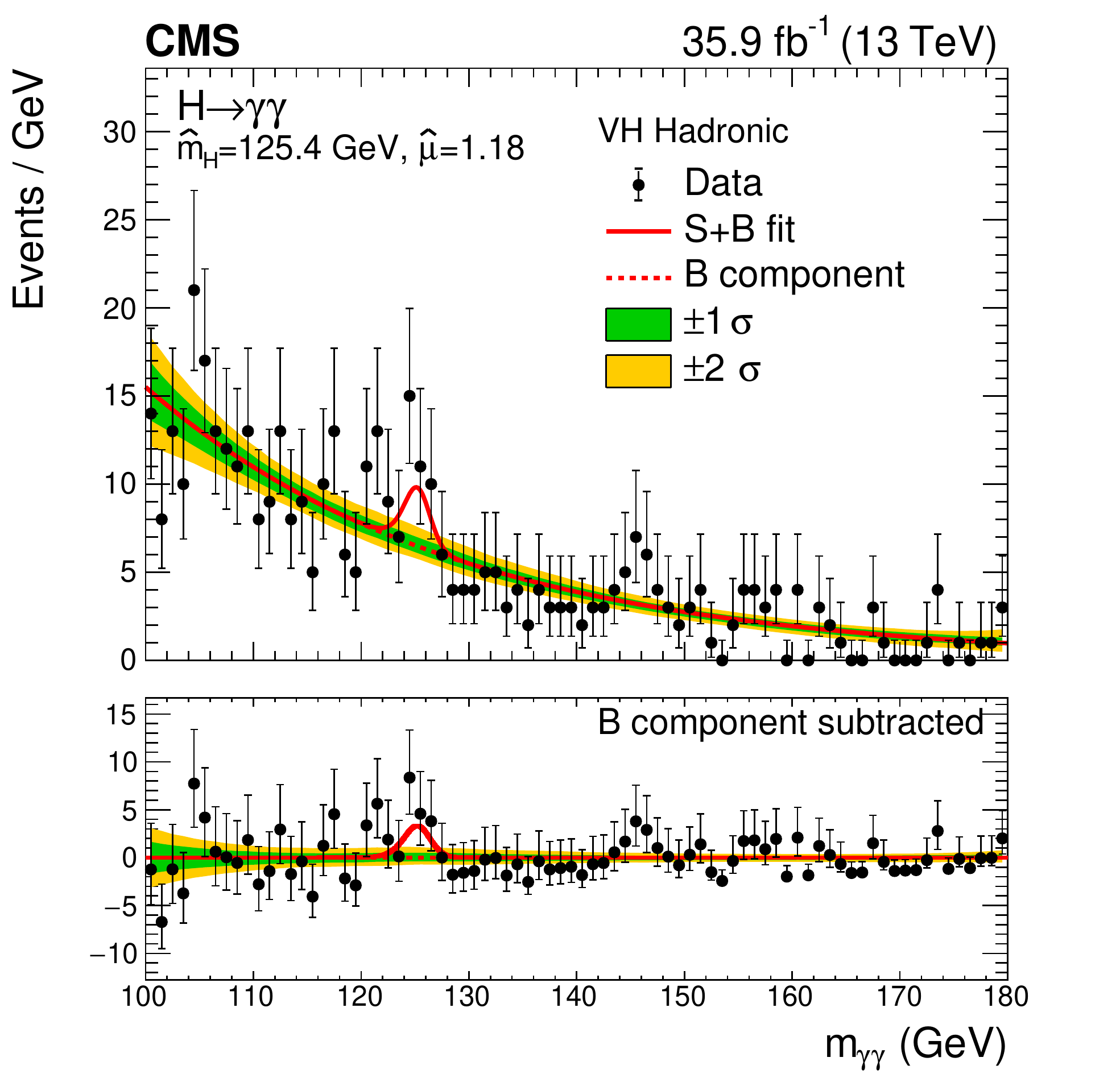}}\\
\
  \caption{Data and signal-plus-background model fits in \VH categories are shown.
  The one (green) and two (yellow) standard deviation bands
  include the uncertainties in the background component of the fit. The lower
  panel in each
  plot shows the residuals after the background subtraction.}
 \label{fig:statAnalysisBkgValidationPlots3}
\end{figure}

\begin{figure}[htbp]
 \centering
 {\includegraphics[width=0.5\textwidth]{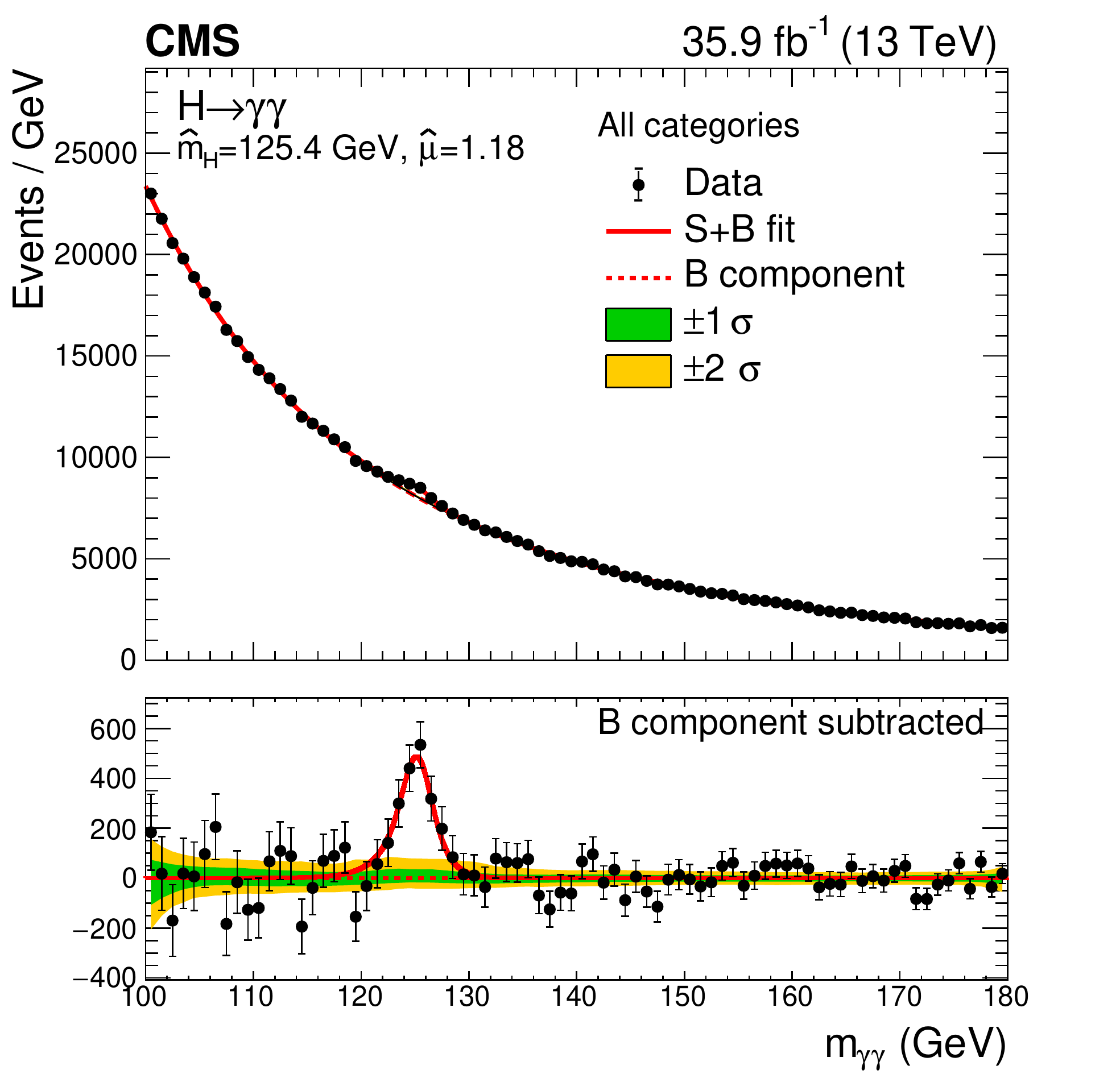}}%
 {\includegraphics[width=0.5\textwidth]{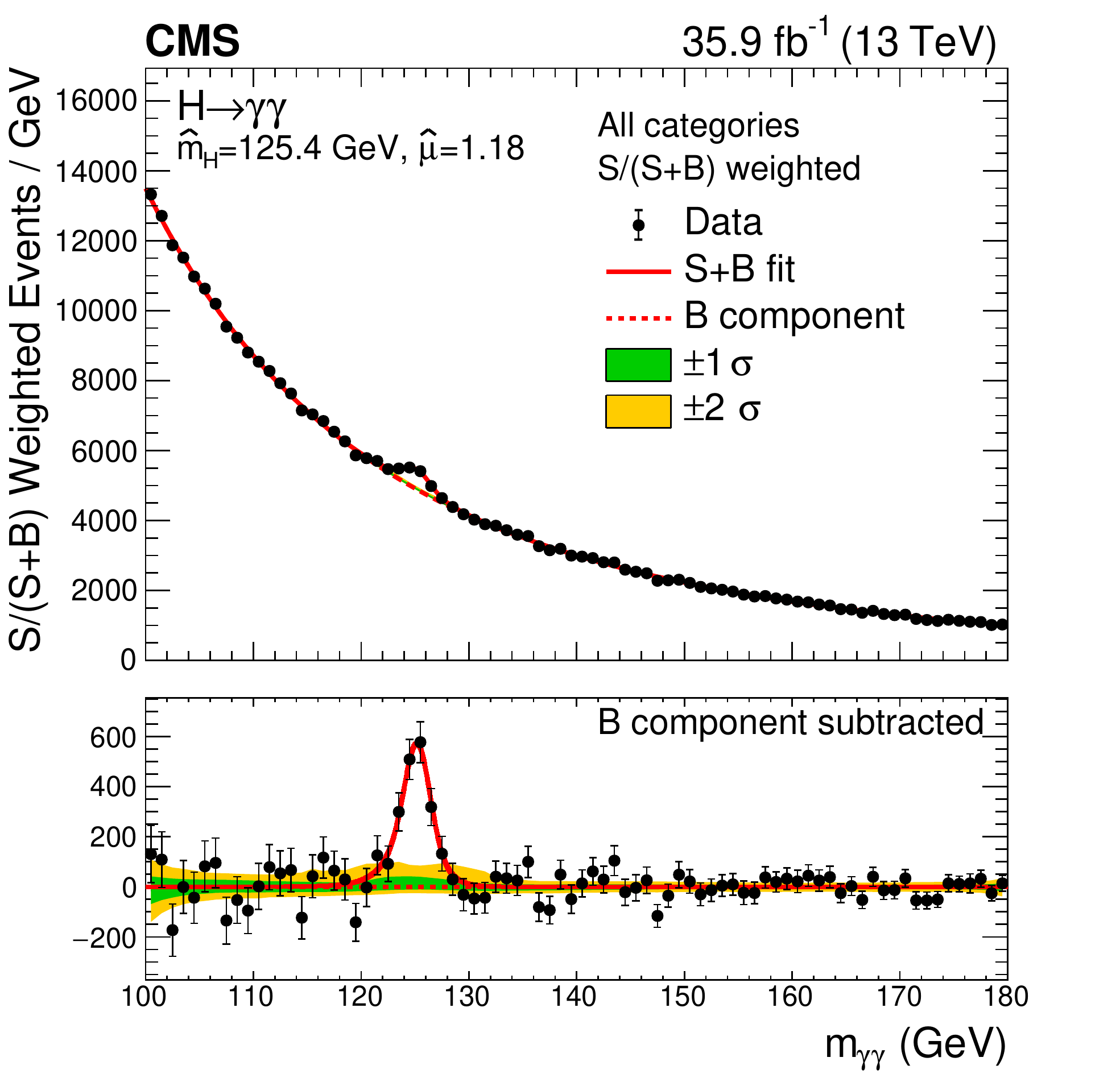}}\\
  \caption{Data and signal-plus-background model fits
for all categories summed (left) and where the categories are summed
weighted by their sensitivity (right). The one (green)
and two (yellow) standard deviation bands include the uncertainties in the
background component of the
fit. The lower panel in each plot show the residuals after the background subtraction.}
 \label{fig:mggOverall}
\end{figure}

To extract the results, binned maximum-likelihood fits are performed to
the $\mgg$ distributions of all categories, in the range $100 < \mgg <
180\GeV$, with a single overall signal strength modifier and a single
value of $\mH$ free to vary in the fit (profiled). Binned fits are
used for speed of computation, and the chosen bin size of 250\MeV is
sufficiently small compared to the mass resolution that no information
is lost.
The signal strength modifier $\mu$ is defined as the ratio of the observed
Higgs boson rate in the \HGG\ decay channel to the SM expectation.
The data and the signal-plus-background model fit for each
category are shown in
Figs.~\ref{fig:statAnalysisBkgValidationPlots1}--\ref{fig:statAnalysisBkgValidationPlots3}.
The $\mgg$ distribution for the sum of all the categories is shown in
Fig.~\ref{fig:mggOverall}.
The one (green) and two (yellow) standard deviation
bands shown for the background component of the fit include
the uncertainty in the fitted parameters.

\begin{figure}
 \centering
 {\includegraphics[width=\textwidth]{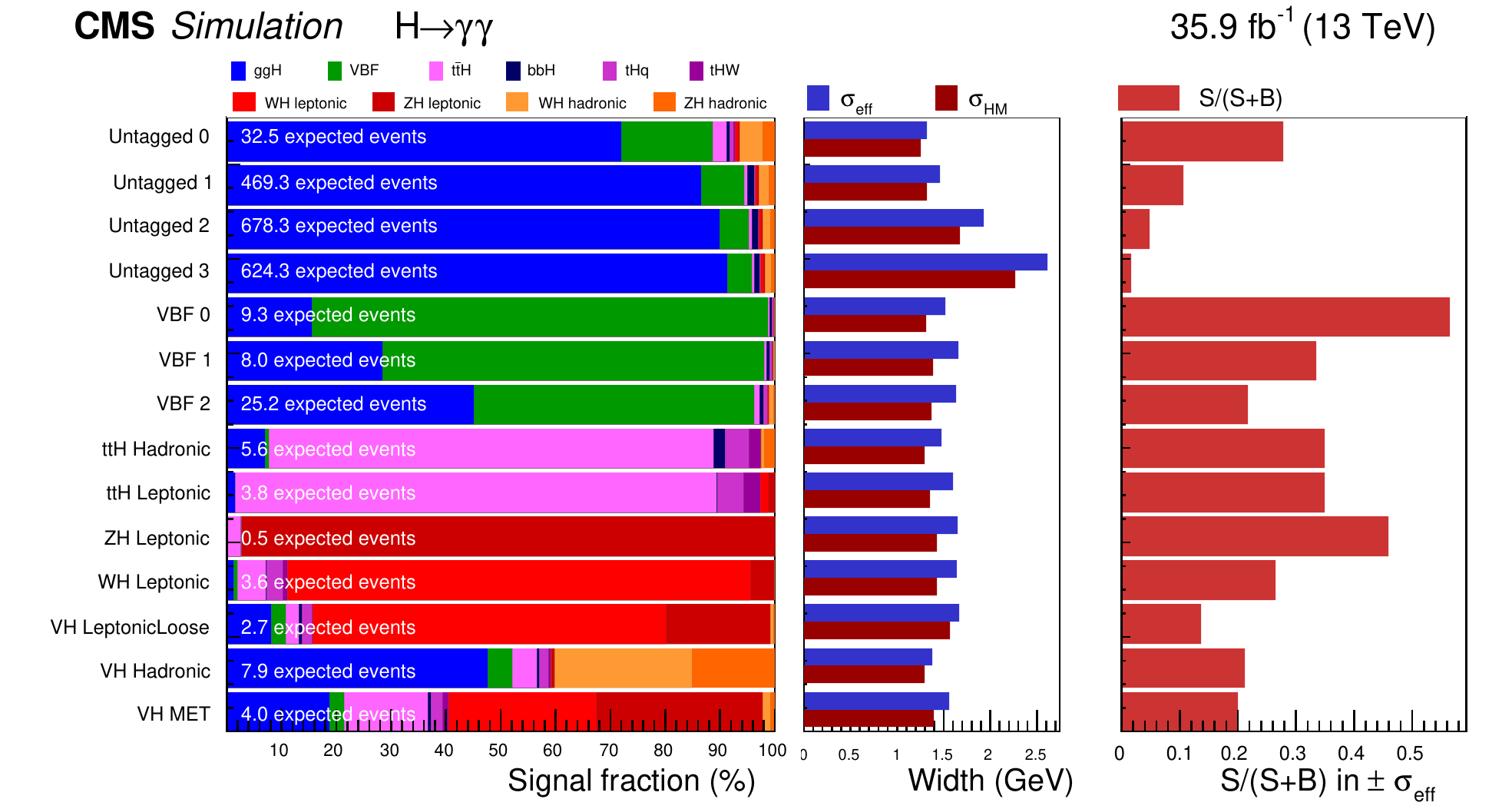}}
 \caption{Expected fraction of signal events per production mode in the
different categories. For each category, the $\sigma_{\text{eff}}$ and
$\sigma_{\text{HM}}$ of the signal model, as described in the text, are
given. The ratio of the number of signal events (S) to the number of
signal-plus-background events (S+B) is shown on the right hand side. }
 \label{fig:statAnalysisColourPlot}
\end{figure}

Table~\ref{tab:statAnalysisSigBkgYields} and
Fig.~\ref{fig:statAnalysisColourPlot} show the expected number of
signal events for each category.
The total number is broken down by the contribution (in percent) of each
production mode to any particular event category.
The $\sigma_{\text{eff}}$ and $\sigma_{\text{HM}}$ are also listed: the
former is defined as the smallest interval containing 68.3\%
of the invariant mass distribution, while the latter represents the
width of the distribution at half of its highest point (FWHM), divided by 2.35.
The table also reports the expected number of background events per \GeV
in the corresponding $\pm\sigma_{\text{eff}}$ window around 125\GeV,
using the best fit background function.

\begin{landscape}
\begin{table}
\topcaption{The expected number of signal events per category and
 the percentage breakdown per production mode in that category. The
 $\sigma_{\text{eff}}$, computed as the smallest interval containing
 68.3\% of the invariant mass distribution, and $\sigma_{\text{HM}}$,
 computed as the width of the distribution at half of its highest point
 divided by 2.35, are also shown as an estimate of the $\mgg$ resolution
 in that category. The expected number of background events per \GeV
 around 125\GeV is also listed.}
 \label{tab:statAnalysisSigBkgYields}
 \resizebox{1.5\textwidth}{!}{
\begin{tabular}{rcccccccccccccc}
\hline
\multirow{2}{*}{Event categories} &\multicolumn{13}{c}{Expected SM 125\GeV Higgs boson signal} & Bkg \\ \cline{2-14}
  &  Total & \ggH & \VBF & ttH & bbH & tHq & tHW & \WH lep & \ZH lep & \WH had & \ZH had &   $\sigma_{\text{eff}}$  & $\sigma_{\text{HM}}$ & ($\GeV^{-1}$) \\
  &  & & & & & & & & & & & (\GeVns) & (\GeVns) & \\
\hline
Untagged 0 &    32.5  &  72.0 \% &  16.6 \% &  2.6 \% &  0.6 \% &  0.7 \% &  0.3 \% &  0.6 \% &  0.3 \% &  4.2 \% &  2.2 \%& 1.32 & 1.26 & 21.8 \\
Untagged 1 &    469.3  &  86.5 \% &  7.9 \% &  0.6 \% &  1.2 \% &  0.1 \% &  $<$0.05 \% &  0.5 \% &  0.3 \% &  1.9 \% &  1.1 \%& 1.46 & 1.32 & 925.1 \\
Untagged 2 &    678.3  &  89.9 \% &  5.4 \% &  0.4 \% &  1.2 \% &  0.1 \% &  $<$0.05 \% &  0.5 \% &  0.3 \% &  1.4 \% &  0.8 \%& 1.93 & 1.67 & 2391.7 \\
Untagged 3 &    624.3  &  91.3 \% &  4.4 \% &  0.5 \% &  1.0 \% &  0.1 \% &  $<$0.05 \% &  0.5 \% &  0.3 \% &  1.2 \% &  0.7 \%& 2.61 & 2.27 & 4855.1 \\
\VBF 0 &    9.3  &  15.5 \% &  83.2 \% &  0.4 \% &  0.4 \% &  0.3 \% &  $<$0.05 \% &  $<$0.05 \% &  $<$0.05 \% &  0.2 \% &  $<$0.05 \%& 1.52 & 1.31 & 1.6 \\
\VBF 1 &    8.0  &  28.4 \% &  69.7 \% &  0.4 \% &  0.6 \% &  0.4 \% &  $<$0.05 \% &  0.1 \% &  $<$0.05 \% &  0.3 \% &  0.1 \%& 1.66 & 1.38 & 3.3 \\
\VBF 2 &    25.2  &  45.1 \% &  51.2 \% &  0.9 \% &  0.8 \% &  0.6 \% &  0.1 \% &  0.2 \% &  0.1 \% &  0.8 \% &  0.3 \%& 1.64 & 1.37 & 18.9 \\
\cttH Hadronic &    5.6  &  7.0 \% &  0.7 \% &  81.1 \% &  2.1 \% &  4.3 \% &  2.1 \% &  0.1 \% &  0.1 \% &  0.7 \% &  1.9 \%& 1.48 & 1.30 & 2.4 \\
\cttH Leptonic &    3.8  &  1.5 \% &  $<$0.05 \% &  87.8 \% &  0.1 \% &  4.7 \% &  3.1 \% &  1.5 \% &  1.2 \% &  $<$0.05 \% &  $<$0.05 \%& 1.60 & 1.35 & 1.5 \\
\ZH Leptonic &    0.5  &  $<$0.05 \% &  $<$0.05 \% &  2.6 \% &  $<$0.05 \% &  $<$0.05 \% &  0.1 \% &  $<$0.05 \% &  97.3 \% &  $<$0.05 \% &  $<$0.05 \%& 1.65 & 1.43 & 0.1 \\
\WH Leptonic &    3.6  &  1.3 \% &  0.6 \% &  5.2 \% &  0.2 \% &  3.0 \% &  0.7 \% &  84.5 \% &  4.3 \% &  0.1 \% &  0.1 \%& 1.64 & 1.43 & 2.1 \\
\VH LeptonicLoose &    2.7  &  8.1 \% &  2.7 \% &  2.4 \% &  0.6 \% &  1.8 \% &  0.1 \% &  64.4 \% &  19.1 \% &  0.6 \% &  0.2 \%& 1.67 & 1.56 & 3.5 \\
\VH Hadronic &    7.9  &  47.6 \% &  4.5 \% &  4.4 \% &  0.4 \% &  1.7 \% &  0.3 \% &  0.2 \% &  0.5 \% &  25.2 \% &  15.1 \%& 1.38 & 1.30 & 7.2 \\
\VH MET &    4.0  &  18.7 \% &  2.6 \% &  15.4 \% &  0.4 \% &  2.1 \% &  1.2 \% &  26.8 \% &  30.4 \% &  1.4 \% &  0.9 \%& 1.56 & 1.39 & 3.5 \\
Total &    1875.0  &  86.9 \% &  7.1 \% &  1.0 \% &  1.1 \% &  0.2 \% &  $<$0.05 \% &  0.8 \% &  0.4 \% &  1.6 \% &  0.9 \%& 1.96 & 1.62 & 8237.8 \\
\hline
\end{tabular}
}
\end{table}
\end{landscape}

\begin{figure}[htbp]
	\centering
		\includegraphics[width=0.7\textwidth]{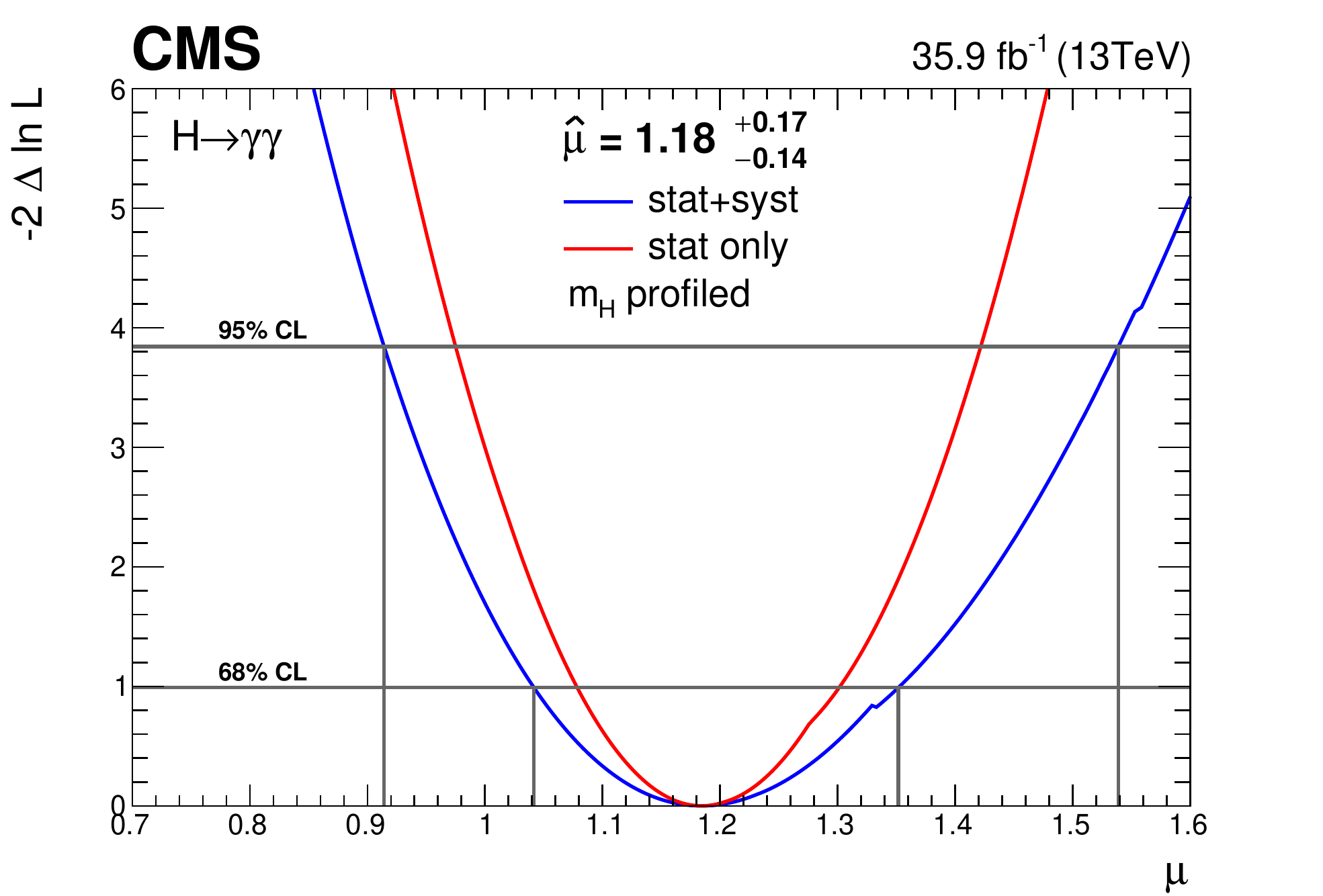}
  \caption{The likelihood scan for the signal strength modifier where
  the value of the SM Higgs boson mass is profiled in the fit.}
		\label{fig:statAnalysisMu}
\end{figure}

A likelihood scan of the signal strength modifier is performed, with
other parameters of the signal and background models allowed to vary.
Systematic uncertainties are included in the form of nuisance parameters
and the results are obtained using an asymptotic
approach~\cite{LHC-HCG,Junk:1999kv,Read:2002hq} with a
test statistic based on the profile likelihood ratio ($q$)~\cite{Cowan:2010st}.
The individual contributions of the statistical and systematic uncertainties
are separated by performing a likelihood scan removing the systematic
uncertainties to determine the statistical uncertainty. The systematic
uncertainty is then taken as the difference in quadrature between the
total uncertainty and the statistical uncertainty.
The results can be found in Fig.~\ref{fig:statAnalysisMu}.
The best fit signal strength modifier measured for all categories
combined using this method is
$\muhat = 1.18^{+0.17}_{-0.14}=1.18\ ^{+0.12}_{-0.11}\stat ^{+0.09}_{-0.07}\syst ^{+0.07}_{-0.06}\thy$.
The best fit mass is found at $\mHhat = 125.4\pm 0.3\GeV = 125.4
\pm 0.2\stat \pm 0.2\syst\GeV$, compatible with the combined mass
measurement from ATLAS and CMS~\cite{ref:2015zhl}. A precise determination of the
systematic uncertainties affecting the best fit mass is not within the
scope of this analysis. The maximum relative variation of $\muhat$ for
$\mH$ within a range of $\pm 1\GeV$ around $125\GeV$ is less than 2\%.

\begin{figure}[htbp]
 \centering
        \includegraphics[width=0.7\textwidth]{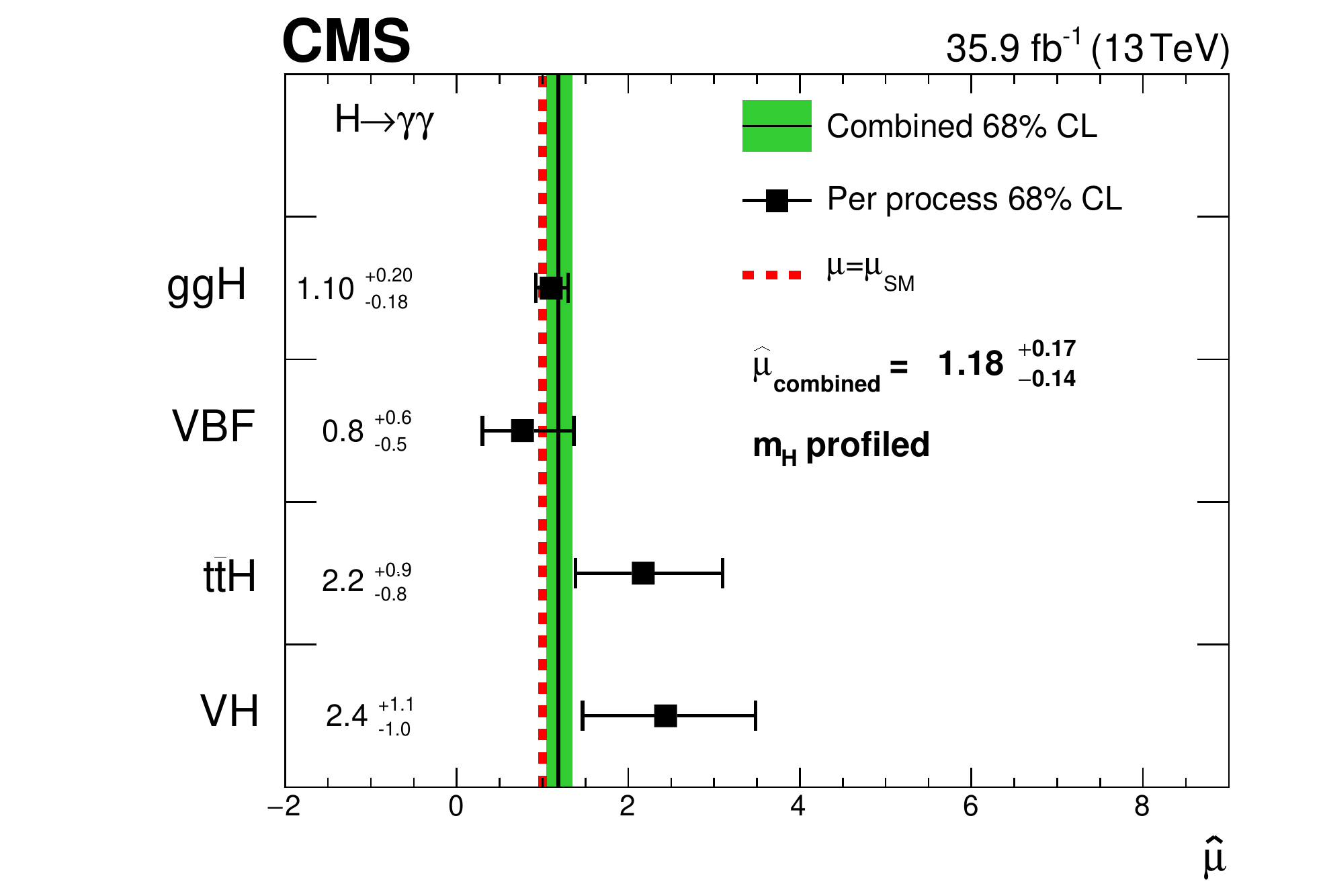}
 \caption{Signal strength modifiers measured for each process (black
 points), with the SM Higgs boson mass profiled, compared to the overall
 signal strength modifier (green band) and to the SM expectation (dashed red
 line). }
 \label{fig:perproc channel compatibility}
\end{figure}

The results of a fit to the signal strength modifier for each production
mode, defined analogously to the overall $\mu$ above, are shown in
Fig.~\ref{fig:perproc channel compatibility} and summarized in Table~\ref{tab:mus}.
The observed rates of the \VBF, \ttH, and \VH production modes correspond
respectively to $p$-values of 4.2, 0.074, and 0.47\%, with respect to the
absence of the considered production mode. The expected $p$-values are 1.8,
7.3, and 12\%, respectively, for an SM Higgs boson, with the current data set.

\begin{table}[htbp]
\centering
\topcaption{Results of the fit to the signal strength modifier for each
production mode. The total uncertainties as well as a their statistical,
systematic, and theory components are shown. The last two columns
report the $p$-value relative to the observed rates and referred to
the abscence of the considered production mode, and its corresponding
estimated significance.}
 \label{tab:mus}
\begin{tabular}{rccccccc}
\hline
\multirow{2}{*}{Process} & \multirow{2}{*}{$\muhat$} & \multicolumn{4}{c}{Uncertainties} & \multirow{2}{*}{$p$-value} & Estimated significance\\
        & & tot & stat & syst & theo   &           & (standard deviations)\\[.15cm]
\hline
\ggH       & $1.10$ & $^{+0.20}_{-0.18}$ & $^{+0.15}_{-0.15}$ & $^{+0.09}_{-0.08}$ & $^{+0.08}_{-0.06}$ & 3.1$\ten{-12}$  & 6.9 \\[.15cm]
\VBF       &  $0.8$ & $^{+0.6}_{-0.5}$   & $^{+0.5}_{-0.4}  $ & $^{+0.3}_{-0.2}$   & $^{+0.2}_{-0.1}$   & 4.2$\ten{-2}$   & 1.7 \\[.15cm]
\ttH       &  $2.2$ & $^{+0.9}_{-0.8}$   & $^{+0.9}_{-0.8}  $ & $^{+0.2}_{-0.1}$   & $^{+0.2}_{-0.1}$   & 7.4$\ten{-4}$   & 3.2 \\[.15cm]
\VH        &  $2.4$ & $^{+1.1}_{-1.0}$   & $^{+1.0}_{-1.0}  $ & $^{+0.2}_{-0.1}$   & $^{+0.2}_{-0.1}$   & 4.7$\ten{-3}$   & 2.6 \\
\hline
\end{tabular}
\end{table}

\begin{figure}[htbp]
 \centering
        \includegraphics[width=0.7\textwidth]{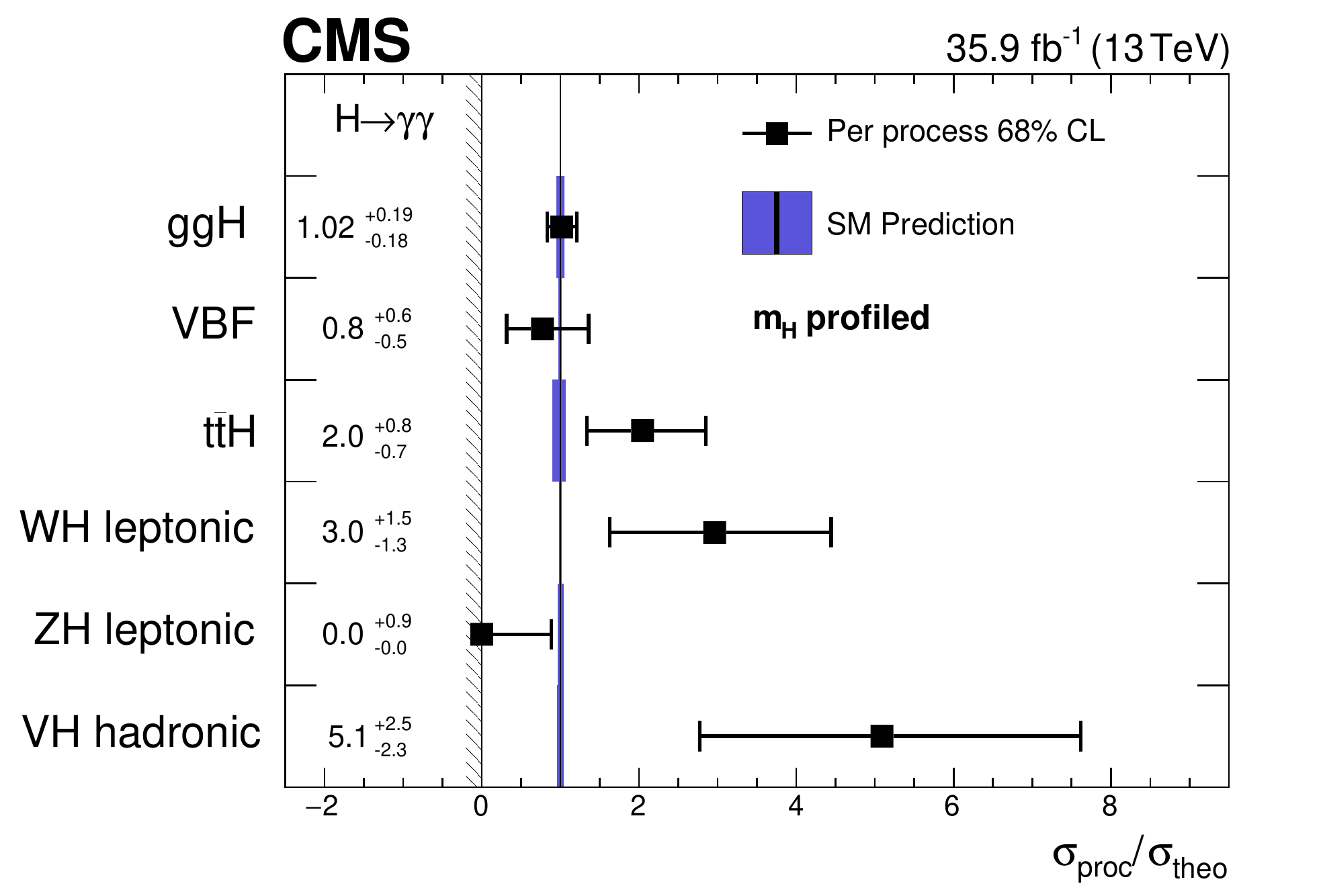}
        \caption{Cross section ratios measured for each process
        (black points) in the Higgs simplified template cross
        section framework~\cite{LHCHXSWG}, with the SM Higgs boson
        mass profiled, compared to the SM expectations and their
        uncertainties (blue band). The signal strength modifiers are
        constrained to be nonnegative, as indicated by the vertical line
        and hashed pattern at zero.
}
 \label{fig:STX perproc channel compatibility}
\end{figure}

A similar fit is performed to extract the ratios of observed cross
sections to the SM prediction in the stage 0 of the simplified template
cross section (STXS) framework~\cite{LHCHXSWG}. These cross sections are
for a reduced fiducial volume, defined by requiring the Higgs boson
rapidity to be less than 2.5. Outside of this volume the analysis has a
negligible acceptance.
The ratios are measured for the \ggH, \VBF, \ttH, and \VH
production processes. \VH is further split considering the decay of the
associated boson into \WH leptonic, \ZH leptonic, and \VH hadronic, which
groups hadronic decays of both the \PW\ and \PZ\ bosons.
The STXS approach differs from the signal strength modifier measurements
in the splitting of the production modes, and reduces the dependence of the
measurements on the theoretical uncertainties in the SM predictions, by
avoiding the sizeable uncertainty associated with the extrapolation to
the full phase space. The measured cross section ratios, where the SM
prediction~\cite{LHCHXSWG} is denoted as $\sigma_\mathrm{theo}$,
are shown in Fig.~\ref{fig:STX perproc channel compatibility}.

\begin{figure}[htbp]
 \centering
   \includegraphics[width=0.7\textwidth]{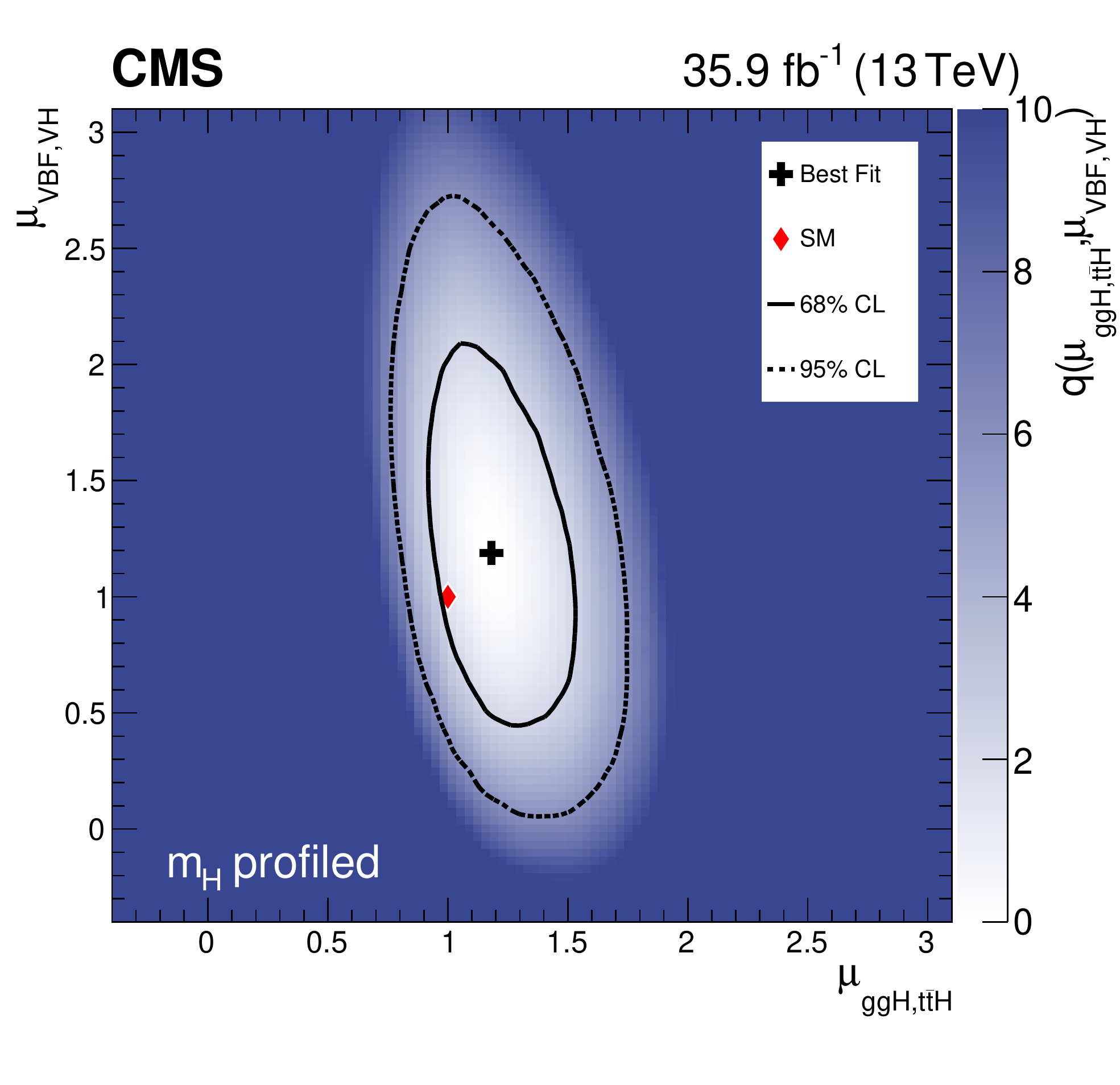}
   \caption{The two-dimensional best fit (black cross) of the signal
	   strength modifiers for fermionic (\ggH, \ttH) and bosonic (\VBF,
	   \ZH, \WH) production modes compared to the SM expectation (red diamond).
	   The Higgs boson mass is profiled in the fit. The solid (dashed) line
	   represents the 68\,(95)\% confidence region.}
 \label{fig:rvrf}
\end{figure}

A two-dimensional likelihood scan of the signal strength modifier
$\mu_{\ggH,\ttH}$ for fermionic production modes (\ggH and
\ttH) and $\mu_{\VBF,\VH}$ for vector boson production
modes (\VBF, \ZH, \WH), with the value of the parameter $\mH$ profiled
in the fit, is performed. Figure~\ref{fig:rvrf} shows the  68 and 95\%
confidence level (\CL) contours.
The best fit values for each modifier are
$\muhat_{\ggH,\ttH} = 1.19^{+0.22}_{-0.18}$
and
$\muhat_{\VBF,\VH}=1.21^{+0.58}_{-0.51}$.

\begin{figure}[h]
\centering
  {\includegraphics[width=0.5\textwidth]{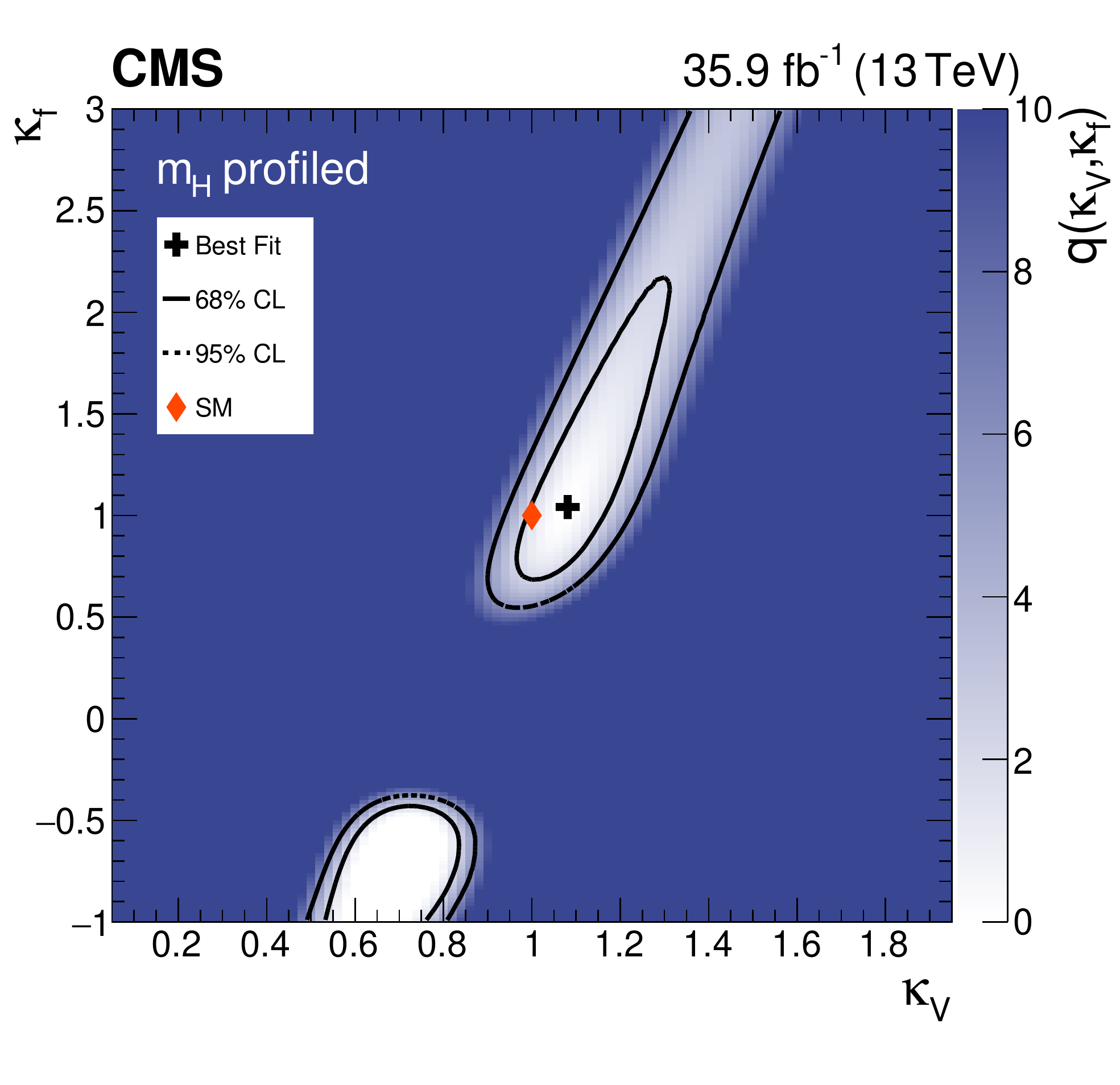}}%
  {\includegraphics[width=0.5\textwidth]{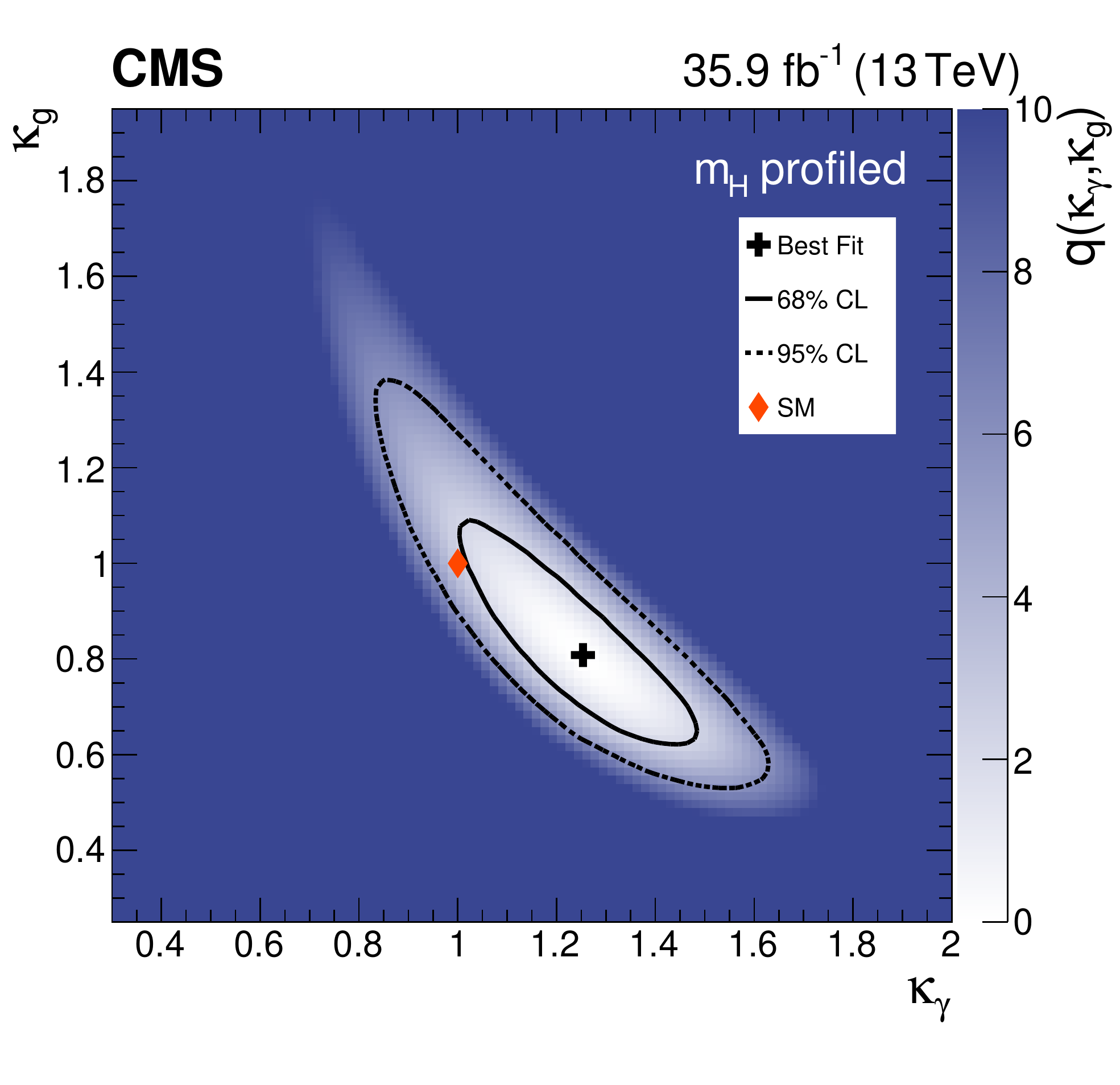}}
\caption{Two-dimensional likelihood scans of $\kappaf$ versus
        $\kappaV$
  (left) and $\kappagluon$ versus $\kappagamma$ (right).
  All four variables are expressed relative to the SM expectations.
  The mass of the Higgs boson is profiled in the fits. The crosses indicate
  the best fit values, the diamonds indicate the standard model
  expectations.
  The colour maps indicate the value of the test statistic $q$
  as described in the text.
}
\label{fig:kVkF_kGlukGam}
\end{figure}

Deviations from the SM expectation in the couplings of the Higgs boson can be parameterized
using coupling modifiers in the so-called $\kappa$ framework~\cite{Heinemeyer:2013tqa}.
Two-dimensional likelihood scans of the Higgs boson coupling modifiers are produced:
$\kappaf$ versus $\kappaV$, the coupling
modifiers to fermions and bosons; and $\kappagluon$ versus
$\kappagamma$, the effective coupling modifiers to gluons and photons.
The $\kappa$ parameters other than those varied are fixed to 1 in each case.
Figure~\ref{fig:kVkF_kGlukGam} shows the test statistic $q$ and the 68\% and
95\% \CL  contours for each scan.
The point $(\kappaV,\kappaf)=(1,-1)$ has an observed (expected) $q$
value of 35.2\,(53.7),
inconsistent with the observed (expected) best fit value at the level of 5.8\,(7.0) standard deviations.

\section{Summary}
\label{sec:summary}

We report measurements of the production cross section and couplings of the
Higgs boson using its diphoton decay:
the overall signal strength modifier;
the signal strength modifier for each production mode separately;
cross section ratios for the stage 0 simplified template cross section
framework;
the best fit rates in the $\mu_{\VBF,\VH}$-$\mu_{\ggH,\ttH}$ plane
with \VBF and \VH production, and \ggH and \ttH production, varied together;
and the best fit coupling modifiers
in the $\kappaf$-$\kappaV$ and $\kappagluon$-$\kappagamma$ planes.
The analysis is based on
proton-proton collision data collected at $\sqrt{s} = 13\TeV$ by the CMS
experiment at the LHC in 2016, corresponding to an integrated luminosity
of 35.9\fbinv.
The best fit signal strength modifier obtained after profiling $\mH$ is
${\muhat = 1.18^{+0.17}_{-0.14}=1.18\ ^{+0.12}_{-0.11}\stat
        ^{+0.09}_{-0.07}\syst ^{+0.07}_{-0.06}\thy }$.
The best fit values in the
$\mu_{\VBF,\VH}$--$\mu_{\ggH,\ttH}$ plane are
$\muhat_{\ggH,\ttH} =1.19^{+0.22}_{-0.18}$
and
$\muhat_{\VBF,\VH}=1.21^{+0.58}_{-0.51}$.
When $\mu_{\ttH}$ is considered separately, the best fit value is
$\muhat_{\ttH} = 2.2^{+0.9}_{-0.8}$, corresponding to a $p$-value of
 0.074\% with respect to the absence of \ttH production.
Stage 0 simplified template cross sections are compatible with the standard model.

\begin{acknowledgments}
\hyphenation{Bundes-ministerium Forschungs-gemeinschaft Forschungs-zentren Rachada-pisek} We congratulate our colleagues in the CERN accelerator departments for the excellent performance of the LHC and thank the technical and administrative staffs at CERN and at other CMS institutes for their contributions to the success of the CMS effort. In addition, we gratefully acknowledge the computing centres and personnel of the Worldwide LHC Computing Grid for delivering so effectively the computing infrastructure essential to our analyses. Finally, we acknowledge the enduring support for the construction and operation of the LHC and the CMS detector provided by the following funding agencies: the Austrian Federal Ministry of Science, Research and Economy and the Austrian Science Fund; the Belgian Fonds de la Recherche Scientifique, and Fonds voor Wetenschappelijk Onderzoek; the Brazilian Funding Agencies (CNPq, CAPES, FAPERJ, and FAPESP); the Bulgarian Ministry of Education and Science; CERN; the Chinese Academy of Sciences, Ministry of Science and Technology, and National Natural Science Foundation of China; the Colombian Funding Agency (COLCIENCIAS); the Croatian Ministry of Science, Education and Sport, and the Croatian Science Foundation; the Research Promotion Foundation, Cyprus; the Secretariat for Higher Education, Science, Technology and Innovation, Ecuador; the Ministry of Education and Research, Estonian Research Council via IUT23-4 and IUT23-6 and European Regional Development Fund, Estonia; the Academy of Finland, Finnish Ministry of Education and Culture, and Helsinki Institute of Physics; the Institut National de Physique Nucl\'eaire et de Physique des Particules~/~CNRS, and Commissariat \`a l'\'Energie Atomique et aux \'Energies Alternatives~/~CEA, France; the Bundesministerium f\"ur Bildung und Forschung, Deutsche Forschungsgemeinschaft, and Helmholtz-Gemeinschaft Deutscher Forschungszentren, Germany; the General Secretariat for Research and Technology, Greece; the National Research, Development and Innovation Fund, Hungary; the Department of Atomic Energy and the Department of Science and Technology, India; the Institute for Studies in Theoretical Physics and Mathematics, Iran; the Science Foundation, Ireland; the Istituto Nazionale di Fisica Nucleare, Italy; the Ministry of Science, ICT and Future Planning, and National Research Foundation (NRF), Republic of Korea; the Lithuanian Academy of Sciences; the Ministry of Education, and University of Malaya (Malaysia); the Mexican Funding Agencies (BUAP, CINVESTAV, CONACYT, LNS, SEP, and UASLP-FAI); the Ministry of Business, Innovation and Employment, New Zealand; the Pakistan Atomic Energy Commission; the Ministry of Science and Higher Education and the National Science Centre, Poland; the Funda\c{c}\~ao para a Ci\^encia e a Tecnologia, Portugal; JINR, Dubna; the Ministry of Education and Science of the Russian Federation, the Federal Agency of Atomic Energy of the Russian Federation, Russian Academy of Sciences and the Russian Foundation for Basic Research; the Ministry of Education, Science and Technological Development of Serbia; the Secretar\'{\i}a de Estado de Investigaci\'on, Desarrollo e Innovaci\'on, Programa Consolider-Ingenio 2010, Plan Estatal de Investigaci\'on Cient\'{\i}fica y T\'ecnica y de Innovaci\'on 2013-2016, Plan de Ciencia, Tecnolog\'{i}a e Innovaci\'on 2013-2017 del Principado de Asturias and Fondo Europeo de Desarrollo Regional, Spain; the Swiss Funding Agencies (ETH Board, ETH Zurich, PSI, SNF, UniZH, Canton Zurich, and SER); the Ministry of Science and Technology, Taipei; the Thailand Center of Excellence in Physics, the Institute for the Promotion of Teaching Science and Technology of Thailand, Special Task Force for Activating Research and the National Science and Technology Development Agency of Thailand; the Scientific and Technical Research Council of Turkey, and Turkish Atomic Energy Authority; the National Academy of Sciences of Ukraine, and State Fund for Fundamental Researches, Ukraine; the Science and Technology Facilities Council, UK; the US Department of Energy, and the US National Science Foundation.

Individuals have received support from the Marie-Curie programme and the European Research Council and Horizon 2020 Grant, contract No. 675440 (European Union); the Leventis Foundation; the A. P. Sloan Foundation; the Alexander von Humboldt Foundation; the Belgian Federal Science Policy Office; the Fonds pour la Formation \`a la Recherche dans l'Industrie et dans l'Agriculture (FRIA-Belgium); the Agentschap voor Innovatie door Wetenschap en Technologie (IWT-Belgium); the F.R.S.-FNRS and FWO (Belgium) under the ``Excellence of Science - EOS" - be.h project n. 30820817; the Ministry of Education, Youth and Sports (MEYS) of the Czech Republic; the Lend\"ulet (``Momentum") Programme and the J\'anos Bolyai Research Scholarship of the Hungarian Academy of Sciences, the New National Excellence Program \'UNKP, the NKFIA research grants 123842, 123959, 124845, 124850 and 125105 (Hungary); the Council of Scientific and Industrial Research, India; the HOMING PLUS programme of the Foundation for Polish Science, cofinanced from European Union, Regional Development Fund, the Mobility Plus programme of the Ministry of Science and Higher Education, the National Science Center (Poland), contracts Harmonia 2014/14/M/ST2/00428, Opus 2014/13/B/ST2/02543, 2014/15/B/ST2/03998, and 2015/19/B/ST2/02861, Sonata-bis 2012/07/E/ST2/01406; the National Priorities Research Program by Qatar National Research Fund; the Programa de Excelencia Mar\'{i}a de Maeztu and the Programa Severo Ochoa del Principado de Asturias; the Thalis and Aristeia programmes cofinanced by EU-ESF and the Greek NSRF; the Rachadapisek Sompot Fund for Postdoctoral Fellowship, Chulalongkorn University and the Chulalongkorn Academic into Its 2nd Century Project Advancement Project (Thailand); the Welch Foundation, contract C-1845; and the Weston Havens Foundation (USA).
\end{acknowledgments}

\bibliography{auto_generated}
\cleardoublepage \appendix\section{The CMS Collaboration \label{app:collab}}\begin{sloppypar}\hyphenpenalty=5000\widowpenalty=500\clubpenalty=5000\vskip\cmsinstskip
\textbf{Yerevan~Physics~Institute, Yerevan, Armenia}\\*[0pt]
A.M.~Sirunyan, A.~Tumasyan
\vskip\cmsinstskip
\textbf{Institut~f\"{u}r~Hochenergiephysik, Wien, Austria}\\*[0pt]
W.~Adam, F.~Ambrogi, E.~Asilar, T.~Bergauer, J.~Brandstetter, E.~Brondolin, M.~Dragicevic, J.~Er\"{o}, A.~Escalante~Del~Valle, M.~Flechl, M.~Friedl, R.~Fr\"{u}hwirth\cmsAuthorMark{1}, V.M.~Ghete, J.~Grossmann, J.~Hrubec, M.~Jeitler\cmsAuthorMark{1}, A.~K\"{o}nig, N.~Krammer, I.~Kr\"{a}tschmer, D.~Liko, T.~Madlener, I.~Mikulec, E.~Pree, N.~Rad, H.~Rohringer, J.~Schieck\cmsAuthorMark{1}, R.~Sch\"{o}fbeck, M.~Spanring, D.~Spitzbart, A.~Taurok, W.~Waltenberger, J.~Wittmann, C.-E.~Wulz\cmsAuthorMark{1}, M.~Zarucki
% ...
\vskip\cmsinstskip
% ...
1:~Also at~Vienna~University~of~Technology, Vienna, Austria\\
% ...
\end{sloppypar}
\end{document}